\pdfoutput=1
\documentclass[11pt,letterpaper]{article}

\usepackage{amsmath}
\usepackage{amsfonts}
\usepackage{amssymb}
\usepackage{graphicx}
\usepackage[hang,small,bf]{caption}

\DeclareRobustCommand{\SkipTocEntry}[4]{}

\def\beq{\begin{equation}}
\def\eeq{\end{equation}}
\def\bea{\begin{eqnarray}}
\def\eea{\end{eqnarray}}
\def\d{{\rm d}}

\newcommand{\WMAP}{\textit{WMAP}}
\newcommand{\Planck}{\textit{Planck}}
\newcommand{\um}{\mu\mathrm{m}}
\newcommand{\bJ}{\mathbf{J}}
\usepackage{epsfig}
\usepackage{url}
\usepackage[hcentering,vcentering,hscale=0.7255,vscale=0.83]{geometry}
\newcommand{\Hammurabi}{\textsc{Hammurabi}}
\newcommand{\rme}{\mathrm{e}}
\newcommand{\rmK}{\mathrm{K}}
\newcommand{\br}{\mathbf{r}}
\usepackage{color}
\newcommand{\red}[1]{\textcolor{black}{#1}}
\newcommand{\rms}{\mathrm{s}}
\newcommand{\bB}{\mathbf{B}}
\newcommand{\dbl}{\mathrm{d}\mathbf{l}}
\newcommand{\rad}{\mathrm{rad}}
\newcommand{\RM}{\mathrm{RM}}
\newcommand{\rmm}{\mathrm{m}}
\newcommand{\rmi}{\mathrm{i}}
\newcommand{\blue}[1]{\textcolor{black}{#1}}
\newcommand{\green}[1]{\textcolor{black}{#1}}
\newcommand{\Ulysses}{\textit{Ulysses}}
\newcommand{\Helios}{\textit{Helios}}
\newcommand{\Cassini}{\textit{Cassini}}
\def \micron {$\mu$m}
\def \cmtwo {${\rm cm}^{-2}$}
\def \gtsim {\gtrsim}
\newcommand{\nred}[1]{\textcolor{black}{#1}}
\newcommand{\nblue}[1]{\textcolor{black}{#1}}
\newcommand{\gdred}[1]{\textcolor{black}{#1}}
\newcommand{\pcfred}[1]{\textcolor{black}{#1}}
\newcommand{\btdred}[1]{\textcolor{black}{#1}}
\newcommand{\mhred}[1]{\textcolor{black}{#1}}
\newcommand{\gdtwo}[1]{\textcolor{black}{#1}}
\newcommand{\alred}[1]{\textcolor{black}{#1}}
\newcommand{\jacbred}[1]{\textcolor{black}{#1}}
\newcommand{\ammred}[1]{\textcolor{black}{#1}}
\newcommand{\awred}[1]{\textcolor{black}{#1}}
\newcommand{\sdred}[1]{\textcolor{black}{#1}}
\newcommand{\rjred}[1]{\textcolor{black}{#1}}
\newcommand{\dnsred}[1]{\textcolor{black}{#1}}

\usepackage{natbib}
\input{journals.cls}

\begin{document}

\vspace{10mm}

\begin{center}

{\large CMBPol Mission Concept Study} \\
\vskip 15pt {\Large Foreground Science Knowledge and Prospects}
\\[1.0cm]

{Aur\'elien~A.~Fraisse$^{\rm a,*}$,
Jo-Anne~C.~Brown$^{\rm b}$, 
Gregory~Dobler$^{\rm c}$,
Jessie~L.~Dotson$^{\rm d}$,
\btdred{Bruce~T.~Draine$^{\rm a}$,}
Priscilla~C.~Frisch$^{\rm e}$,
Marijke~Haverkorn$^{\rm f,g,h}$, 
Christopher~M.~Hirata$^{\rm i}$, 
Ronnie~Jansson$^{\rm j}$, 
Alex~Lazarian$^{\rm k}$, 
\ammred{Antonio}~Mario~Magalh\~aes$^{\rm l}$,
Andr\'e~Waelkens$^{\rm m}$, and
Maik~Wolleben$^{\rm n}$}
\\[0.5cm]

{\small
\textit{$^{\rm a}$Princeton University Observatory, Peyton Hall, Princeton, NJ 08544, USA}}

{\small
\textit{$^{\rm b}$\jacbred{Centre for Radio Astronomy}, University of Calgary,
Calgary, AB T2N 1N4, Canada}}

{\small
\textit{$^{\rm c}$Harvard-Smithsonian Center for Astrophysics, 60 Garden Street, Cambridge, MA 02138, USA}}

{\small
\textit{$^{\rm d}$NASA Ames Research Center, Moffet Field, CA 94035, USA}}

{\small
\textit{$^{\rm e}$University of Chicago, 5460 South Ellis Avenue, Chicago, IL 60637, USA}}

{\small
\textit{$^{\rm f}$\mhred{Jansky Fellow,} National Radio Astronomy Observatory, Charlottesville, VA 22903, USA}}

{\small
\textit{$^{\rm g}$Astronomy Department, UC Berkeley, 601 Campbell Hall, Berkeley, CA 94720, USA}}

{\small
\textit{\mhred{$^{\rm h}$ASTRON, P.O. Box 2, 7990 AA Dwingeloo, the Netherlands}}}

{\small
\textit{$^{\rm i}$Caltech, Mail Code 130-33, Pasadena, CA 91125, USA}}

{\small
\textit{$^{\rm j}$Center for Cosmology and Particle Physics, Department of Physics, NYU, NY, NY 10003, USA}}

{\small
\textit{$^{\rm k}$Department of Astronomy, University of Wisconsin-Madison, Madison, WI 53706, USA}}

{\small
\textit{$^{\rm l}$\ammred{IAG,} Universidade de S\~{a}o Paulo, Rua do Mat\~{a}o 1226, S\~{a}o Paulo, SP 05508-900, Brazil}}

{\small
\textit{$^{\rm m}$Max-Planck-Institut f{\"u}r Astrophysik, Garching 85741, Germany}}

{\small
\textit{\nred{$^{\rm n}$Covington~Fellow,}~NRC~Herzberg~Institute~of~Astrophysics,~DRAO,~Penticton,~BV~V2A~6J9,~Canada}}
\\[0.5cm]

{\small $^*$\url{fraisse@astro.princeton.edu}}

\vspace{1cm}

\end{center}

\hrule \vspace{0.3cm}
{\small  \noindent \textbf{Abstract} \\[0.3cm]
\noindent
Detecting ``B-mode" (i.e., divergence free) polarization in the Cosmic Microwave Background (CMB) would open a new window on the very early Universe.  However, the polarized microwave sky is dominated by polarized Galactic dust and synchrotron emissions, which may hinder our ability to test inflationary predictions.  In this paper, we report on our knowledge of these ``Galactic foregrounds," as well as on how a CMB satellite mission aiming at detecting a primordial B-mode signal (``CMBPol") will contribute to improving it. 
We review the observational and analysis techniques used to constrain the structure of the Galactic magnetic field, whose presence is responsible for the polarization of Galactic emissions.  Although \jacbred{our} current \jacbred{understanding} 
of the magnetized interstellar medium is
\jacbred{somewhat limited},
dramatic improvements in our knowledge of its properties are expected by the time CMBPol flies.
Thanks to high resolution and high sensitivity instruments observing the whole sky at frequencies between 30~GHz and 850~GHz, CMBPol will not only improve this picture by observing the synchrotron emission from our galaxy, but also help constrain dust models.  Polarized emission from interstellar dust indeed dominates over any other signal in CMBPol's highest frequency channels.  Observations at these wavelengths, combined with ground-based studies of starlight polarization, will therefore enable us to improve our understanding of dust properties and of the mechanism(s) responsible for the alignment of dust grains with the Galactic magnetic field.
CMBPol will also shed new light on observations
\jacbred{that are presently not well} understood.
\jacbred{M}orphological studies of anomalous dust and synchrotron emissions will \jacbred{indeed}
constrain
their natures and properties, while \pcfred{searching for}
fluctuations in the emission from heliospheric dust will test our understanding of the circumheliospheric interstellar medium.  Finally, 
acquiring more information on the properties of
extra-Galactic sources will be necessary in order to maximize the cosmological constraints extracted from CMBPol's observations of CMB lensing.}
\vspace{0.5cm}  \hrule
\def\thefootnote{\arabic{footnote}}
\setcounter{footnote}{0}
\vspace{0.9cm}

\tableofcontents

\vspace{1.5cm}
\hrule

\section{Introduction}

Over the course of the last two decades, studies of the Cosmic Microwave Background (CMB) have revolutionized our understanding of the composition, structure and evolution of the Universe.  After \jacbred{sixteen} years of observations of the CMB temperature anisotropies first detected by the COBE satellite in the early 1990's \citep{Smoot+Bennett+Kogut_etal_1992}, we now have a standard cosmological model fully described by a handful of parameters, and in spectacular agreement with a broad range of independent astrophysical observations \citep{Komatsu+Dunkley+Nolta_etal_2008}.

The cornerstone of this achievement is doubtless
inflation.  This paradigm\jacbred{,} initially designed to explain the observed flatness and isotropy of the Universe \citep{Guth_1981}\jacbred{,} provides a natural explanation for several of its other properties, including the origin of the primordial density fluctuations that led to structure formation.
A generic prediction of the most natural inflationary models is the existence of primordial gravitational waves that should have left their imprint on the CMB anisotropies in the form of a divergence-free polarization pattern known as ``B-mode polarization."  However, this signal is expected to have a very low amplitude, typically more than three to four orders of magnitude smaller than the measured CMB temperature fluctuations \citep[for a review of inflationary predictions, see the companion paper by][]{CMBPol_Baumann_2008}.  Detecting \jacbred{primordial B-modes}
is therefore a challenging enterprise made \jacbred{only more difficult}
by polarized emissions from our galaxy and others.  Among these so-called ``polarized foregrounds" are polarized synchrotron and dust emissions from the Galaxy, which dominate 
the expected primordial B-mode signal
over the whole sky
 at all frequencies of interest \citep[see, e.g., the discussion in][]{Page+Hinshaw+Komatsu_etal_2007}.  \jacbred{Thus, b}eing able to properly handle them is central to being able to detect low amplitude B-modes in the~CMB.
 
In this document, we do not address the question of how to look behind this Galactic (and extra-Galactic) screen, but focus on our knowledge of the foregrounds themselves. 
\jacbred{Our ability to use the data provided by a  satellite mission 
sensitive enough to detect B-modes of inflationary origin 
(hereafter, CMBPol) is indeed limited by our understanding of foregrounds \citep[see the companion paper by][]{CMBPol_Dunkley_2008}.
We therefore review what we know now, how we know it, and how we hope to improve it by the time CMBPol flies.}
\jacbred{Moreover, since} Galactic foregrounds
\jacbred{are}
the dominant signal CMBPol will observe, we can expect our understanding of these emissions and their underlying physical processes to significantly improve through an analysis of the CMBPol data. 
Although our current knowledge of foregrounds makes it difficult to quantify many of these possible improvements, we
review
what we think are the main ways in which CMBPol will help us explore the magnetized~Galactic~interstellar~medium.

Four companion papers complement the contents of this document\jacbred{:}  \citet{CMBPol_Baumann_2008} review the inflationary science case for CMBPol\jacbred{;}
\citet{CMBPol_Smith_2008} report on CMB lensing\jacbred{;}
\citet{CMBPol_Zaldarriaga_2008} \jacbred{discuss}
reionization\jacbred{;} and
\citet{CMBPol_Dunkley_2008} provide a detailed analysis of how and how well a primordial B-mode signal can be separated from the polarized emissions from the Galaxy.  An executive summary of these documents is also available \citep{CMBPol_Summary_2008}.
For the purposes of this paper and all companion documents, we assume that CMBPol will have specifications close to those provided in Tab.~\ref{t:epic_specs}, which fall between the specifications of the two EPIC designs proposed in the \citet{Bock+Cooray+Hanany_etal_2008} report.

~

\begin{table}[h]
\begin{center}
\caption{\label{t:epic_specs} Possible CMBPol$^{\rm a}$ Specifications}
\begin{tabular}{ll|c|c|c|c|c|c|c|c|c}
\hline
\hline
Frequency & (GHz) & 30 & 45 & 70 & 100 & 150 & 220 & 340 & 500 & 850\\
\hline
Resolution & (FWHM in arcmin) & 26 & 17 & 11 & 8 & 5 & 3.5 & 2.3 & 1.5 & 0.9\\
\hline
Sensitivity$^{\rm b}$ & (nK CMB) & 160 & 69 & 35 & 27 & 26 & 40 & 180 & 150$^{\rm c}$ & 400$^{\rm d}$\\
\hline
\hline
\end{tabular}
\vspace{0.15cm}

{\small$^{\rm a}$Expected mission lifetime of 4 years}
\vspace{-0.15cm}

{\small$^{\rm b}$CMB temperature fluctuations detectable at 1$\sigma$ in a $2^\circ \times 2^\circ$ pixel}
\vspace{-0.15cm}

{\small$^{\rm c}$Could be improved by a factor of 5 by cooling the telescope to 4~K}
\vspace{-0.15cm}

{\small$^{\rm d}$Could be improved by a factor of 40 by cooling the telescope to 4~K}
\end{center}
\end{table}

This document is organized as follows. In Sec.~\ref{s:gmf}, we review the observational techniques used to probe the Galactic magnetic field (Sec.~\ref{s:obs}), as well as what they teach us about its main properties (Sec.~\ref{s:bfld-current}), while Sec.~\ref{s:mim} presents an analysis technique aimed at improving our constraints on the magnetized interstellar medium\jacbred{.}
Sec.~\ref{s:cmbpolera} discusses the progress one can expect from upcoming radio (Sec.~\ref{s:radio_next}) and CMB (Sec.~\ref{ss:gmf_cmb}) observations, including those by CMBPol.  
Our knowledge of the polarized emission from interstellar dust is then summarized in Sec.~\ref{s:pde}, and followed by a description of the physical mechanism responsible for it in Sec.~\ref{s:gam}.  Heliospheric dust emission, and ``anomalous emissions" detected in the \WMAP~data, are the topics of, respectively, Sec.~\ref{s:hde} and Sec.~\ref{s:had}.  Finally, Sec.~\ref{s:egs} gives an overview of the foreground situation in the context of CMB lensing studies, and our conclusions are presented in Sec.~\ref{s:conclusion}.

\section{Probing the Galactic Magnetic Field}
\label{s:gmf}

The idea that the Galaxy has a magnetic field, and that it may play an important role in Galactic physics, dates back to \citet{Fermi_1949}.  He proposed that cosmic rays might be generated outside of the solar system as a result of energetic particles colliding with moving irregularities in this field.  In his view, the magnetic field would not only be a cosmic ray generator, but also a containment factor preventing the rays from escaping the Galaxy.

It is now believed that magnetic fields and cosmic rays indeed contribute to the vertical support of the gas in the Galaxy, with the magnetic field energy density dominating the turbulent and thermal gas energy densities at high Galactic latitudes \citep{Boulares+Cox_1990}.  However, astronomical magnetic fields also play important roles in many other processes affecting the structure of the Galaxy.  In star formation, they can inhibit the gravitational collapse of interstellar clouds (primary star formation regions) and help removing prestellar angular momentum \citep{Zweibel+Heiles_1997}, thereby affecting the distribution of stars in the Galaxy.  It is also believed that magnetic fields influence galaxy formation and evolution by causing large density fluctuations \jacbred{which} result
in structures within galaxies \citep{Kim+Olinto+Rosner_1996}.

Yet
despite the Galactic magnetic field's importance, many mysteries remain about its generation and evolution.
\jacbred{A} dynamo mechanism is believed to be necessary to increase galactic magnetic field strengths from that of a tiny primordial seed field to the observed values
\citep[see, e.g.,][]{Beck+Brandenburg+Moss_etal_1996}, but neither the exact dynamo mechanism nor the generation and strength of the seed field are 
\jacbred{significantly}
constrained by \jacbred{available} observations.  External galaxies appear to have a wide range of magnetic field characteristics \citep{Beck_2001}, indicating that different dynamos must be acting in different galaxies.  Furthermore, in some cases, rapid field amplification seems to be required
\citep{Gaensler+Haverkorn+Staveley-Smith_etal_2005}.
\jacbred{B}efore we can hope to fully explore these aspects of the field, we must first understand the field's general topology.

While observations of polarized emission from interstellar dust grains, polarization of starlight, and Zeeman splitting can provide in depth probing of the field in localized regions, most of the knowledge we have about the magnetic field across the bulk of the Galaxy comes from radio observations of linearly-polarized radiation.  At frequencies above a few GHz, maps of radio polarization angles directly yield the direction of the magnetic field component in the plane of the sky, while the total emission indicates its strength.  At lower frequencies, Faraday rotation of polarized emission traveling through the interstellar medium (ISM) can be used to probe the magnetic field component parallel to the line-of-sight.  We discuss these techniques in Sec.~\ref{s:obs}, before detailing the insights gained from these observations in Sec.~\ref{s:bfld-current}.

\subsection{Observational Techniques}
\label{s:obs}

\subsubsection{Radio Synchrotron Radiation}
\label{s:synch}

\btdred{R}elativistic \btdred{electrons} moving
in the Galactic magnetic field
emit synchrotron radiation.
\jacbred{O}pti\-cal\-ly thin synchrotron emission, $\epsilon_\rms$, at frequency $\nu$ depends on the component \jacbred{of the} uniform magnetic field perpendicular to the line-of-sight\jacbred{,} $B_\perp$\jacbred{,} according to
$\epsilon_\rms \propto {B_\perp}^{(p+1)/2} \, \nu^{-(p-1)/2}$
\citep{Rybicki+Lightman_1979}.
\jacbred{The exponent $p$ describes the} power law energy spectrum of \jacbred{the}
ensemble of relativistic electrons
\jacbred{through the relation} $N(E) \propto E^{-p}$.
This emission therefore provides information about the slope of the electron energy spectrum, as well as an estimate of the strength of the magnetic field if assumptions are made about the volume of the source and the energy density of cosmic rays.  The minimum combined energy density of cosmic ray particles and magnetic field necessary to reproduce the emission from an observed source is similar to the equipartition energy density between magnetic field and cosmic rays.  Magnetic field estimates from the brightness of synchrotron emission usually assume this minimum energy condition, or equipartition, though often without clear observational evidence (through $\gamma$- or X-ray data) that it applies \citep[see, e.g., the discussion in][]{Beck+Krause_2005}.

Synchrotron emission from a region with a uniform magnetic field and an electron distribution with $p=3$ (respectively, $p=2$) has a theoretical polarization of $75\%$ (respectively, 60\%), with the plane of polarization perpendicular to the direction of $B_\perp$ in the plane of the sky
\citep{Rybicki+Lightman_1979}.  Polarization of synchrotron emission therefore also gives information on the magnetic field component perpendicular to the line-of-sight.  In practice, the observed emission is integrated over large regions in space with a complicated magnetic field structure and an ionized plasma present in addition to the relativistic electrons.  Consequently, the integrated emission is usually much less polarized than the theoretical \jacbred{limit}.

Although maps of the synchrotron emission can inform us on the strength and direction of the component of the Galactic magnetic field perpendicular to the line-of-sight, they do not encode any information related to the component parallel to the line-of-sight.  The latter can be provided by Faraday rotation studies (Sec.~\ref{ss:faraday}) or rotation measure synthesis
(Sec.~\ref{s:rmsyn}).

\vspace{0.3cm}
\subsubsection{Faraday Rotation of Compact Sources}
\label{ss:faraday}

When a linearly polarized electromagnetic wave propagates through a region of free electrons permeated by a magnetic field (e.g., a magnetized plasma such as the ISM), its plane of polarization rotates.  In the quasi-longitudinal (QL) approximation \citep{Ratcliffe_1959}, which assumes the field direction is roughly along the line-of-sight, the Faraday rotation angle (in radians) at wavelength $\lambda$ (in meters) is $\Psi \simeq \lambda^2\,\RM$, where
$\RM$, the ``rotation measure," is
\begin{equation}
\label{eq:ql}
\RM \equiv 0.812 \int n_\rme \, \bB \cdot \dbl \;\; (\rad \, \rmm^{-2})\,,
\end{equation}
with $n_\rme$ the electron density in units of  cm$^{-3}$, $\bB$ the magnetic field in units of $\mu$G, and $\dbl$ the incremental pathlength \jacbred{from the source to the receiver} in units of pc.
At radio frequencies\jacbred{, with typical Galactic magnetic field strengths and electron densities},
the QL approximation \jacbred{can be shown to} hold
for \jacbred{fields which are essentially perpendicular to}
the line-of-sight.  \jacbred{Consequently, Eq.~(\ref{eq:ql}) is actually valid for most ISM applications.}

If we assume that at all wavelengths 
the polarized emission 
from a given source is emitted at the same \jacbred{initial} polarization angle $\btdred{\varphi}_\circ$, and that
the radiation is only affected by Faraday rotation, the detected polarization
angle $\btdred{\varphi}$ is given by the linear relationship
$\btdred{\varphi} = \btdred{\varphi}_\circ + \lambda^2 \, \RM$. 
Measurements of $\btdred{\varphi}$ at multiple wavelengths
can therefore determine the RM for a given source as the slope of the graph of $\btdred{\varphi}$ vs.~$\lambda^2$.  In practice, at least three frequency channels are needed to
properly do this exercise.  If many more
are available, RMs may also be calculated by using a RM transfer function (RMTF)
\citep[see][and Sec.~\ref{s:rmsyn}]{Brentjens+de-Bruyn_2005}.     

The simplicity with which RMs can be determined, coupled with the significance
of their signs, makes them powerful tools with which one can probe
the ISM magnetic field.  Pulsars and extragalactic sources (EGS) emit
linearly polarized radiation and are
\jacbred{typically} used to study the Galactic magnetic field.  Since it is possible 
to estimate their distances to us,
we can work backwards and find out what the field looks like along
lines-of-sight to them, provided we make some assumptions about the electron density distribution within
the Galaxy.

When pulsar\mhred{s}
are used, studying the dispersion of the emitted pulses \mhred{provides information to replace some of these assumptions}.
Different frequencies propagate at different group velocities in the interstellar medium.  As a result, a pulse emitted by a given pulsar is not received at the same time at different frequencies.  The arrival time $\tau$ at frequency $\nu$ is such that
\begin{equation}
\frac{\d\tau}{\d\nu} \propto -\frac{1}{\nu^3}\,\int n_\rme\,\dbl\,,
\end{equation}
which is a measurable quantity.  Since the positive proportionality constant is a combination of known physical constants, it is possible to compute the value of the integral, called dispersion measure (DM), and therefore to estimate the integrated electron density along the line-of-sight.

By measuring
rotation measures for the highest density of sources possible, and over a wide range of Galactic longitudes, one can thus accurately reconstruct 
the large-scale Galactic magnetic field in the disk.  
Much effort has been put into these studies, and experiments such as the Canadian Galactic Plane Survey \citep{Taylor+Gibson+Peracaula_etal_2003} and the Southern Galactic Plane Survey \citep{McClure-Griffiths+Dickey+Gaensler_etal_2005,Haverkorn+Gaensler+McClure-Griffiths_etal_2006} have now observed hundreds of square degrees near the Galactic plane with arcminute resolution.

\subsubsection{Rotation Measure Synthesis}
\label{s:rmsyn}

Rotation measure synthesis, also called Faraday tomography, is a
novel technique for 3-D mapping of RM structures in the ISM.  The idea
was described by \citet{Burn_1966} decades ago, but only now are spectropolarimetric capabilities
advanced enough that we can use this technique
\citep{Brentjens+de-Bruyn_2005}.
The Faraday depth
of a rotating medium is defined as 
\begin{equation}
\phi \equiv 0.812 \int_{x_1}^{x_2}
n_\rme\, {\mathbf B}\cdot\dbl \;\; (\rad \, \rmm^{-2})\,,
\end{equation}
where the integral runs over a particular \jacbred{segment}
of the line-of-sight,
from $x_1$ to $x_2$, and $n_\rme$, $\bB$, and $\dbl$ have the same meaning and units as in Sec.~\ref{ss:faraday}.
In general, $\phi \neq \RM$.
In addition, one can define the observed polarized flux density function as
$\bar{P}(\lambda^2) \equiv W(\lambda^2)\,P(\lambda^2)$, where $P(\lambda^2)$
is the complex polarized surface brightness of the source as a function of
wavelength, and $W(\lambda^2)$ is a sampling
function which is non-zero only at the wavelengths where measurements
are taken. Fourier transformation of the polarized surface
brightness as a function of $\lambda^2$ gives $F(\phi)$, the
complex polarized surface brightness per unit Faraday depth $\phi$ through
\begin{equation}
F(\phi) = \frac{1}{\pi} \int_{-\infty}^{+\infty} P(\lambda^2)\,\rme^{-2\rmi\phi\lambda^2} \d(\lambda^2)\,.
\end{equation}

Thus, provided one can account for the effect of the sampling function, which is possible when many wavelengths are observed \citep{Brentjens+de-Bruyn_2005}, and make reasonable assumptions as to the definition of $P$ for $\lambda^2 \leq 0$ \citep{Burn_1966}, RM synthesis provides a 3-D map of polarized intensity, where the third dimension is Faraday depth. However, Faraday depth does not
necessarily increase monotonically with distance, as significant changes in field
direction, electron density, or magnetic field strength will also impact its value, and 
there is therefore no direct distance information in these maps.  Morphological correspondences with structures indicating a dramatic change in the magneto-ionic medium along the observed line-of-sight, such as spiral arms, supernova remnants, or external
galaxies, can provide this missing link.
Since the traditional way of measuring RMs only provides
one RM value for each line-of-sight, and none where depolarization is
severe, this new technique provides a tremendous increase in the
amount of obtainable information from radio spectropolarimetric
surveys.

\subsubsection{Other Techniques}
\label{s:other}

\paragraph{\it Polarized Dust Emission}
\label{s:dust}

Non-spherical dust grains can partially align under the influence of
an external magnetic field, which causes the emission from interstellar dust at far-infrared and submm wavelengths to be polarized.  Although
tremendous progress has been made in recent years \citep[see, e.g., the review by][]{Lazarian_2007}, the grain alignment mechanisms are not yet fully understood. As the degree of polarization of this emission
depends not only on the magnetic field strength but also on dust
properties and alignment mechanisms, observations of polarized emission
from interstellar dust do not currently allow the magnetic field
strength to be directly inferred. However, the magnetic field direction averaged over the line-of-sight can be measured, as the polarization angle is perpendicular to the component of the magnetic field in the plane of the sky.  This method has so far been mostly used in denser
regions of the Galaxy, such as molecular clouds and cores \citep[see, e.g.,][]{Vaillancourt_2007}.  In these environments, a rough estimate of the strength of the magnetic field
can be obtained with the Chandrasekhar-Fermi method \citep{Chandrasekhar+Fermi_1953}
by comparing the dispersion in polarization angle to the
turbulent velocity dispersion. Numerical simulations suggest that
results from this technique should be corrected by factors of order unity \citep{Heitsch+Zweibel+Mac-Low_etal_2001}.

\paragraph{\it Starlight Polarization}
\label{s:starlight_gmf}

When non-spherical dust grains are (partially) aligned with the Galactic magnetic field,  the dusty region they are part of acts like a polarizer.  As absorption of radiation by dust grains is more efficient in the direction of their long axes, initially unpolarized starlight illuminating the region becomes polarized in the direction of the grains' short axes, i.e., mostly parallel to the Galactic magnetic field \citep[see][for a more detailed discussion]{Lazarian_2007}.  By observing the polarization of the remaining
starlight, inferences can therefore be made about the Galactic magnetic field required to
produce the observed polarization close to the Sun, or, more generally, in regions of low dust extinction \citep[see, e.g.,][]{Fosalba+Lazarian+Prunet_etal_2002}.

\paragraph{\it Zeeman Splitting}

In the presence of a 
magnetic field, degeneracies in atomic or molecular 
energy levels can be lifted.  The energy separation between the \jacbred{previously degenerate} Zeeman sub-levels
is proportional to the strength of the magnetic field.  As a result,
a sufficiently large field must be present for the corresponding
separation in a given radio-frequency line for a given species 
of atom or molecule to be detectable.  In general, the magnitude of the interstellar magnetic 
field ($\sim 10$~$\mu$G) is not strong enough to generate detectable Zeeman splitting \citep[see, e.g.,][]{Davies+Slater+Shuter_etal_1960}
outside of localized interstellar clouds where the field is at least $\sim 50$~$\mu$G \citep[see, among others, the review by][]{Davies_2007}.

That being said, recent studies of Zeeman splitting from OH masers in massive star forming 
regions \citep{Fish+Reid+Argon_etal_2003} and within molecular
clouds \citep{Han+Zhang_2007} have suggested that the fields
within these regions are influenced by, and even reminiscent of, the 
large-scale Galactic field.  While these results seem to be consistent
with observations from other methods, more data is required before it can be considered a new way of constraining the Galactic field.

Although Zeeman splitting can provide information on the strength of the Galactic magnetic field, it does not encode information related to its direction.  Observations of the Gold\-reich-Ky\-la\-fis effect \citep{Goldreich+Kylafis_1981}, which leads to a small linear polarization of interstellar radio-frequency lines, could complement Zeeman constraints from molecular clouds and star forming regions by providing the direction of the sky-projected component of the field.

\paragraph{\it UHECR Multiplets}

The deflection angle of a ultrahigh energy cosmic ray (UHECR) is independent of the Galactic electron density, and is proportional to its charge $Z$ and the transverse magnetic field, while being inversely proportional to its energy $E$. Thus, UHECRs originating from the same source will have their arrival directions appear, roughly, as a string on the sky, with the cosmic rays with higher value of $Z/E$ further from the true source direction.  By detecting these multiplets, one can therefore estimate the direction of the source and the strength of the integrated transverse magnetic field along the line-of-sight \citep{Jansson+Farrar+Waelkens_etal_2007}.  This offers a direct way of constraining models of the Galactic magnetic field without having to rely on a model for the Galactic electron density (on this topic, see also Sec.~\ref{ss:ismconstraints}).

Although using this technique will have to wait until a bigger UHECR data set is available, it is worth noting that the Pierre Auger collaboration \citep{Auger_2007}, after 1.2 years of observations, found correlations between 28 cosmic ray events above 60 EeV and the positions of nearby AGNs.  In ten years, over 200 events at that energy, including many multiplets, can therefore be expected in the Auger data, which could lead to potentially interesting constraints.

\subsection{Current Knowledge}
\label{s:bfld-current}

For practical reasons, the Galactic magnetic field is usually discussed
as consisting of two components: the large-scale and the small-scale
field. The large-scale, or ``regular," magnetic field is predominantly
azimuthal in the plane of the Milky Way, as clearly shown by the
direction of starlight polarization by interstellar dust \citep{Heiles_2000} and
confirmed with the lowest-frequency maps in the \WMAP~3-year polarization data \citep{Page+Hinshaw+Komatsu_etal_2007}. The
small-scale, or ``random," component is of comparable strength as the regular
field, exhibits structure over a wide range of physical scales, and is
presumably turbulent (see discussion in Sec.~\ref{ss:randomgmf}).

The average strength of the total magnetic field, i.e., regular and
random components together, can be derived from observations of the 
Galactic synchrotron emission, as discussed in Sec.~\ref{s:synch}.  Under the assumption of
equipartition, the total magnetic field strength is estimated at
$6 \pm 2~\mu$G in the solar neighborhood, and is found to increase to $10 \pm 3~\mu$G
at a Galactocentric radius of 3~kpc \citep{Beck_2001}.  The equipartition
assumption used in these estimates has been validated by measurements of the local $\gamma$-ray
density, which yields comparable field strengths as a function of radius \citep{Strong+Moskalenko+Reimer_2000}.  These results are consistent with those of 
\citet{Heiles_1996_a}, who combined various methods to conclude that the total magnetic field strength is about 6~$\mu$G in the
spiral arms and 4~$\mu$G elsewhere.  Near the Galactic center, the
magnetic field strength increases to several mG \citep{Yusef-Zadeh+Roberts+Goss_etal_1996}.

In Sec.~\ref{ss:lsgmf}, we review the properties of the regular component of the Galactic field, whereas our knowledge of the random field is presented in Sec.~\ref{ss:randomgmf}.  A Galactic halo field is also required to explain observations at low radio frequencies, which we discuss in Sec.~\ref{ss:hgmf}.

\subsubsection{The Large-Scale Magnetic Field in the Galactic Disk}
\label{ss:lsgmf}

The local regular field strength estimated from pulsar RMs
combined with their dispersion measures (DMs) is about $1.4\, \mu$G
\citep[see, e.g.,][]{Rand+Lyne_1994}, while estimates from radio synchrotron
intensity give higher values of $\sim 4\,\mu$G.  Because
of uncertainties and assumptions in both methods \citep{Beck+Shukurov+Sokoloff_2003},
these results are not necessarily in contradiction.

The large-scale magnetic field is observed to be concentrated in the Galactic
disk \citep[see, e.g.,][]{Simard-Normandin+Kronberg_1980, Han+Qiao_1994}, and is generally assumed to be 
closely aligned with the Galactic spiral arms, consistent with observations
of external galaxies \citep{Beck_2001}.  
One feature that is critical in the determination of the dynamo mode(s)
operating in the Galaxy is the presence of magnetic field reversals, regions 
of magnetic shear where the field direction changes by roughly $180^\circ$.
One large-scale field
reversal unquestionably exists between the local arm and
the Sagittarius-Carina spiral arm of our galaxy, in the first quadrant \citep{Simard-Normandin+Kronberg_1979,Thomson+Nelson_1980}.
There are indications that this reversal slices through the
Sagittarius-Carina arm near $\ell = 0$ \citep{Weisberg+Cordes+Kuan_etal_2004,Brown+Haverkorn+Gaensler_etal_2007},
suggesting a significantly different inclination of the field from the
spiral arms, at least in this region.  The existence of additional
reversals is highly controversial
\citep[see, e.g.,][]{Vallee_2002,Weisberg+Cordes+Kuan_etal_2004,Han+Manchester+Lyne_etal_2006,Brown+Haverkorn+Gaensler_etal_2007}, and recent modeling
based on RM data \citep{Noutsos+Johnston+Kramer_etal_2008,Men+Ferriere+Han_2008}, as well as on a combination of Faraday
rotation and diffuse Galactic
synchrotron emission measurements (see Sec.~\ref{ss:ismconstraints}), all conclude that the current data
is not sufficient to conclusively distinguish between several
large-scale magnetic field models predicting differing numbers and
locations of reversals. It is worth pointing out that magnetic field reversals have
yet to be observed in external \jacbred{spiral} galaxies \citep{Beck_2007}.  It is uncertain if this is a 
consequence of observation techniques and resolution, or
a true indication of some uniqueness in the properties~of~the~Galaxy.

\begin{figure}[h]
\begin{center}
\includegraphics[width=7.5cm]{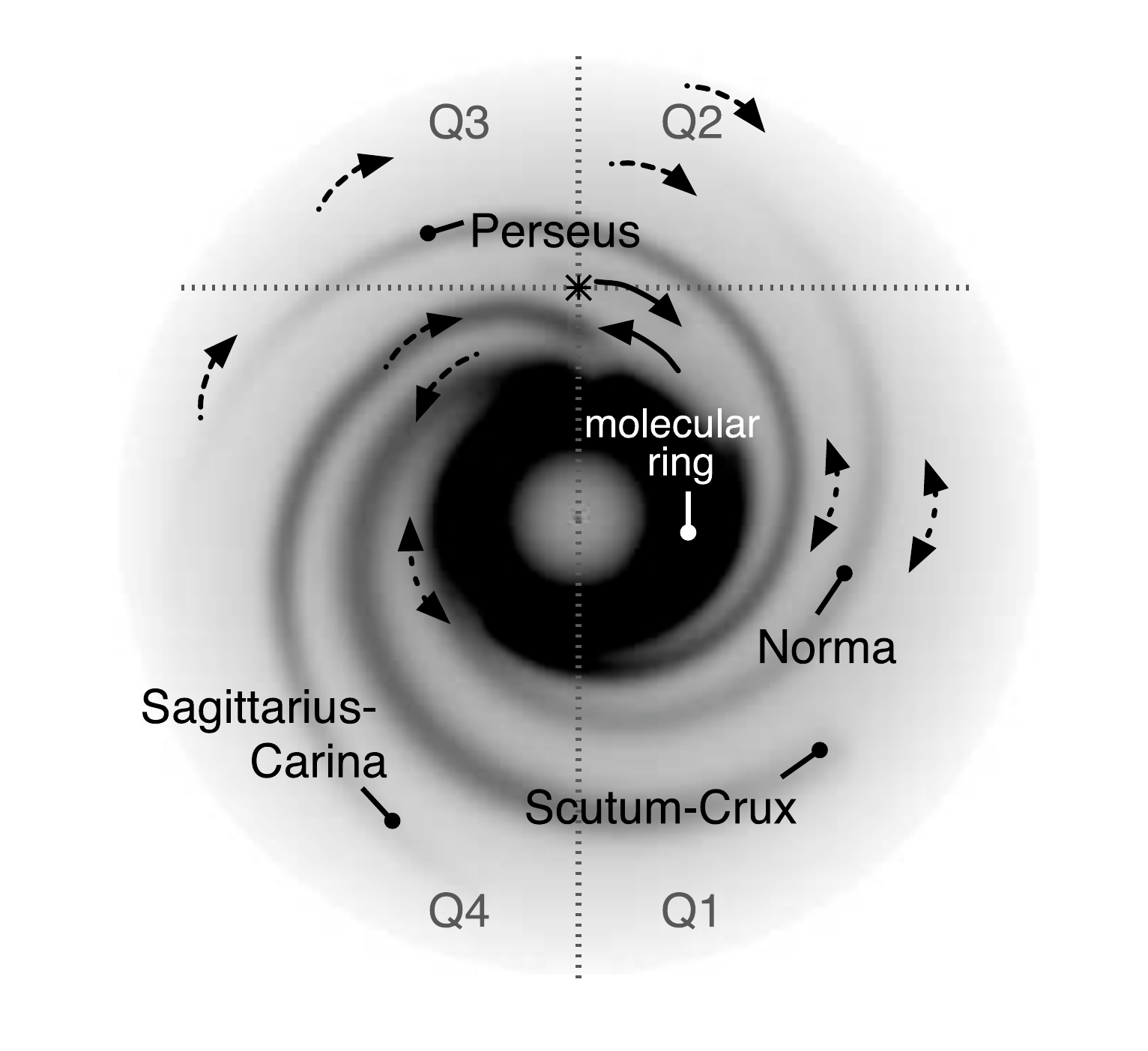}
\caption{Directions of the large-scale Galactic magnetic field as 
viewed from the North Galactic pole.  The 
grey scale represents the electron density model of \protect\citet{Cordes+Lazio_2002},  Q1--4 indicate
the four Galactic quadrants, and the asterisk shows the location of the
Sun.  Solid arrows correspond to universally accepted field
directions, while single-ended dashed arrows show field directions as supported
by the Canadian or the Southern Galactic Plane Survey (CGPS and SGPS, respectively), though not necessarily universally
accepted~\protect\citep{Brown+Taylor+Wielebinski_etal_2003, Brown+Haverkorn+Gaensler_etal_2007}.  Finally, double-ended arrows indicate
regions remaining highly debated with either no data or no clear indications currently available from the CGPS or the SGPS.}\label{arrows}
\end{center}
\end{figure}

Figure~\ref{arrows} illustrates the directions of the large-scale Galactic magnetic field
superposed on the electron density model of \citet{Cordes+Lazio_2002}.  Both universally accepted (solid arrows) and still debated (dashed arrows) directions are shown (see caption of Fig.~\ref{arrows} for more details).

\subsubsection{The Small-Scale Magnetic Field in the Galactic Disk}
\label{ss:randomgmf}

The ratio of the regular to the small-scale Galactic magnetic field strength is
estimated from synchrotron emission to be approximately 0.6 -- 1
\citep[see, e.g.,][]{Phillipps+Kearsey+Osborne_etal_1981, Heiles_1995}, in agreement with estimates from the
variance in pulsar RMs \citep{Rand+Kulkarni_1989}.  

As the energy density in the magnetic field is comparable to the
(turbulent) gas density \citep{Boulares+Cox_1990}, interstellar turbulence is
expected to be magnetic, although details about its structure,
strength and variability are still largely unknown. Interstellar
turbulence is traditionally measured through power spectra of either velocity
fluctuations in emission or absorption lines, or density variations.  Similarly, one can get an idea of the role of the magnetic field by studying power spectra of RM measurements.  To avoid the
problem of depolarization along the line-of-sight, this is usually done using RMs of pulsars
and polarized extragalactic sources.  The corresponding data points being irregularly
spaced, the computation of a power spectrum is usually replaced by a structure
function analysis, with the second order structure function of RM given by
$D_\RM(\d x) = \langle \RM(x+\d x) - \RM(x)\rangle_x$,
where $\d x$ is the distance between two sources, and $\langle\cdot\rangle_x$
denotes averaging over all positions $x$.  These analyses
show that RM fluctuations exist over a wide range of scales
and exhibit a power law behavior \citep[see, e.g.,][]{Minter+Spangler_1996}. The largest observed fluctuations are on scales of
50 -- 100~pc when averaged over the sky
\citep{Rand+Kulkarni_1989,Ohno+Shibata_1993}, and of order 1~pc in the spiral arms
\citep{Haverkorn+Brown+Gaensler_etal_2008}. At least part of this structure is caused by magnetic
field fluctuations as opposed to variations in the electron density
\citep{Minter+Spangler_1996}.

~

\subsubsection{The Magnetic Field in the Galactic Halo}
\label{ss:hgmf}

Significant synchrotron emission is observed at high Galactic latitudes,
indicating the presence of a magnetic field in the Galactic thick disk or halo.  The regular magnetic field \jacbred{has a}
scale height
\jacbred{of} about 1.5~kpc \jacbred{and is an order of magnitude weaker than the disk field}, as
derived from pulsar RMs \citep{Han+Qiao_1994} and supported by extra-Galactic
RMs (Brown et al., in preparation). In the thin disk of the Galaxy, the
field points in the same direction above and below the Galactic plane
\citep{Frick+Stepanov+Shukurov_etal_2001}.  However, there is evidence for large-scale reversals of the
field's direction in the thick disk/halo, which are
likely caused either by a dipole structure of the halo magnetic field
\citep{Andreasyan+Makarov_1988, Han+Manchester+Berkhuijsen_etal_1997}, or by local structures that appear large on the plane of the sky.

\vspace{0.4cm}
\section{\blue{Global Constraints on the Magnetized ISM}}
\label{s:mim}

As reviewed in Sec.~\ref{s:obs}, numerous techniques can be used to probe the structure of the Galactic magnetic field on various scales.  However, most of these studies do not lead to direct constraints on the latter, but on the magnetized interstellar medium, of which the Galactic magnetic field is only one component along with, e.g., thermal and cosmic ray electrons.  Given a model for the physical properties of these other components, each technique can then be used to characterize the field in our galaxy.  These models are usually themselves derived from other observations, and their reliability is often limited, which, in turn, weakens the constraints one can infer for the Galactic magnetic field.  An example can be found in the controversy surrounding the existence of field reversals in the outer parts of the Galaxy (see Sec.~\ref{ss:lsgmf}).

Although this approach has several advantages, in particular insofar as it offers the possibility of comparing independent constraints on the various components of the field~\citep[see, e.g.,][]{Crutcher+Heiles+Troland_2003}, it fails to account for the possible degeneracies between properties of the field and characteristics of other components of the magnetized interstellar medium.  This can be avoided by simultaneously fitting multiple observations,  which also has the advantage of leading to a unique model able to describe a diverse set of data.  \red{Such a model} is all the more desirable \blue{since} the same physical quantities are involved in various probes of the field.

There are multiple data sets one can combine to build such a model, but one first needs a simulation code able to generate the corresponding theoretical maps from a set of parameters describing the relevant components of the magnetized interstellar medium. The \Hammurabi{} software~\citep{Waelkens+Jaffe+Reinecke_etal_2008} was developed to provide such a tool for the Galactic synchrotron emission (total and polarized), free-free emission, and the effect of Galaxy-induced Faraday rotation on the emission from extra-galactic sources. In this section, we report on the efforts by two groups: \citet{Jansson+Farrar+Waelkens_etal_2007} and \citet{Sun+Reich+Waelkens_etal_2008}.  The former used \Hammurabi{} to constrain the magnetized interstellar medium with the three-year \WMAP~23~GHz polarization data~\citep{Page+Hinshaw+Komatsu_etal_2007}.  \citet{Sun+Reich+Waelkens_etal_2008} \blue{then performed an extensive} analysis taking into account the total synchrotron emission \blue{observed} by~\citet{Haslam+Klein+Salter_etal_1981, Haslam+Salter+Stoffel_etal_1982}, the 1.4~GHz synchrotron map of~\citet{Wolleben+Landecker+Reich_etal_2006}, the \WMAP~three-year free-free emission template~\citep{Hinshaw+Nolta+Bennett_etal_2007}, and the rotation measures \green{of} \citet{Brown+Taylor+Wielebinski_etal_2003,Brown+Haverkorn+Gaensler_etal_2007}.

\vspace{0.4cm}
\subsection{ISM Modeling}
\label{ss:ismmodel}

Assumptions about three components of the magnetized ISM, namely, thermal electrons (Sec.~\ref{sss:thelec}), cosmic ray electrons (Sec.~\ref{sss:crelec}), and the Galactic magnetic field (Sec.~\ref{sss:bfield}), are necessary to make it possible for \Hammurabi~to generate the simulated maps mentioned in Sec.~\ref{s:mim}.
Comparison to data then enables~\citet{Jansson+Farrar+Waelkens_etal_2007} and \citet{Sun+Reich+Waelkens_etal_2008} to constrain the parameters of these models by making sure that they can reproduce their selected sets of observations.  We review each of these models and their associated parameters below.  Some of the constraints obtained by the Jansson and Sun groups  are reviewed in Sec.~\ref{ss:ismconstraints}.

\subsubsection{Thermal Electron Density and Temperature}
\label{sss:thelec}

\Hammurabi~uses the NE2001 model developed by \citet{Cordes+Lazio_2002, Cordes+Lazio_2003} to describe the Galactic thermal electron density $n_\rme$.  This model, based on dispersion measures, contains both a thick disk and a thin disk, as well as spiral arms and some local ISM features.

\citet{Sun+Reich+Waelkens_etal_2008} combined the results of \citet{Reynolds+Haffner+Tufte_1999}, \citet{Peterson+Webber_2002}, and \citet{Quireza+Rood+Bania_etal_2006} to show that the temperature distribution of the thermal electrons is well described by $T_\rme(\br)/\rmK=5780 + 287\,r - 526\, |z| + 1770\, z^2$, where $r$ and $z$ (cylindrical coordinates) are in units of kpc.  Moreover, the free-free emission intensity is proportional to $\langle {n_\rme}^2 \rangle$, where the average is along a given line-of-sight, which is a priori different from the quantity $\langle {n_\rme} \rangle^2$ that can be inferred from the NE2001 model.  This discrepancy can be accounted for by the filling factor $f_\rme (z) \equiv \langle {n_\rme} \rangle^2/\langle {n_\rme}^2 \rangle$ provided by \citet{Berkhuijsen+Mitra+Mueller_2006}.

\vspace{0.2cm}
\subsubsection{Cosmic Ray Electron Density}
\label{sss:crelec}

Models of the cosmic ray electron density such as those used by~\citet{Page+Hinshaw+Komatsu_etal_2007} and \citet{Sun+Reich+Waelkens_etal_2008} \blue{are composed of a simple spatial distribution characterized by a scale height and a scale length, and also assume a power law energy distribution for the electrons.  The latter has} so far been sufficient to model foregrounds in the context of CMB data analysis~[see, e.g., \citet{Tegmark+Efstathiou_1996}, \citet{de_Oliveira-Costa+Tegmark+Gaensler_etal_2008}, \citet{Dunkley+Komatsu+Nolta_etal_2008}, and \citet{Bottino+Banday+Maino_2008}].  \blue{However,} simulation codes, such as \textsc{GalProp}, lead to models much closer to the state-of-the-art in cosmic ray modeling~\citep[as in][]{Strong+Moskalenko+Ptuskin_2007} and might have to be used in conjunction with more sophisticated analyses in the future (we refer the reader to the companion paper by~\citet{CMBPol_Dunkley_2008} for details on this issue).

\blue{The} \citet{Sun+Reich+Waelkens_etal_2008} \blue{model assumes a} scale height (respectively, \blue{a} radial scale length) \blue{of} 1~kpc (respectively, 8~kpc), whereas the normalization is chosen to reproduce observations at $\br=\br_\odot$~\citep[see, e.g.,][]{Strong+Moskalenko+Ptuskin_2007}.  \blue{Furthermore,} a truncation is introduced for $|z|>1$~kpc to explain the low synchrotron emission observed at high Galactic latitudes.  Finally, the energy distribution is given by the power law $N(E) \propto E^{-p}$, where $p$ is 3 for $\nu > 408$~MHz, and $2$ at lower frequencies.  No local features of the Galactic interstellar medium are \blue{modeled}.

\vspace{0.2cm}
\subsubsection{Galactic Magnetic Field}
\label{sss:bfield}

\red{Any} large-scale magnetic field model can be tested with \Hammurabi\red{. \citet{Jansson+Farrar+Waelkens_etal_2007} looked at axisymmetric and bisymmetric spiral models, with symmetry or antisymmetry in the $z$ direction, while \citet{Sun+Reich+Waelkens_etal_2008}}\blue{, after demonstrating the need to improve models found in the literature,} \red{considered} an axisymmetric spiral field with a ring, an axisymmetric spiral field with a field reversed arm, and a bisymmetric spiral field.  In addition, a halo magnetic field, necessary to fit high-latitude rotation measure profiles, and a random field, required by the synchrotron \blue{and RM} observations, are included \red{in the \citet{Sun+Reich+Waelkens_etal_2008} analysis.}

As for the cosmic ray electron model, no attempt has been made at modeling the Galactic magnetic field to explain local features such as the North Polar Spur.  Although they occupy a large fraction of the sky, and are fascinating objects to explore on their own \citep[see, e.g., the Loop~I model developed by][]{Wolleben_2007}, they are beyond the scope of a first global description of the diffuse Galactic emission, and are therefore not included in \Hammurabi.

\subsection{Simulation Outputs}

\begin{figure}[h]
\begin{tabular}{c c}
\includegraphics[width=4.5cm,angle=90]{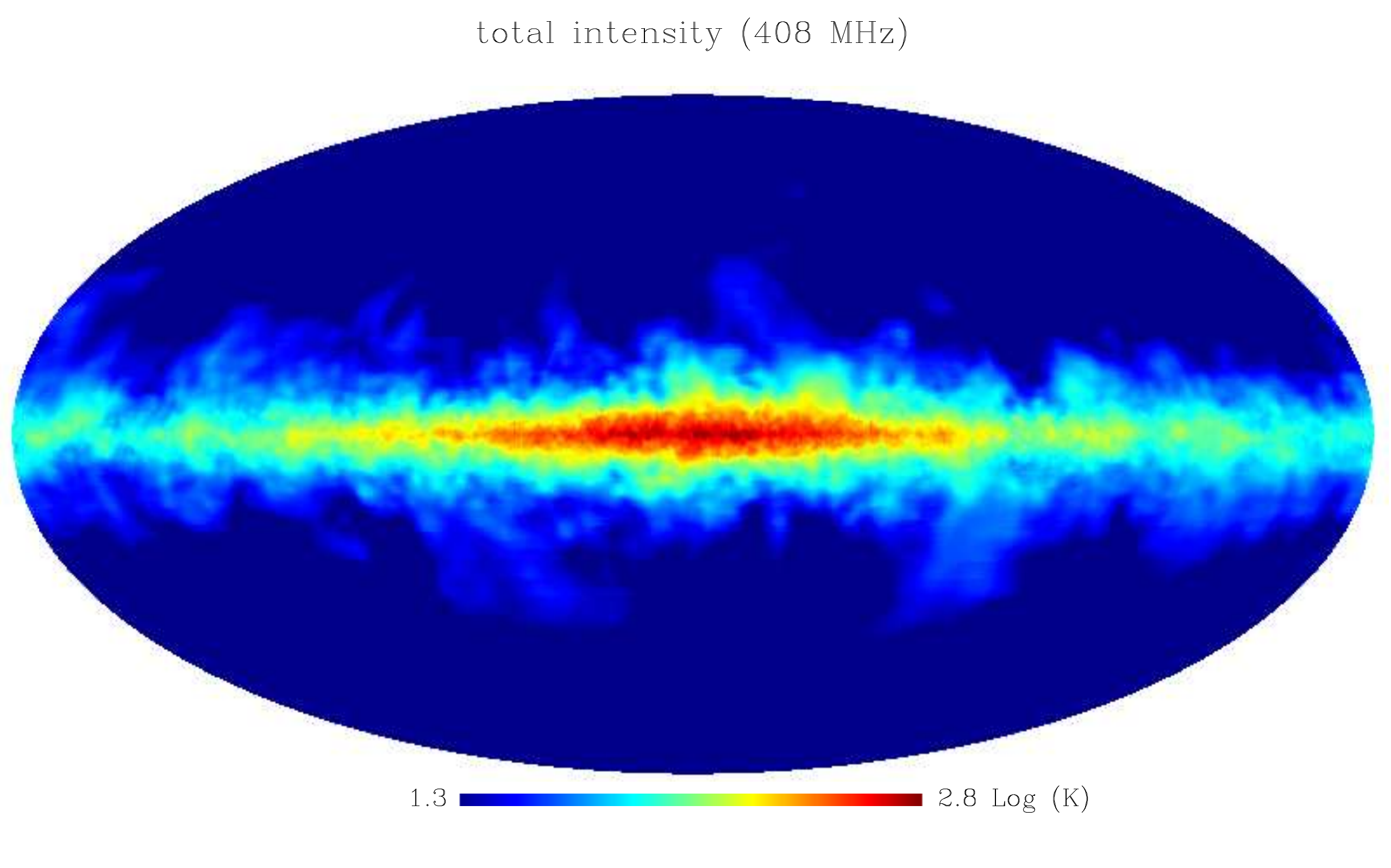}&
\includegraphics[width=4.5cm,angle=90]{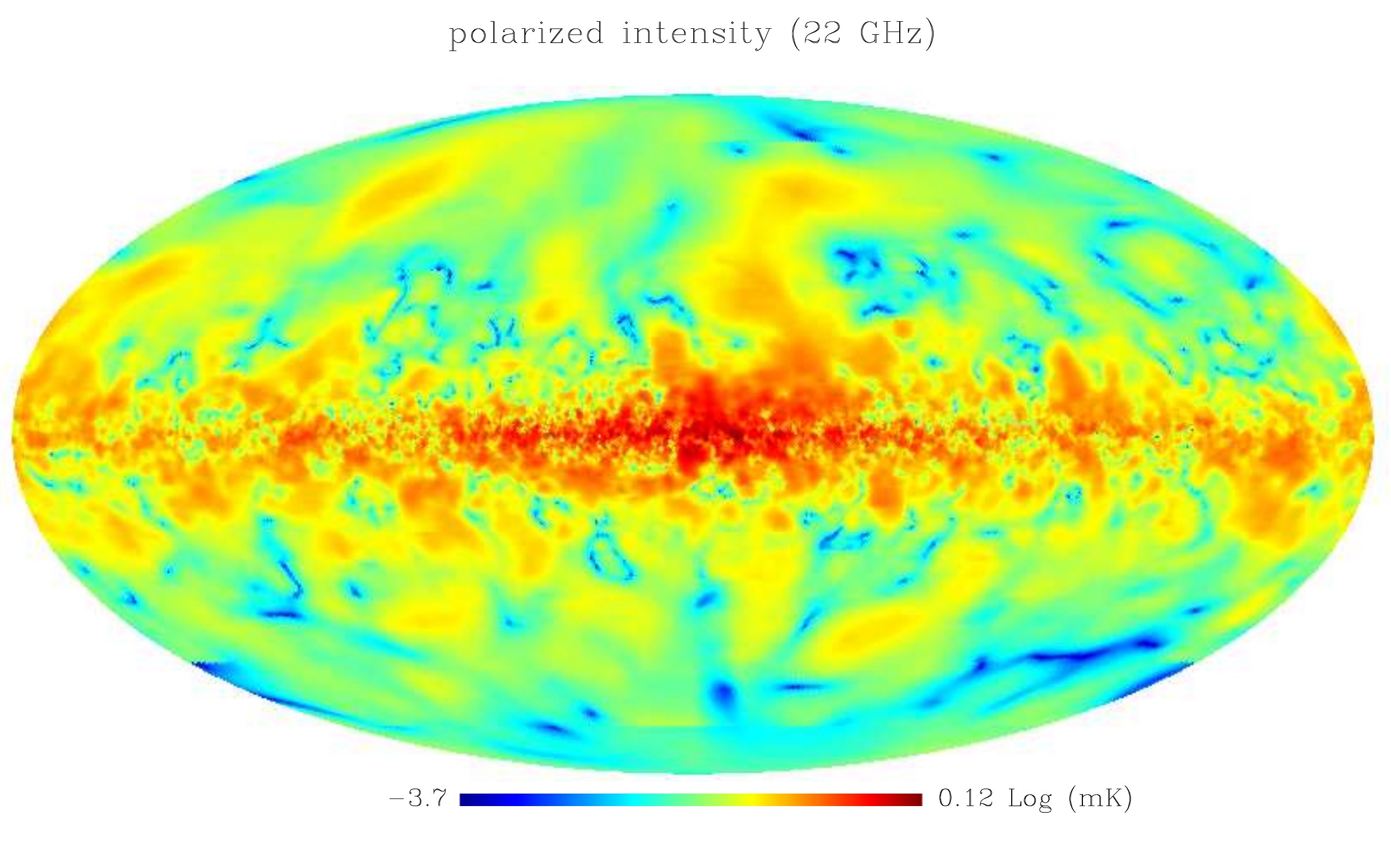}\\
\includegraphics[width=4.5cm,angle=90]{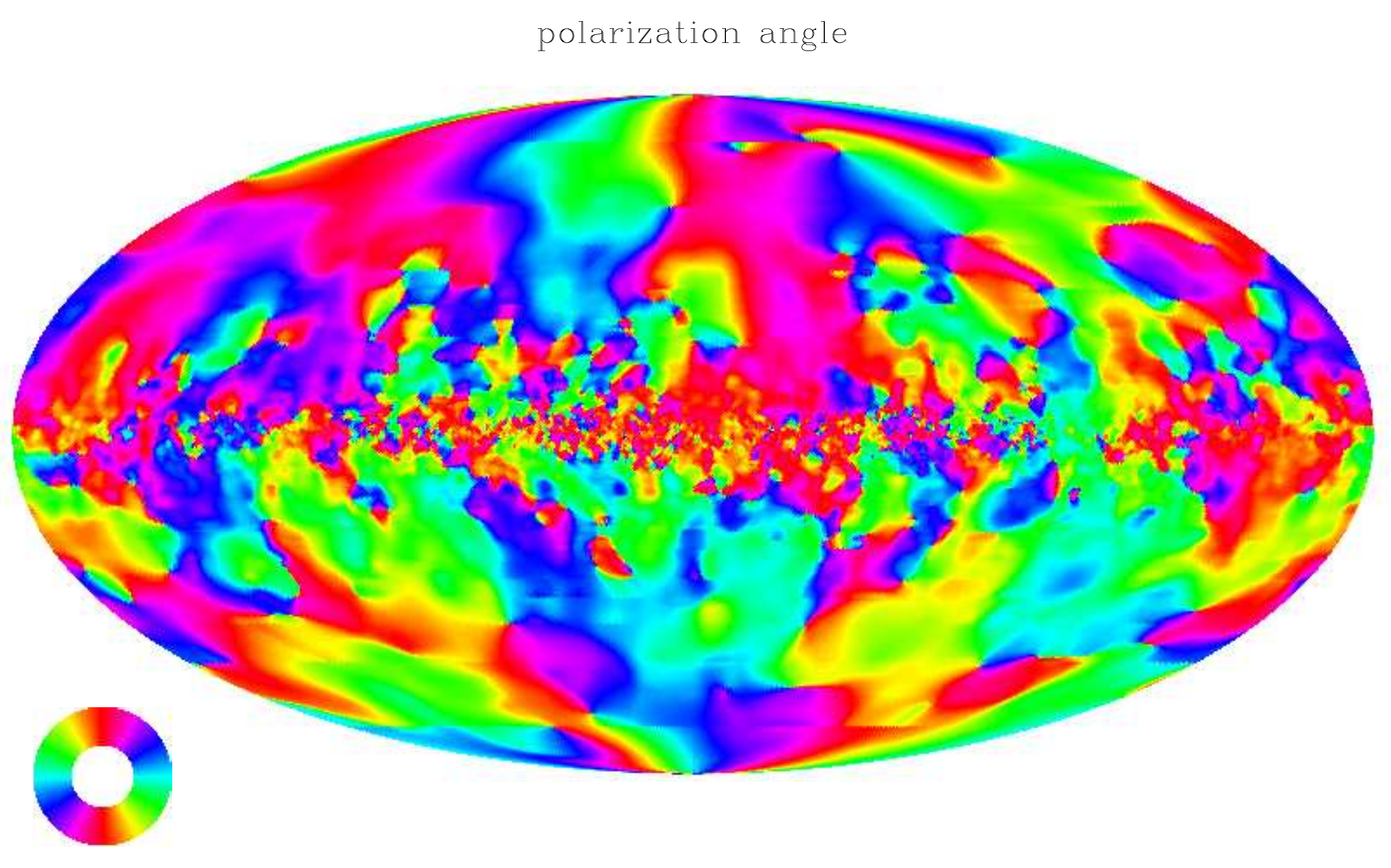} &
\includegraphics[width=4.5cm,angle=90]{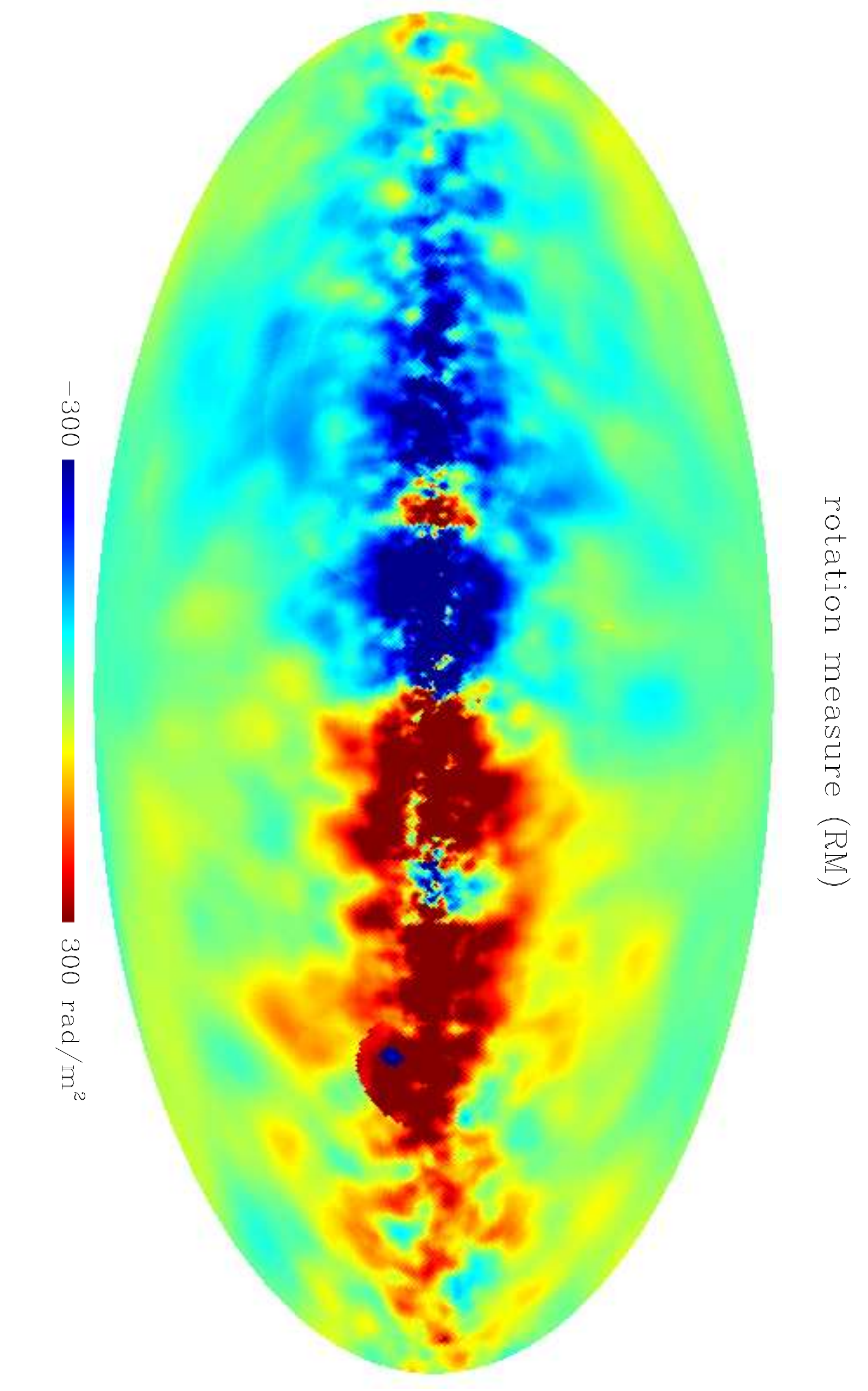}
\end{tabular}
\caption{Mock observations of the \protect\citet{Sun+Reich+Waelkens_etal_2008} model.  The top row shows full-sky maps of the total (left) and polarized (right) synchrotron emission, whereas the bottom row provides a map of the corresponding polarization angle (left) and a rotation measure map (right).  Unlike in \protect\citet{Sun+Reich+Waelkens_etal_2008}, the small-scale magnetic field is modeled as a Gaussian random field.  The simulated map of free-free emission is not shown here.}
\label{f:hammurabi}
\end{figure}

Based on the model described in Sec.~\ref{ss:ismmodel}, \Hammurabi~can generate Galactic simulated maps of synchrotron emission (total and polarized), free-free emission, and rotation measure (Fig.~\ref{f:hammurabi}).  \awred{S}tructures like the \red{North Polar Spur}, or any of the \red{other} Loops, are not reproduced by these simulations \awred{since the magnetic field and cosmic ray models only reproduce the large-scale features of the Galaxy}.
However, local features in the electron density distribution, which are part of the NE2001 model (see Sec.~\ref{sss:thelec}), can be recognized in the rotation measure map.

\subsection{Comparison to Observations}
\label{ss:ismconstraints}

Using a $\chi^2$ minimization technique, \citet{Jansson+Farrar+Waelkens_etal_2007} compare the polarized synchrotron maps produced by \Hammurabi~ to the three-year \WMAP~K-band (23~GHz) map smoothed to 4$^\circ$.  They find that the minimized $\chi^2$ is close to identical for all models of the Galactic magnetic field \red{they consider (see Sec.~\ref{sss:bfield})}.  However, for each of them, large regions of the parameter space can be excluded, even though there are no obvious best-fit parameters as a number of combinations yield a value of $\chi^2$ very close to the minimum.  In particular, the vertical and radial scale-lengths can take unphysically large values, which is explained by the drop in synchrotron emission due to the reduction in the relativistic electron density at large radial and vertical distances. Thus, knowing the polarized synchrotron emission is not sufficient to strongly constrain the parameters of \red{these} Galactic magnetic field models.

By combining synchrotron emission data~\citep{Page+Hinshaw+Komatsu_etal_2007} with rotation measure information~\citep{Brown+Taylor+Wielebinski_etal_2003,Brown+Haverkorn+Gaensler_etal_2007} and knowledge of the full-sky free-free emission~\citep{Hinshaw+Nolta+Bennett_etal_2007}, \citet{Sun+Reich+Waelkens_etal_2008} showed that this degeneracy can be at least partially lifted.  They indeed demonstrated that increasing the vertical scale-length of the \blue{NE2001} thermal electron density, removing the unphysical cut in the cosmic ray electron density (see Sec.~\ref{sss:crelec}), and decreasing the strength of the halo magnetic field, leads to a better global fit to the observations.  \blue{The need for a thermal electron scale height about twice as large as that of the NE2001 model} was confirmed by an independent data analysis by \citet{Gaensler+Madsen+Chatterjee_etal_2008}.

\citet{Sun+Reich+Waelkens_etal_2008} were also able to find statistical evidence that the Galactic magnetic field is best fit by an axisymmetric spiral field with a ring \red{located in the inner part of the Galaxy}, thereby adding to the case that combining observations probing a broad range of physical processes is a powerful way of constraining the various components of the magnetized ISM.

\subsection{Future Developments}

In addition to the probes combined by \citet{Sun+Reich+Waelkens_etal_2008}, other observations could already be used to improve our knowledge of the magnetized interstellar medium.  An example is pulsar RM -- DM plots~\citep[see, e.g.,][and Sec.~\ref{ss:lsgmf}]{Brown+Taylor+Wielebinski_etal_2003}, which trace the three-dimensional structure of the Galactic magnetic field.  \Hammurabi's design allows it to perform the corresponding mock observations, and the Square Kilometer Array (SKA) is expected to provide thousands of data points for these plots by 2020~\citep{Beck+Gaensler_2004}.

The \citet{Sun+Reich+Waelkens_etal_2008} model does not contain any dust related inputs, and, as a result, cannot make starlight polarization or dust emission predictions.  Dust emission mechanisms are difficult to model given the sophisticated and sometimes unknown physics associated with them.  Moreover, models developed to match all available observations can lead to significantly different predictions for what the polarized emission from interstellar dust might look like (see Sec.~\ref{ss:firpol}).
\awred{A}dding a dust component to \Hammurabi~\awred{will therefore be a challenging task}.

Upcoming radio telescopes, such as the SKA, the LOw Frequency ARray (LOFAR), and the Atacama Large Millimeter Array (ALMA), associated to upcoming CMB experiments, such as \Planck, will greatly increase the number and the quality of available observations.  A tool like \Hammurabi~will allow us to make the most of these improvements, and, in particular, to generate valuable information regarding the structure and properties of the magnetized interstellar medium.  This will be a welcomed source of inputs for the CMBPol data analysis.

\section{The Galactic Magnetic Field in the CMBPol Era}
\label{s:cmbpolera}

\subsection{Radio Surveys in the Next Decade}
\label{s:radio_next}

Narrow frequency and spatial coverage have so far been the main limiting factors in studying the low frequency polarized emission from the Galaxy.  This issue is expected to be addressed by large radio polarimetric surveys currently planned or underway, which will provide a wealth of new data by the time CMBPol flies.  The S-PASS polarimetric survey (PI: E.~Carretti) will observe the entire Southern
sky 
\mhred{in a 256~MHz frequency band centered on}
2.3~GHz by 2009.  \mhred{The}
GALFACTS
polarized continuum survey
(PI: A.~R.~Taylor)
will map the entire sky visible from Arecibo \mhred{in a 300~MHz frequency band centered on}
1.4~GHz by 2012\mhred{, whereas the STAPS survey (PI: M.~Haverkorn) will cover 500~MHz around the same frequency in the Southern sky}.  Radio
polarimetric observations are also planned with the Australian SKA Pathfinder
\citep{Johnston+Bailes+Bartel_etal_2007}, which should be fully operational by 2013.  Finally, the Galactic Magneto-Ionic
Medium Survey (GMIMS, PI: M.~Wolleben), a major project consisting
of six independent radio polarimetric surveys, 
\mhred{is} collect\mhred{ing} high-resolution ($\sim 30'$) all-sky polarization data in a near-continuous
frequency band ranging from 300~MHz to 1.8~GHz.

These and similar surveys  will provide ``RM grids''
\citep{Beck+Gaensler_2004} gathering numerous measurements of the Galactic Faraday rotation of polarized emission from
extragalactic sources across the whole sky.  Although EGSs have a small intrinsic RM
component, it can be statistically eliminated by averaging the RMs of enough point sources in
a given patch of the sky.  The current average RM source density is about 1 RM
source per 10 square degrees at high latitudes, and approximately 1
source per square degree in parts of the Galactic plane \citep{Brown+Taylor+Jackel_2003,Brown+Haverkorn+Gaensler_etal_2007}.  As a result, inferred Galactic RM maps are very low resolution over most of the sky.
However, the radio surveys described above are expected to fill in the 
Galactic plane to the level of 1 source per square degree, as well 
as to increase the source density at high latitudes by a factor of 100.  This will allow averaging of a sufficient number of RM
sources to remove intrinsic RM contributions on scales
of about 1$^\circ$.
Detailed modeling of the
pitch angle of the Galactic magnetic field, the number and locations
of large-scale field reversals, and the vertical component of the magnetic field will therefore soon become
possible.

Moreover, RM grids will enable detailed study of the turbulent
field component over a wide range of scales and large parts of the
sky.  Given their expected high source density, it will become possible to
construct a structure function every few square degrees on the sky,
which will result in unprecedented detailed knowledge of the turbulent
behavior of the magnetized ISM on degree scales.  In particular, possible changes in
the turbulence characteristics from the Galactic plane, via the
disk-halo connection to the Galactic halo, should become visible.

Finally, RM synthesis of polarized emission from diffuse
synchrotron emitting gas in the Galaxy can be used to obtain a
3-dimensional map of Galactic RM structure.  All of the above
mentioned surveys have the capability to perform RM synthesis on their
data, with various limitations, which will be used to improve
modeling of the large-scale components of the Galactic magnetic
field.  As the technique allows separation of magnetic
foregrounds and backgrounds, it will also be possible to study magnetic
field structures in  Galactic objects such as supernova
remnants, HII regions, and planetary nebulae, as well as in external
galaxies.

\subsection{The Contribution of Upcoming CMB Experiments}
\label{ss:gmf_cmb}

To study the global character of the Galactic magnetic field through
synchrotron emission and (de)polarization, observations over a
significant part of the sky are needed, at a frequency high enough
that Faraday rotation is negligible. \WMAP~provides polarization maps
that allow study of the Galactic magnetic field through synchrotron
radiation \citep{Page+Hinshaw+Komatsu_etal_2007}. However, the polarized emission in the lowest
frequency \WMAP~maps has a signal to noise ratio below 3 in
about $55\%$ of the sky when smoothed to 2$^\circ$.  The only polarized
structures that are clearly visible in the \WMAP~maps are the Galactic
plane, the region of high and regular polarization called the Fan
region \citep{Berkhuijsen+Brouw+Muller_etal_1964, Brouw+Spoelstra_1976}, and the North Polar Spur \citep{Hanbury-Brown+Davies+Hazard_1960},
generally believed to be an old, nearby supernova remnant.  Therefore,
to be able to study the behavior of the synchrotron radiation away
from the Galactic plane in the ``typical'' interstellar medium, higher
sensitivity observations are needed.
\WMAP~polarization maps show that although in both the
Galactic plane and discrete structures the polarized intensity can
go up to $\sim100~\mu$K, at intermediate and high latitudes the
polarized intensity is primarily below $20~\mu$K. \dnsred{T}his will allow a
nominal detection with \Planck~\dnsred{or nine years of \WMAP~observations},  \dnsred{but with} a typical signal-to-noise lower
than 2\dnsred{, which} will lead to an error in polarization angle determination of at least
1~rad. CMBPol, which could be 10 times more sensitive than \Planck~\dnsred{at the same frequencies},
would have much higher signal-to-noise across the whole sky, but in particular in these intermediate and high Galactic latitude regions.
This data would complement the information provided by the radio surveys described in Sec.~\ref{s:radio_next}, in particular with respect to the vertical component of the field, as well as the
strength and structure of the halo field, both of which are necessary knowledge
to test dynamo theories of the origin and evolution of Galactic
magnetism.

In addition, the resolution of CMBPol might be higher than that of
\Planck, which would significantly decrease depolarization.  At the
frequencies currently considered for CMBPol, there is no significant Faraday
depolarization, but small-scale structure in the magnetic field will
cause small-scale structure in polarization angles of the synchrotron
emission, resulting in depolarization of the radiation within one
telescope beam. If the satellite's resolution is not high enough, the polarized emission will be below its detection threshold, destroying all diagnostic capability
of the data for Galactic magnetic fields.
However, at high latitudes, turbulent
cells are expected to be at $\sim100$~pc scales \citep{Dumke+Krause+Wielebinski_etal_1995}, which, at the distances of interest, corresponds to a much larger angle than the resolution currently considered for CMBPol.  As a result, depolarization is not
expected to be severe there.  On the other hand, the typical structure
of the magneto-ionized medium in spiral arms \alred{could} be on scales as small as 1~pc \citep{Haverkorn+Brown+Gaensler_etal_2008},
which makes depolarization resolution dependent.  In particular, higher
resolution than \Planck's $33'$ beam at 30~GHz will be necessary to observe 
turbulent structures in the magneto-ionized medium in and close to the
Galactic plane.

The same holds for discrete polarized structures, such as supernova remnants (SNRs).  Statistical studies
of supernova rates \citep[see, e.g.,][]{Li+Wheeler+Bash_etal_1991} indicate that there should be
many more SNRs in the Galaxy than are currently known. This ``missing 
SNRs" problem is likely due to a
selection effect of many radio surveys against large low-brightness
SNRs as well as small remnants.  However, low-brightness remnants might be
observable through their magnetic field structure, which remains
visible through polarized radio emission even when synchrotron intensities
are too low to be observed \citep{Haverkorn_2005}. A sensitive all-sky survey like
CMBPol with the same resolution as necessary to study the turbulent structure of the field in spiral arms therefore opens the possibility of detecting the magnetic
field structures of old and faint SNRs, which could help
solve the long-standing ``missing SNRs" problem.

\section{Polarized Dust Emission (PDE)}
\label{s:pde}

Over the course of the last 60 years, the linear polarization of starlight has been observed at wavelengths ranging from the ultraviolet~\citep{Clayton+Anderson+Magalhaes_etal_1992, Anderson+Weitenbeck+Code_etal_1996, Martin+Clayton+Wolff_1999} to the far-infrared~\citep[see, e.g., the review by][]{Whittet_2003} along numerous lines-of-sight.  The Archeops experiment first released detailed studies of the emission properties of dust at CMB wavelengths based on observations of 20\% of the sky at 353~GHz~\citep{Benoit+Ade+Amblard_etal_2004, Ponthieu+Macias-Perez+Tristram_etal_2005}, and warned that the polarized dust emission would be a major contaminant of the primordial E- and B-mode signals unless removed properly.  These results were largely confirmed by the \WMAP~mission~\citep{Kogut+Dunkley+Bennett_etal_2007}, although the fact that the highest \WMAP~frequency (94~GHz) is below 100~GHz made it easier to handle this emission in the polarization analysis~\citep{Page+Hinshaw+Komatsu_etal_2007, Gold+Bennett+Hill_etal_2008}.

The \Planck~satellite will soon be on its way to L2, where it will map the sky at frequencies ranging from 30~GHz to 857~GHz, with 5 of its bands covering frequencies higher than 100~GHz.  Compared to \WMAP, \Planck~will have twice as many frequency channels, 3 times higher resolution, and 10 times higher sensitivity, all of which will, by design, be of tremendous help to distinguish between polarized foregrounds and candidate inflationary signals.

However,  a number of assumptions had to be made to reach this ``optimal" configuration.  In particular, the dust emission, which will be brighter in the 5 \Planck~frequency bands mentioned above  than in any \WMAP~map, was assumed to have the same polarization degree at all wavelengths, with the 857~GHz observations providing a way of determining, among other information, this polarized fraction.  Although so far consistent with the degrees of polarization inferred by the Archeops and \WMAP~groups~\citep{Benoit+Ade+Amblard_etal_2004, Kogut+Dunkley+Bennett_etal_2007} at, respectively, 353~GHz and 94~GHz, there is no strong observational evidence that this will hold at all \Planck~frequencies.  If the level of polarized dust emission varies as a function of frequency, it will reduce our ability to remove this emission from the \Planck~maps.  A CMBPol mission is likely to have between 4 and 6 frequency bands in which the brightest signal will be dust emission, and might therefore have to deal with the same problem.

In this section, we follow~\citet{Draine+Fraisse_2008}, and show how our current knowledge of the extinction and polarization of starlight can be used to predict the degree of polarized dust emission one could observe at CMB wavelengths with a high-resolution and high-sensitivity mission such as \Planck~or CMBPol, as well as its potential dependence on frequency.

\subsection{Observed Extinction and Polarization of Starlight}
\label{s:obs_ext_pol_star}

Although observations of the extinction and polarization of starlight exist at numerous wavelengths [see references in the first paragraph of Sec.~\ref{s:pde}, as well as~\citet{Martin+Adamson+Whittet_etal_1992}, and, e.g., the review by~\citet{Fitzpatrick_2004}], they are well fit by simple parametric \red{functions}.  Fitting the average Galactic extinction curve is discussed in \citet{Fitzpatrick_1999} for $0.1<\lambda/\um<2.65$, whereas~\citet{Anderson+Weitenbeck+Code_etal_1996} showed that the fractional polarization curve $p(\lambda)$ is well described by a Serkowski law~\citep{Serkowski_1973} for $0.15<\lambda/\um<1.39$.  Finally, for $1.39<\lambda/\um<5$, $p(\lambda) \propto \lambda^{-1.7}$ is a good match to the observations by \citet{Martin+Adamson+Whittet_etal_1992}.  It is therefore possible to represent the observed extinction and polarization of starlight by smooth continuous functions in the range of wavelengths for which data is available.

The fact that we observe starlight polarization implies that dust grains cannot be spherical, and that they have to be at least partially aligned with the Galactic magnetic field.  However, the data currently in hand does not provide direct constraints on the shape of grains, their size distribution(s), or their alignment function(s).  As a result, modeling interstellar dust requires relying on theory to predict the general behavior expected for the latter.  One can then assume a shape and try to fit the observed extinction and polarization of starlight.
Of course, this process does not guarantee that there is a unique good fit as only a few shapes can be tested in a reasonable amount of time.  But the hope is that, given the many constraints imposed on candidate models, if several of them can accurately reproduce observations of the extinction and polarization of starlight over a wide range of wavelengths, they will lead to similar predictions for what the polarized dust emission should look like at CMB wavelengths, at least as long as the non-spherical interstellar grains are made of the same material(s).

\subsection{Modeling Interstellar Dust}
\label{ss:dust_models}

Dust grains are believed to be made of the numerous elements that have been depleted from the gas phase~\citep{Jenkins_2004}.  However, observed spectroscopic features in the Galactic extinction curve and in observations of emission by interstellar dust in the mid-infrared indicate that the bulk of the mass in interstellar grains is in the form of amorphous silicates and carbonaceous materials, among which a population of Polycyclic Aromatic Hydrocarbons~(PAHs) is required to explain several emission features~\citep[see, e.g., the review by][]{Draine_2003}.

In order to explain the polarization observed in some of the spectroscopic features that have been studied in the far-infrared, amorphous silicates have to be present in non-spherical dust grains aligned with the Galactic magnetic field\btdred{, $\bB$} \citep[see, e.g.,][]{Whittet_2003}.  As of today, there is however no compelling evidence that the same \red{is} true for carbonaceous materials~\citep{Chiar+Adamson+Whittet_etal_2006}.
\btdred{Finally, PAHs are expected to have their angular momenta only slightly aligned with~$\bB$, and a PAH's principal axis of largest moment of inertia will be only partially aligned with its angular momentum \citep{Lazarian+Draine_2000}. Furthermore, PAHs are thought to contribute only a small fraction of the total emission by dust at frequencies $\nu \lesssim 10^3$~GHz.  As a result, PAHs should contribute negligibly to the polarized emission at those frequencies.}

\subsection{Far-Infrared Emission from Interstellar Dust}
\label{ss:firpol}

Based on the considerations discussed in Sec.~\ref{ss:dust_models}, \citet{Draine+Fraisse_2008} fit the observed extinction and polarization of starlight with dust models \btdred{consisting of (1) a population of (nearly) randomly-oriented PAHs, (2) oblate spheroidal silicate grains, and (3)}
graphite grains assumed to be either spheres (in which case their emission is not polarized) or oblate spheroids with the same axial ratio \btdred{$b/a$ of their equatorial to their polar axes} as the silicates.

~

\begin{table}[h]
\begin{center}
\caption{\label{t:dust_models} \citet{Draine+Fraisse_2008} Dust Models}
\begin{tabular}{ccc}
\hline
\hline
Model number & $(b/a)_{\rm silicate}$ & $(b/a)_{\rm carbon}$ \\ 
\hline
1            & 1.4        & 1.0 \\
2            & 1.4        & 1.4 \\
3            & 1.6        & 1.0 \\
4            & 1.6        & 1.6 \\
\hline
\hline
\end{tabular}
\end{center}
\end{table}

\begin{figure}[h]
\includegraphics[width=7.73cm]{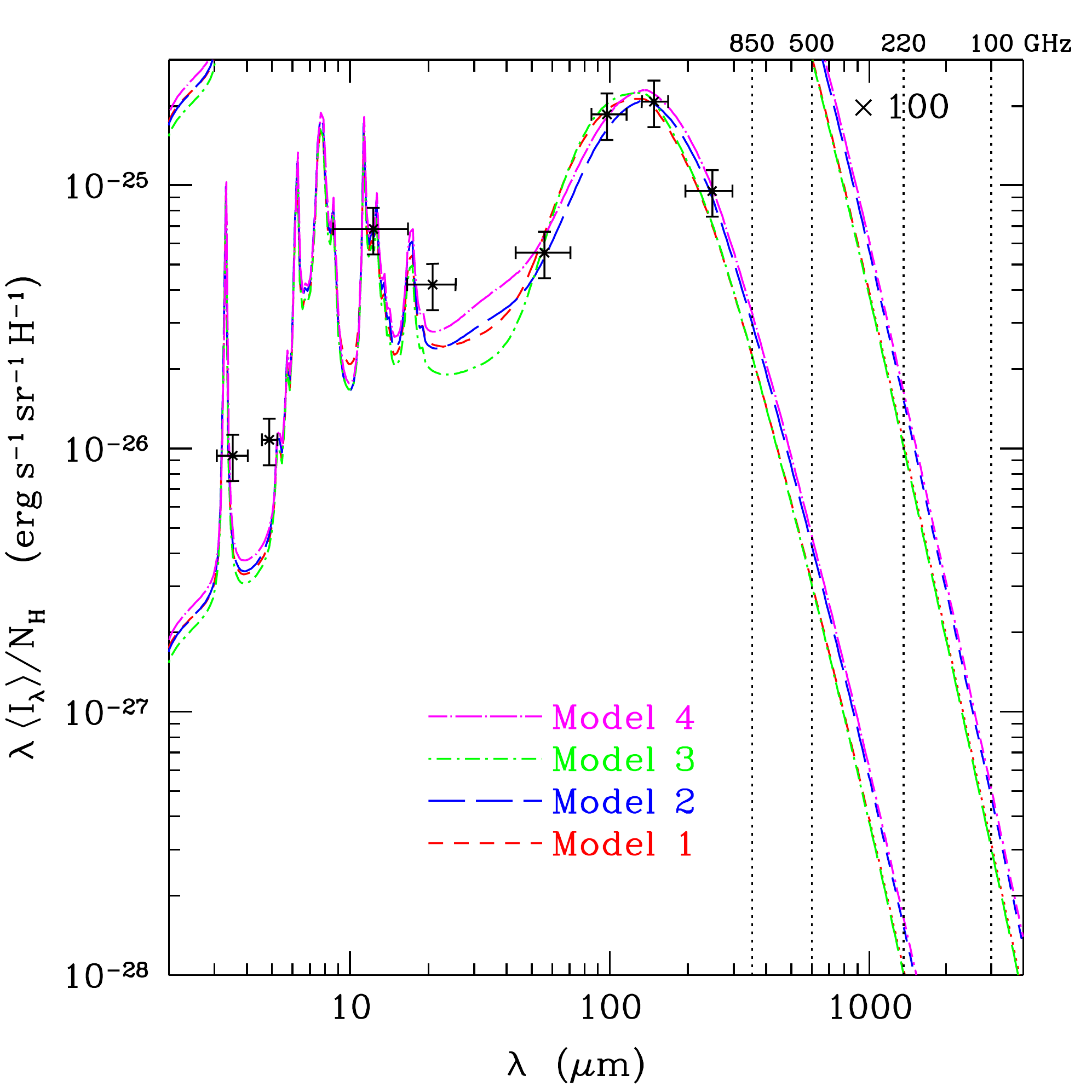}
\includegraphics[width=7.73cm]{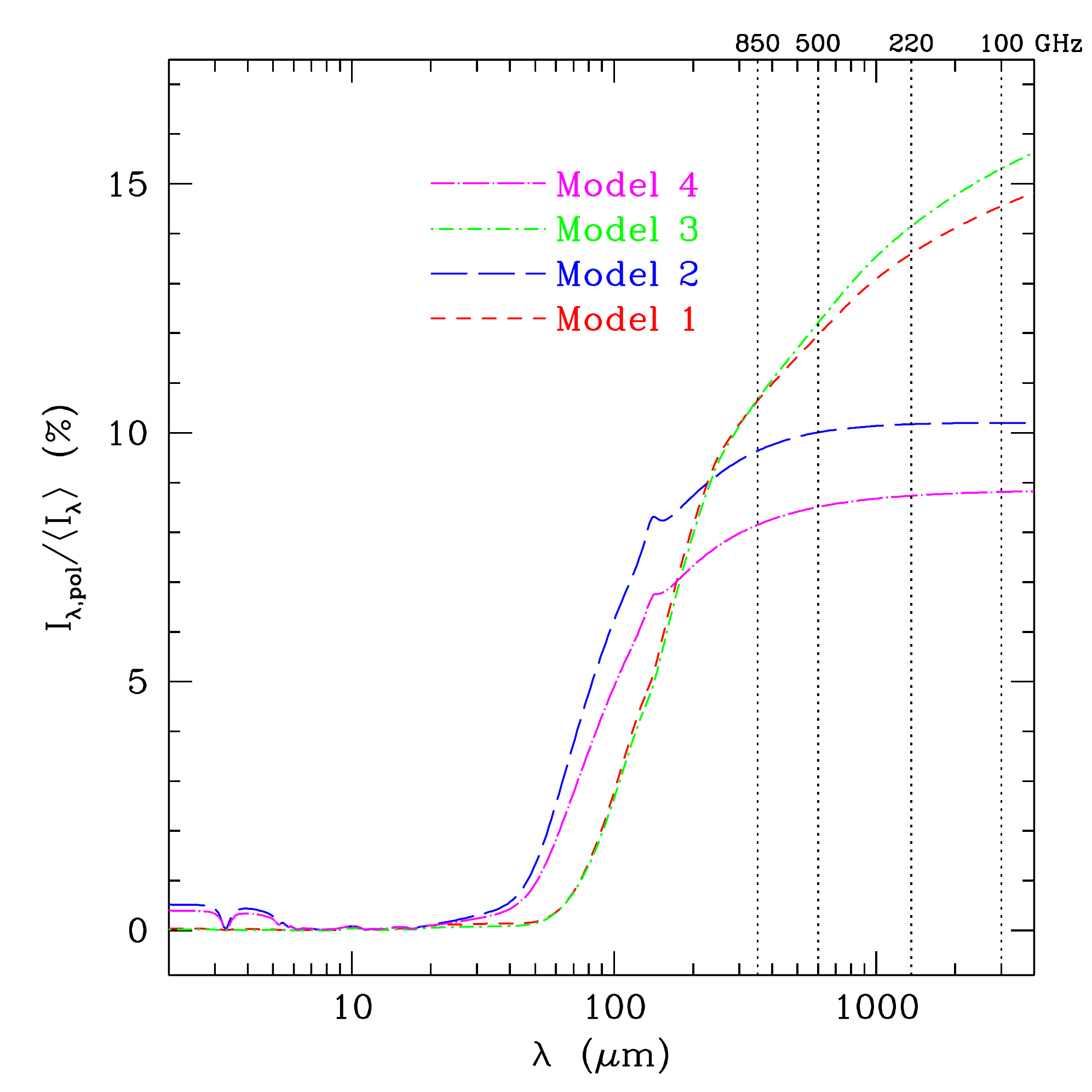}
\caption{\label{f:firpol_predictions}({\it Left}) Far-infrared emission spectra and ({\it Right}) corresponding degrees of polarization calculated by~\protect\citet{Draine+Fraisse_2008} for the models in Tab.~\ref{t:dust_models}.  On each plot, the vertical dotted lines indicate 4 possible CMBPol frequencies ranging from 100~GHz to 850~GHz.  The symbols on the emission spectra represent the HI correlated emission as measured by DIRBE at high Galactic latitudes~\protect\citep{Dwek+Arendt+Fixsen_etal_1997, Arendt+Odegard+Weiland_etal_1998}.}
\end{figure}

Table~\ref{t:dust_models} provides the values of this ratio for each component of the \citet{Draine+Fraisse_2008} dust models, while Fig.~\ref{f:firpol_predictions} shows the corresponding predicted far-infrared emission spectra, as well as its expected degree of polarization as a function of wavelength, under the assumption that all dust grains are heated by the~\citet{Mathis+Mezger+Panagia_1983} Galactic radiation field, and that the observer is looking in a direction perpendicular to the Galactic magnetic field.

Despite the wide range of axial ratios probed by the 4 models listed in Tab.~\ref{t:dust_models} for both amorphous silicate and carbonaceous grains, \citet{Draine+Fraisse_2008} showed that they all provide adequate fits (not shown here) to the observed extinction and polarization of starlight, and give far-infrared emissions compatible with the DIRBE measurements (see Fig.~\ref{f:firpol_predictions}).  The emission predicted at CMB wavelengths is also remarkably similar between all models.

~

However, the expected degree of polarization of the far-infrared emission from interstellar dust and its wavelength dependence are strong functions of the class of models one considers.  When both silicate and carbonaceous grains are non-spherical and partially aligned with the Galactic magnetic field, the fractional polarization of the dust emission is essentially independent of frequency at CMB wavelengths, and of order 10\%.  But if only the emission by silicates contributes to the polarized signal, the polarized fraction of the dust emission can vary by a factor of 1.4 between 850 and 100~GHz, from around 11\% to over 15\%.  For a given class of models, the level of polarization calculated by~\citet{Draine+Fraisse_2008} is not very sensitive to the value of the ratio $b/a$ considered, which makes their predictions relatively robust.

\subsection{Polarized Fraction of Dust Emission for CMB Forecasts}
\label{s:dust_forecasts}

The polarized fractions shown in Fig.~\ref{f:firpol_predictions} may seem high in light of the degrees of polarization reported by Archeops~\citep{Benoit+Ade+Amblard_etal_2004} or \WMAP~\citep{Kogut+Dunkley+Bennett_etal_2007} for the diffuse Galactic emission.   However, there should be very few lines-of-sight in the sky close to perpendicular to the Galactic magnetic field, so the average numbers these groups report, ranging from 4\% to 7\%, \emph{should} indeed be smaller.  That said, on a given sightline, the ratio of the observed polarization to  the polarization calculated for any of the models in Tab.~\ref{t:dust_models} should not be a function of frequency.  Therefore, if one believes that the Galactic magnetic field models used by Archeops and \WMAP~are accurate enough to get a good estimate of the polarized fraction of dust emission \red{of} 5\% at 100~GHz, we can easily rescale the~\citet{Draine+Fraisse_2008} models to reproduce this observation.  All of them are then consistent with a polarized fraction between 3\% and 5\% at all frequencies CMBPol would potentially observe.

Depending on the precision to which the polarized fraction $\Pi$ of dust emission needs to be determined in order to achieve a mission's science goals, such as detecting B-modes, several scenarios might therefore have to be considered in a CMB forecasting pipeline:
\begin{itemize}
\item $\Pi=x\%$ and is independent of frequency for $\nu \lesssim 900$~GHz [as is approximately the case for models 2 and 4 of \citet{Draine+Fraisse_2008}],
\vspace{-0.15cm}
\item $\Pi$ is a decreasing function of frequency, smoothly varying from $\Pi = x\%$ at $\sim 100$~GHz  to $\sim 0.7\,x\%$ at $\sim 850$~GHz [as in models 1 and 3 of \citet{Draine+Fraisse_2008}],
\end{itemize}
\vspace{0.05cm}
where $x$ is chosen either as having a value corresponding to one of the models shown in Fig.~\ref{f:firpol_predictions}, or from an estimate obtained by the Archeops or the \WMAP~groups. In the former case, one should keep in mind that the chosen value will likely be an overestimate of the observable degree of polarization, whereas in the latter case, the polarized fraction might be underestimated compared to its true value due to the coarse resolution and low sensitivity of the Archeops and \WMAP~instruments compared to the \Planck~or the CMBPol detectors.

\vspace{0.2cm}
\subsection{Constraining Dust Models with CMBPol}
\label{s:constraints_pde}

With its many channels and its high resolution, high sensitivity detectors, CMBPol should be able to average the polarized fraction of dust emission on much smaller patches than \WMAP~\citep{Kogut+Dunkley+Bennett_etal_2007}, with potentially higher precision, and to detect variations of the degree of  polarization between frequency channels if it exists, and if the polarized fraction of dust emission is high enough.  How well this can be achieved will of course depend on various factors, including the quality of the Galactic magnetic field models that we will be able to put together by the time the CMBPol data is analyzed.  As a result, it is hard, if not impossible, to know what CMBPol will bring to the debate on what interstellar dust grains look like, and how they behave.  However, we can envision two main game-changing scenarios:
\begin{itemize}
\item Consistent observations of low polarized fractions of dust emission (typically, $\lesssim 5\%$) at all CMB frequencies would suggest that \red{the \citet{Draine+Fraisse_2008}} models in which only silicate grains contribute to the polarized emission are not an obvious good match to the measurements.
\red{In this event, we might have to conclude that} carbonaceous grains \red{are} significantly non-spherical and somewhat aligned with the Galactic magnetic field \red{(as in their models 2 and 4)}, a fact for which we have currently no evidence (see \citet{Chiar+Adamson+Whittet_etal_2006} and discussion in Sec.~\ref{ss:dust_models})\red{. Alternatively, it could mean that the adopted opacities of the silicate material have to be reduced (in a wavelength-dependent fashion), while those of the carbonaceous grains should be correspondingly increased.}
\vspace{-0.15cm}
\item However, if the polarized fraction of dust emission is found to be significantly decreasing when frequency increases, and potentially higher than what the Archeops and \WMAP~results suggest, this would be another indication that \red{amorphous silicate grains and carbonaceous grains indeed form separate populations, with the polarized emission being dominated by the former.  This would pose a challenge to theories of grain alignment.}
\end{itemize}
\vspace{0.05cm}
We also have to mention the possibility that a behavior completely different from the results shown in Fig.~\ref{f:firpol_predictions} could be observed.  Although the models of \citet{Draine+Fraisse_2008} appear to be consistent with all current observations, our understanding of interstellar dust is far from perfect.  Such observations would therefore likely lead us to revisit some of the assumptions of our models for interstellar dust, opening a new window on dust properties and dust physics.

\subsection{Tracing PDE with Starlight Polarization}
\label{s:starlight}

The \citet{Draine+Fraisse_2008} results predict the polarized fraction of far-infrared and submillimeter  dust emission, and its dependence with wavelength, for lines-of-sight perpendicular to the Galactic magnetic field.  
Going from observed degrees of polarization to a function directly useable to constrain dust models (see Sec.~\ref{s:constraints_pde}), and, in particular, theories of grain alignment (see Sec.~\ref{predictions}), therefore requires as good a large-scale magnetic field template as possible.

As mentioned in Sec.~\ref{s:other}, optical or near-infrared starlight polarization gives us the sky-projected direction of the Galactic magnetic field integrated along lines-of-sight going through regions of low dust extinction (typically, $A_V$ or $A_K\lesssim 5$~mag).  As a result, not much can be learned from this technique when it comes to studying the large-scale magnetic field \mhred{in the Galactic disk},
since it cannot probe the field in or near the
\mhred{P}lane outside~of~the~solar~neighborhood.  However, at intermediate and high Galactic latitudes, $|b| \gtrsim10^\circ$, dust extinction is relatively low \citep[see, e.g.,][and references therein]{Burstein+Heiles_1982}, with $A_V \simeq 0.03$~mag near the Galactic poles (the estimated error on this number is of order the extinction itself).
Since these latitudes correspond to the regions of main interest to the CMB community, useful magnetic field templates can therefore be constructed from starlight polarization~observations.

\citet{Heiles_2000} compiled a catalog, the largest to date, of the polarization of starlight from nearly 9300 stars.
However, among those, \citet{Fosalba+Lazarian+Prunet_etal_2002} only found $\sim1400$ stars with reliable distance and extinction estimates lying sufficiently far from the Galactic plane, i.e., with $|b| > 10^\circ$, to be useful for CMB purposes.  Moreover,  the vast majority of these stars, about 1300, are less than 1~kpc away from the Sun.
Despite this lack of in\-for\-ma\-ti\-on outside of the plane, starlight polarization data has already been useful to CMB polarization studies.  \citet{Page+Hinshaw+Komatsu_etal_2007} and \citet{Dunkley+Komatsu+Nolta_etal_2008} both used the  \citet{Heiles_2000} catalog to inform their fit to the dust emission component of their signal, and \citet{Kogut+Dunkley+Bennett_etal_2007} showed that using the polarization angle provided by starlight polarization observations leads to a (marginally) better fit to this component than a model based on
\WMAP's low frequency polarized synchrotron observations.  Since both starlight polarization and polarized dust emission are processes closely related to how grains align with the interstellar magnetic field and to the optical properties of interstellar dust grains, this improvement is not totally unexpected.

The scarcity of optical and near-infrared starlight polarization data has led several groups to try and remedy the situation, especially in view of the accuracies now achievable \citep[see, e.g.,][]{Carciofi+Magalhaes+Leister_etal_2007}.
An example is the Galactic Plane Infrared Polarization Survey~(GPIPS), which aims at
measuring nearly half a million $H$-band stellar polarizations for $H$ down to 12~mag
across a 72 square degree region of the inner Northern Galactic plane by 2011 \citep{Clemens_Pinnick_Pavel_etal_2007}. 
But neither the GPIPS nor other surveys currently planed or underway \citep[see, e.g.,][]{Magalhaes+Pereyra+Melgarejo_etal_2005, Nishiyama+Tamura+Hatano_etal_2008} will lead to a significant increase in the number of available data points in the diffuse interstellar medium at high Galactic latitudes.
Given the implications this new data would have for our understanding of dust processes, but also for our ability to separate polarized dust emission from a potential gravity wave signal,
we believe that acquiring it should be a high priority goal achieved by the time CMBPol flies.

\vspace{0.6cm}
\section{Constraining Grain Alignment Mechanisms}
\label{s:gam}

The mechanism explaining how non-spherical dust grains align with a magnetic field has been sought for ever since \citet{Hall_1949} and \citet{Hiltner_1949} discovered dust-induced starlight extinction polarization.
This quest has been highly controversial, and numerous mechanisms have been proposed and developed to various degrees over the years \citep[see, e.g., the review by][]{Lazarian_2007}.
Among those is the paramagnetic Davis-Greenstein mechanism that has been matured through intensive work  \citep{Jones+Spitzer_1967, Purcell_1979, Spitzer+McGlynn_1979, Mathis_1986, Roberge+Degraff+Flaherty_1993, Lazarian_1997, Roberge+Lazarian_1999} since its introduction by \citet{Davis+Greenstein_1951}.  However, this ``textbook solution" to the grain alignment problem has several shortcomings, among which the fact that thermal flipping prevents grains with effective radii $a \lesssim 1~\um$ from aligning~\citep{Lazarian_2007}, which is hard to reconcile with fits to starlight polarization data~\citep{Draine+Fraisse_2008}. Radiative torques appear to not only solve this problem, but to also explain why, and not only how, large grains ($a \gtrsim 0.1~\um$) align while small grains don't~\citep[as first established by][]{Kim+Martin_1995}.

This promising mechanism was first discovered by \citet{Dolginov+Mitrofanov_1976}. Considering a grain exhibiting a difference in optical cross-section for right-handed and left-handed photons, they noticed that scattering of unpolarized light by the grain resulted in its spin-up, which leads to its alignment with the surrounding magnetic field. However, they did not have the necessary tools to quantify this effect and, as a result, their pioneering work was mostly ignored in the following 20 years. \citet{Draine+Weingartner_1996} realized that this mechanism could be studied with a modified version of the DDSCAT
code by \citet{Draine+Flatau_1994}.  In particular, this work, along with follow-up studies by \citet{Draine+Weingartner_1997} and \citet{Weingartner+Draine_2003}, demonstrated that the magnitude of radiative torques is very substantial for the grain shapes they study.  It is worth pointing out that the spin-up of grains by radiative torques was later observed in laboratory conditions by \citet{Abbas_Craven_Spann_etal_2004}. 

However, radiative torques have the inconvenien\btdred{ce} of depending on many parameters, such as grain shape, grain size, radiation wavelength, grain composition, and the angle between the radiation direction and the local magnetic field.  It is therefore difficult to compute their effect in general, 
and empirical studies have so far been focused on demonstrating the radiative torque effect for specific values of these parameters.  As a result, it has been claimed that the predictive power of this theory was limited, which is true if a ``brute force" approach is used to explore the relevant parameter space.
An alternative is to use analytical models, such as the one proposed by \citet{Lazarian+Hoang_2007a}, to describe the effects of radiative torques.

\subsection{Analytical Model of Radiative Torques}
\label{s:model}

\begin{figure}[h]
\begin{center}
\includegraphics[width=3.1in]{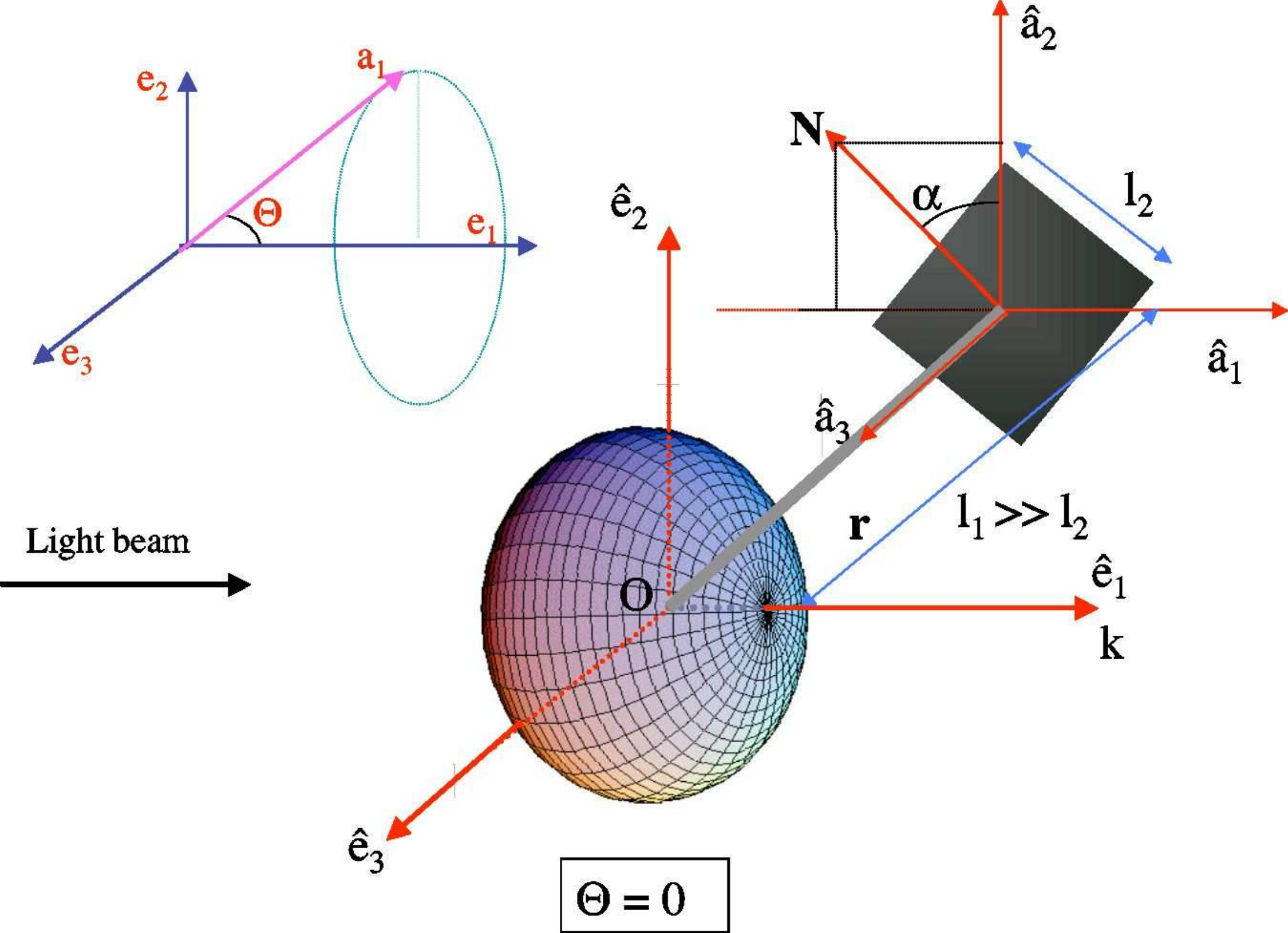}
\caption{
\protect\citet{Lazarian+Hoang_2007a} model of a ``helical'' grain consisting of a spheroidal grain with an inclined mirror attached to it.  This model can be used to simplify the calculation of radiative torques (see text). ${\bf k}$ indicates the direction of propagation of the incoming radiation, and $\Theta$ the angle between the incident light and the polar axis of the ellipsoidal~grain.
}
\label{AMO}
\end{center}
\end{figure}

\citet{Lazarian+Hoang_2007a} demonstrated that the simple model shown in Fig.~\ref{AMO} can be used to accurately calculate the radiative torques exerted by an incoming radiation on a dust grain.  In this model, grains are assumed to be ellipsoidal, and each grain is equipped with a mirror attached to
its side.  
A grain can be both  
``left-handed'' and ``right-handed" (from the point of \nred{view} of the incoming radiation) depending on the orientation of its mirror with respect to the incident light -- the grain shown in Fig.~\ref{AMO} is left-handed, and would become right-handed if the mirror were rotated by $90^\circ$ in $\alpha$.
Radiative torques are then computed under the assumption that geometric optics apply, and compared to numerical calculations performed by DDSCAT.

~

\begin{figure}[h]
\begin{center}
\includegraphics[width=3.2in]{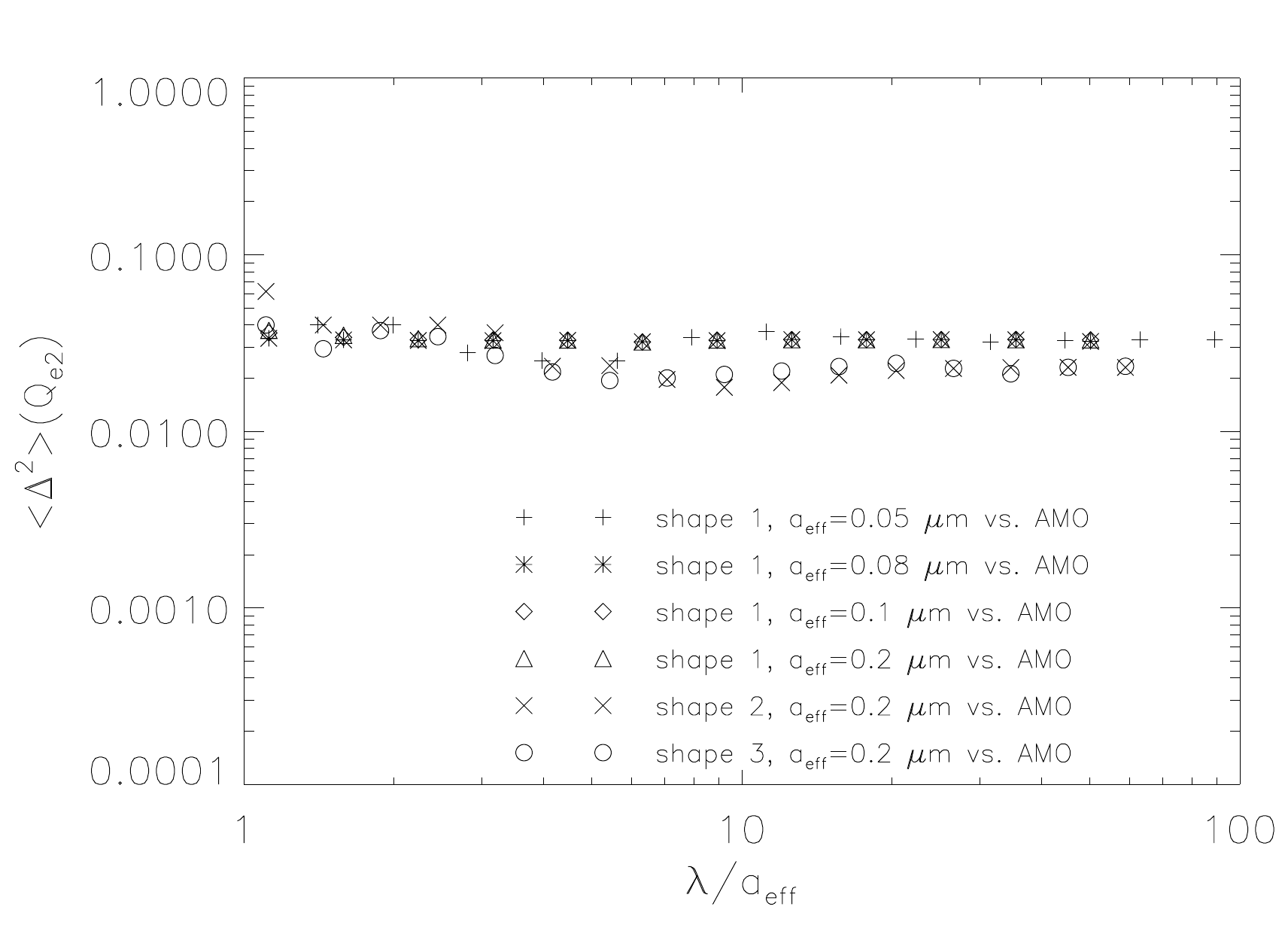}
\caption{
Numerical comparison of the radiative torques calculated with DDSCAT for irregular grains with the results provided by the analytical model (``AMO") described in Sec.~\ref{s:model} using the wavelength-dependent $\chi^2$ estimator defined by Eq.~(\ref{chi_eq}). $a_{\rm eff}$ refers to the radius of a sphere with the same volume as the ellipsoid used in the analytical model (the index is omitted in the text),
and the considered shapes are DDSCAT shapes \protect\citep[see][]{Lazarian+Hoang_2007a}.
}
\label{comparison}
\end{center}
\end{figure}

The comparison is done for light propagating in the direction of the grain's short axis ${\bf \hat{e}_1}$, while the grain's angular momentum $\bJ$ is in the $({\bf \hat{e}_1}, {\bf \hat{e}_2})$ plane.  For the purpose of modeling grain alignment, only torques in that plane then matter -- a torque in the ${\bf \hat{e}_3}$ direction would induce grain precession negligible compared to the Larmor precession of the grain in the interstellar magnetic
field (see Tab.~1 of \citet{Lazarian_2007} for the time scales involved).  As a result, the model only needs to accurately reproduce the components
$Q_{\rm e1}$ and $Q_{\rm e2}$ of the torque ${\bf Q}$ along ${\bf \hat{e}_1}$ and ${\bf \hat{e}_2}$, which it does.  Defining $\Theta$ as the angle between the
magnetic field and the radiation direction, \citet{Lazarian+Hoang_2007a} indeed showed that the deviation of the torques calculated numerically for irregular grains by DDSCAT from their analytical predictions is always modest at the wavelengths of interest. As an illustration, Fig.~\ref{comparison} shows 
\begin{equation}
\langle \Delta^2\rangle(Q_{\rm e2})\equiv\frac{1}{\pi (Q_{\rm e2}^{\rm max})^2} \int^{\pi}_{0} \left[Q_{\rm e2}^{\rm DDSCAT}(\Theta) - Q_{e2}^{\rm model} (\Theta)\right]^2 \d\Theta\,,
\label{chi_eq}
\end{equation}
which characterizes the deviation in the ${\bf \hat{e}_2}$ direction,
as a function of wavelength for various grain shapes and sizes.  \nblue{T}he value of $\Delta^2$ \nblue{remains small no matter the values} of these parameters, \nblue{which indicates a remarkable} agreement between analytical model and numerical calculations.

In fact, \citet{Lazarian+Hoang_2007a} showed that, for a given grain, radiative torque alignment is fully determined by  
the ratio $q\equiv Q_{\rm e1}^{\rm max}/Q_{\rm e2}^{\rm max}=
Q_{\rm e1}(0)/Q_{\rm e2}(\pi/4)$\nred{. This} enormously simplifies the calculations
of radiative torques \nred{since} one does not need to calculate {\it two functions}, $Q_{\rm e1}(\Theta)$ and $Q_{\rm e2}(\Theta)$, but only {\it two numbers}, $Q_{\rm e1}(0)$ and $Q_{\rm e2}(\pi/4)$, to characterize the alignment.
$q$ is therefore as important to grain alignment as grain axial
ratios are to determining the polarized radiation by aligned \nred{interstellar dust} grains.

~

\vspace{-0.7cm} 
\vspace{-0.9cm} 

\subsection{Alignment Configuration}
\label{s:why}

Observations indicate that interstellar grains tend to align with long axes perpendicular to the local magnetic field, a fact once frequently used to argue that the Davis-Greenstein mechanism had to be responsible for the alignment.  Radiative torques also explain this configuration.

Interstellar grains indeed experience internal relaxation that tends to make them rotate about their axis of maximal moment of inertia, which, in the case of an ellipsoidal grain, is the grain's short axis. Thus, the grain's angular momentum $\bJ$ is perpendicular to its long axis, and it is therefore sufficient to follow its dynamics to determine the alignment configuration.
Let us call the component of the torque parallel (respectively, perpendicular) to ${\bf J}$ the {\it spin-up torque} ${\bf H}$ (respectively, the {\it alignment torque} ${\bf F}$).  The angular momentum ${\bf J}$ is precessing about the magnetic field $\bB$ due to the magnetic moment of the grain \citep{Dolginov+Mitrofanov_1976}. 
\nred{It is possible to show that, when the angle $\xi$ between $\bJ$ and $\bB$ tends to zero (or, by symmetry,~$\pi$), the alignment torque ${\bf F}$ averaged over one Larmor precession of $\bJ$ about $\bB$ also tends to zero \citep[see][]{Lazarian+Hoang_2007a}.}
In other words, $\langle {\bf F}\rangle_{\xi}$ vanish as $\xi\to 0$ \nred{ or $\pi$} irrespectively of the radiative torques' functional form, and \nred{the points} $\xi=0$ \nred{and $\pi$ are attractors}.

This qualitative argument is quite general, but it does not address the question of whether there are other stationary points, and in particular of whether the alignment can also happen with ${\bf J}$ perpendicular to ${\bf B}$.
To answer this question, one has to consider the behavior of the functions $Q_{\rm e1}(\Theta)$ and $Q_{\rm e2}(\Theta)$. The analysis in \citet{Lazarian+Hoang_2007a} shows that there is, indeed, a range of angles between the direction of radiation and the magnetic field for which grains tend to align in the ``wrong" way, i.e., with long axes parallel to the magnetic field.
However, this range is very narrow around $\pi/2$, and therefore not typical of conditions in the interstellar medium.  Moreover, such an alignment corresponds to positions for which the spin-up torque ${\bf H}$ is negative, which induces grain alignment with low angular momentum.  Grain thermal wobbling at low-$J$ attractor points leads to variations
in $\xi$ that typically exceed the range of angles in which ``wrong alignment" happens \citep{Lazarian_1994, Lazarian+Roberge_1997} . Therefore, grains {\it always} align with long axes perpendicular to the magnetic field.

\subsection{Testing Quantitative Predictions with CMBPol}
\label{predictions}

In Sec.~\ref{s:model}, we pointed out that, when induced by radiative torques, grain alignment is fully characterized by the value of the parameter $q$. For typical interstellar conditions, this process dominates over all other alignment-inducing torques, and alignment of interstellar dust grains should therefore be governed by $q$ independently of the magnitude
of radiative torques.

~

\vspace{-0.5cm} 
\begin{figure}[h]
\begin{center}
\includegraphics[width=3.2in]{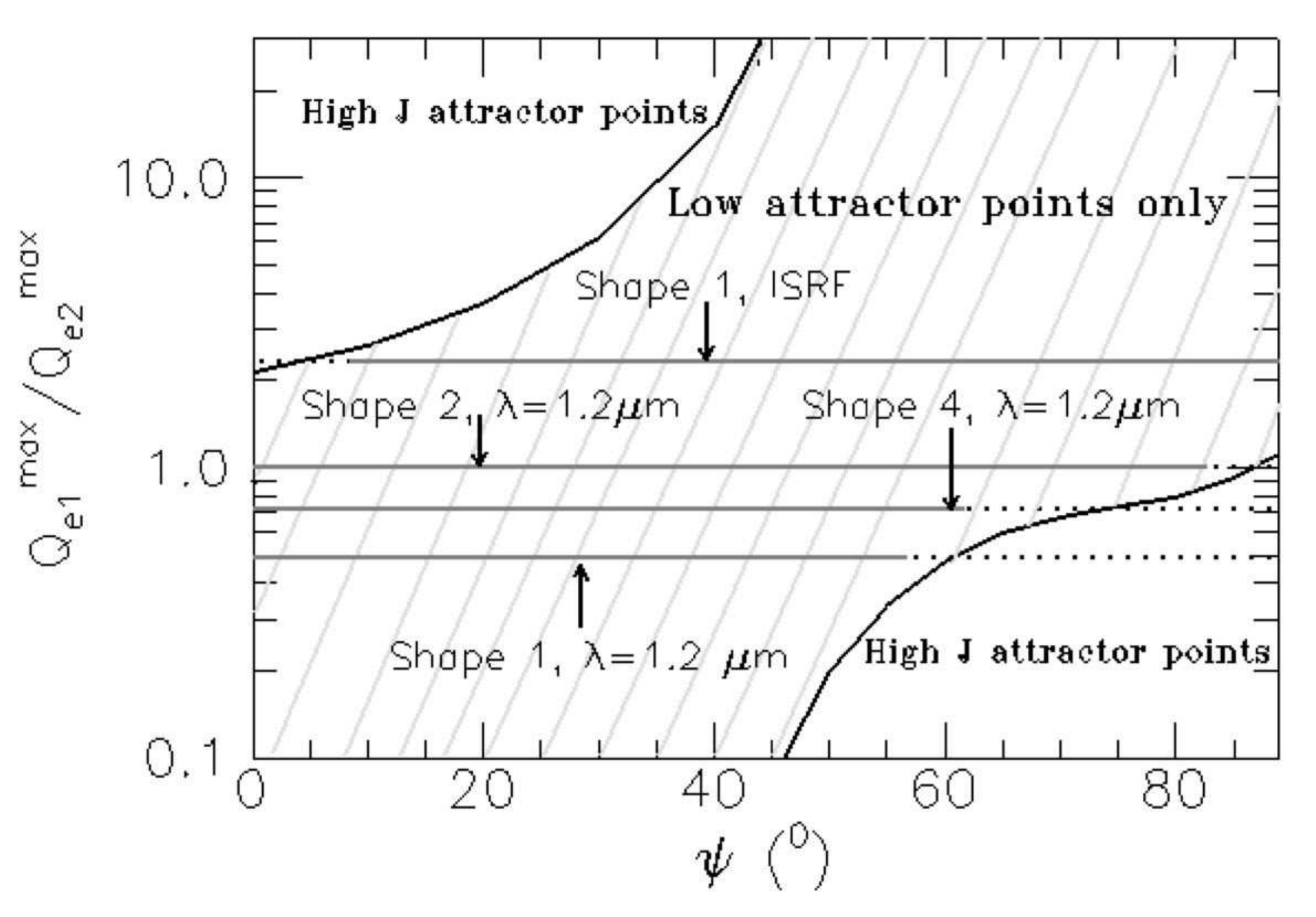}
\caption{Type of available attractor points in the $(q,\psi)$ parameter space (see text for the definition of these parameters).  When both high-$J$ and low-$J$ attractor points are present, grains eventually perfectly align in the state of highest $J$.  If only low-$J$ attractor points are present, the alignment is partial. ``ISRF" refers to the local interstellar radiation field, and the various shapes considered here are DDSCAT shapes \protect\citep[see][]{Lazarian+Hoang_2007a}.}
\label{chi}
\end{center}
\end{figure}
\vspace{-0.7cm} 

Fig.~\ref{chi} shows how a given value of $q$ can be translated into information about grain alignment.
In particular, depending on the angle $\psi$ between the incident radiation and the local magnetic field, grains align with $\xi$ taking the value of high-$J$ or low-$J$ attractor points.
In the former case, alignment with the local magnetic field is perfect.  However, 
when $\xi$ takes the value of a low-$J$ attractor point, the alignment is unstable, which leads to \nred{a degree of alignment of} only $\sim 20\%$ to 50\%. 
This situation is actually the most common for relevant values of $q$, meaning that most grains align {\it subthermally}.  This conclusion is quite different from the assumption by \citet{Draine+Weingartner_1996} that in the presence of radiative torques
most interstellar grains should rotate with $T_{\rm rot}\gg T_{\rm gas}$, and therefore align suprathermally with $\bB$.

When superparamagnetic grains, i.e., grains with enhanced paramagnetic relaxation, are considered, the conclusion is quite different.
These grains, invoked by \citet{Jones+Spitzer_1967} in the context of paramagnetic alignment \citep[see also arguments in support of their existence in][]{Bradley_1994, Martin_1995,Goodman+Whittet_1995}, indeed exhibit 
low-$J$ attractor points when subjected to the diffuse interstellar radiation field, as for ordinary paramagnetic grains, but a more stable high-$J$ attractor point also always exists.
Therefore, if a grain aligns \nred{at} a low-$J$ attractor point, gaseous collisions ensure that its state will change to that of the high-$J$ attractor point.
Thus, rather
unexpectedly, intensive paramagnetic relaxation changes the rotational state of dust grains, enabling them to
rotate more {\it rapidly}, and leading to perfect suprathermal alignment with the local interstellar magnetic~field.

Although we mentioned that radiative torques dominate over all other alignment-inducing torques in the diffuse interstellar medium, this might not be the case for paramagnetic grains undergoing ${\rm H}_2$ formation at their surface.  The resulting ``pinwheel" torques \citep{Purcell_1979}
can indeed create new high-$J$ attractor points \citep{Hoang+Lazarian_2008b}.  As a result, provided that H$_2$ torques are as strong
as considered in \citet{Purcell_1979} and subsequent papers by \citet{Spitzer+McGlynn_1979}, \citet{Lazarian_1995}, and \citet{Lazarian+Draine_1997}, 
higher degrees of polarization are expected in regions of higher atomic hydrogen content.
Detection of these variations would be highly interesting since it would rule out 
the superparamagnetic dust grains hypothesis for which the alignment should be perfect irrespectively of other factors.

Thanks to the existence of these quantitative predictions, \nred{the theory of} grain alignment \nred{is} now becoming testable. \citet{Lazarian_2007} reviews some of the first tests the theory of radiative torque alignment has already passed.  Since then, it has also been tested in dense environments such as molecular clouds \citep{Andersson+Potter_2007,Whittet+Hough+Lazarian_etal_2008}.
The current agreement between theory and observations is encouraging, and the mere fact that quantitative predictions exist that can be tested shows the tremendous progress made in the study of grain alignment over the course of the last few years.  However, additional tests are required, and CMBPol offers a great opportunity to further constrain theories of grain alignment.  By improving our understanding of the structure of the Galactic magnetic field (Sec.~\ref{ss:gmf_cmb}), and our ability to measure polarized dust emission (Sec.~\ref{s:constraints_pde}) and morphologically trace foregrounds (which we discuss in Sec.~\ref{s:morphol_trace}), CMBPol will indeed make it possible to test radiative torque alignment predictions at multiple frequencies and in various environments.

In the meantime, observations at smaller scales, and later on with the \Planck~satellite, will improve our knowledge of grain alignment.  With a better understanding of the degree of alignment of various types of grain, it will become possible to make more robust predictions of the expected polarized dust emission at CMB frequencies by combining fits to data at other wavelengths with theory, therefore improving on the \citet{Draine+Fraisse_2008} results.  These improved templates will be helpful in generating more accurate sky models used as input to study how to best separate foreground emissions from cosmological signal \citep[for more details on this topic, we refer the reader to the companion paper by][]{CMBPol_Dunkley_2008}.

\section{Heliospheric Dust Emission}
\label{s:hde}

\subsection{Basic Properties from In Situ Observations}

Circumheliospheric interstellar material (CHISM) flows through the
heliosphere with a heliocentric velocity of 26.3~km$/$s from the
upwind direction  $ \ell \sim 3.5^\circ$, $b \sim 15.2^\circ$.
Approximately 0.3\% to 0.9\% of the CHISM mass is contained in
interstellar dust grains with masses ranging from $10^{-15}$ to $10^{-10}$~g
\citep{Grun+Gustafson+Mann_etal_1994,Frisch+Dorschner+Geiss_etal_1999,Landgraf+Baggaley+Grun_etal_2000,Witte_2004,Slavin+Frisch_2008}, corresponding to radii $a$ between $\sim 0.04$~\micron~and $2$~\micron~for spherical grains with density 2.5 g/cm$^3$.  The dynamic interaction of an
interstellar dust grain (ISDG) with the heliosphere depends on
the ratio of radiation pressure to gravitational force ($\beta$), as well as on the Lorentz force, which is a function of grain velocity,
charge-to-mass ratio ($Q/m$), and solar wind magnetic field.
Consequently, the distribution of ISDGs in the heliosphere varies
between upwind and downwind, i.e., between Northern and Southern ecliptic
regions, and with the 22-year solar magnetic activity cycle. \Ulysses~observed a varying flux of grains through the inner heliosphere of
$0.6$ -- $2 \times 10^{-8}$ \cmtwo~s$^{-1}$ \citep{Landgraf+Kruger+Altobelli_etal_2003}.
The composition and optical properties of ISDGs interacting with
the heliosphere are consistent with astronomical silicates such as
olivine; organic particles \btdred{may} be excluded by the
overabundance of carbon in CHISM gas
\citep{Landgraf+Augustsson+Grun_etal_1999,Slavin+Frisch_2008}.  However, \Helios~data
inside 1 AU showed large ISDGs (radii between 1 and 2 \micron), with mixed
composition grains containing both silicates and carbonaceous material
\citep{Altobelli+Grun+Landgraf_2006}.

~

\vspace{-0.2cm}
\begin{figure}[h]
\begin{center}
\includegraphics[scale=0.17]{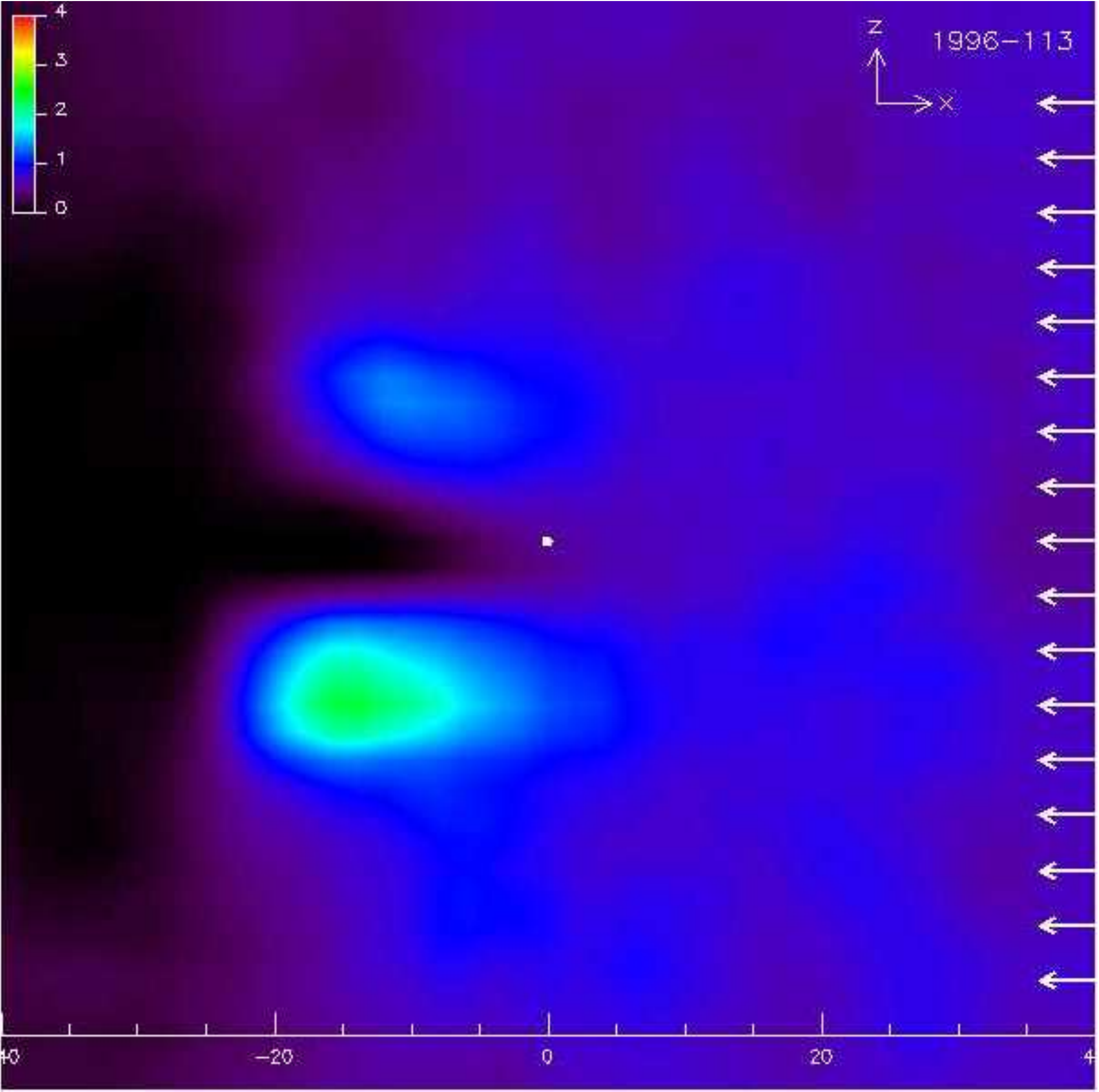}
\includegraphics[scale=0.17]{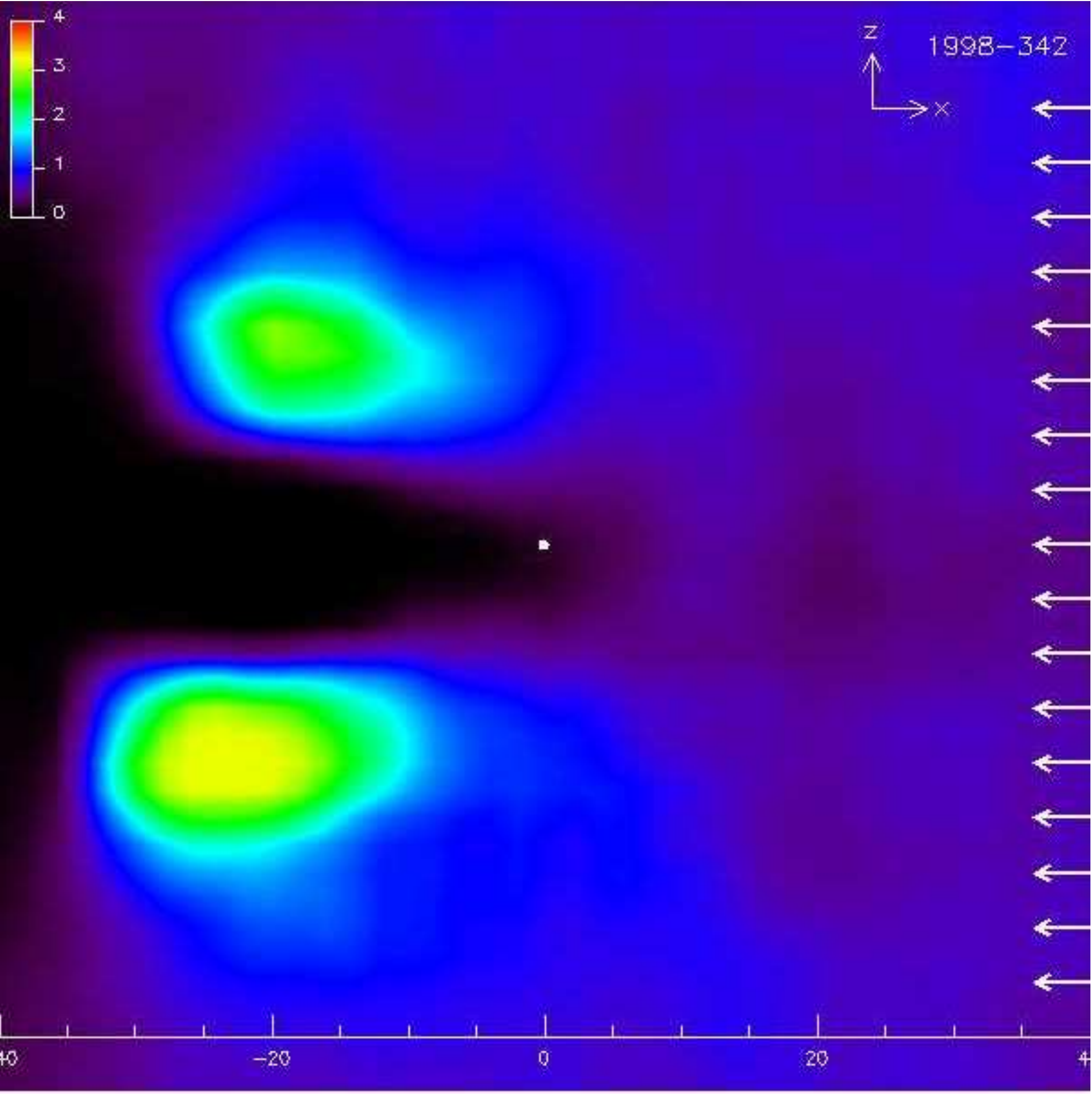}
\includegraphics[scale=0.17]{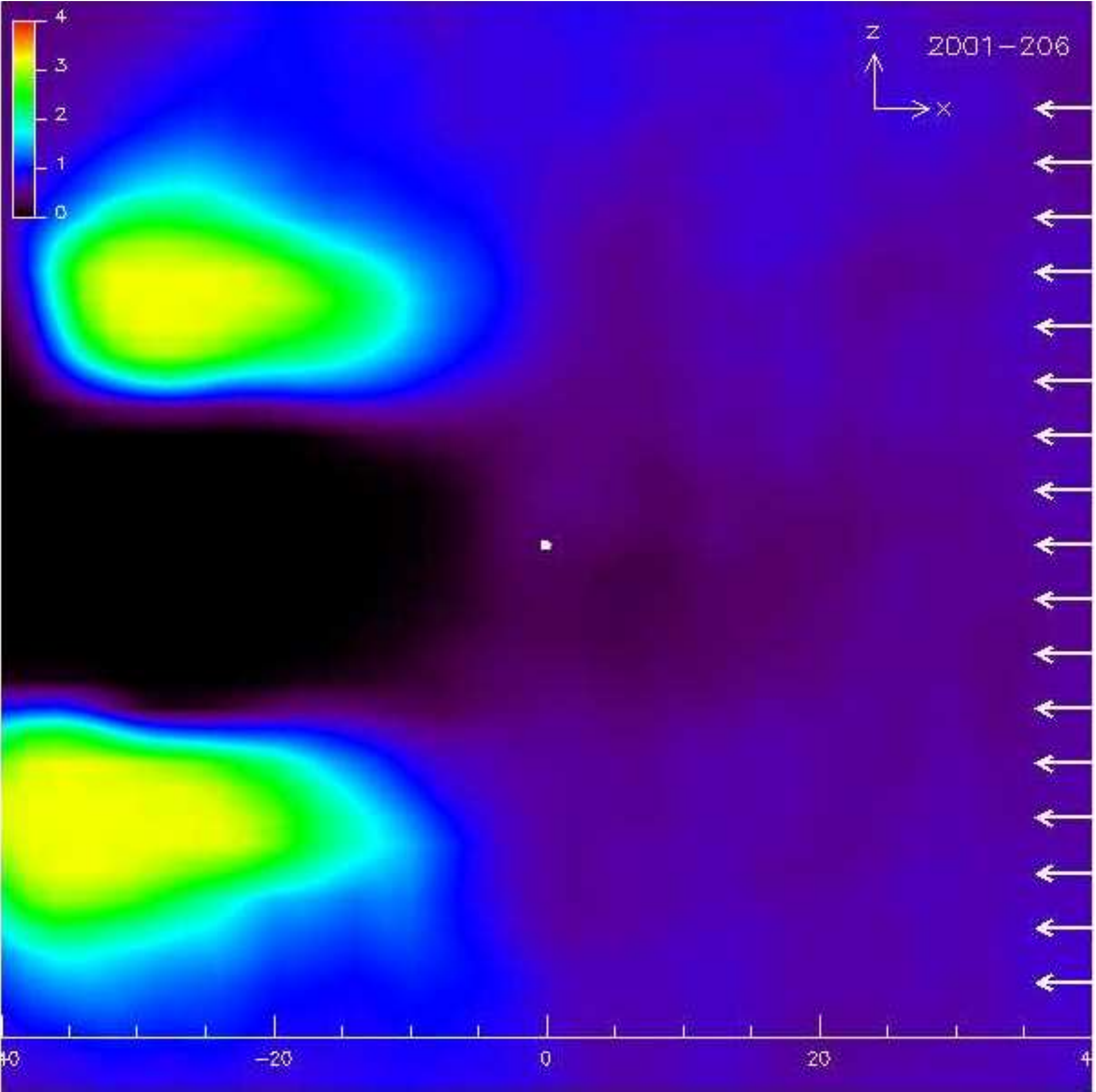}

\includegraphics[scale=0.17]{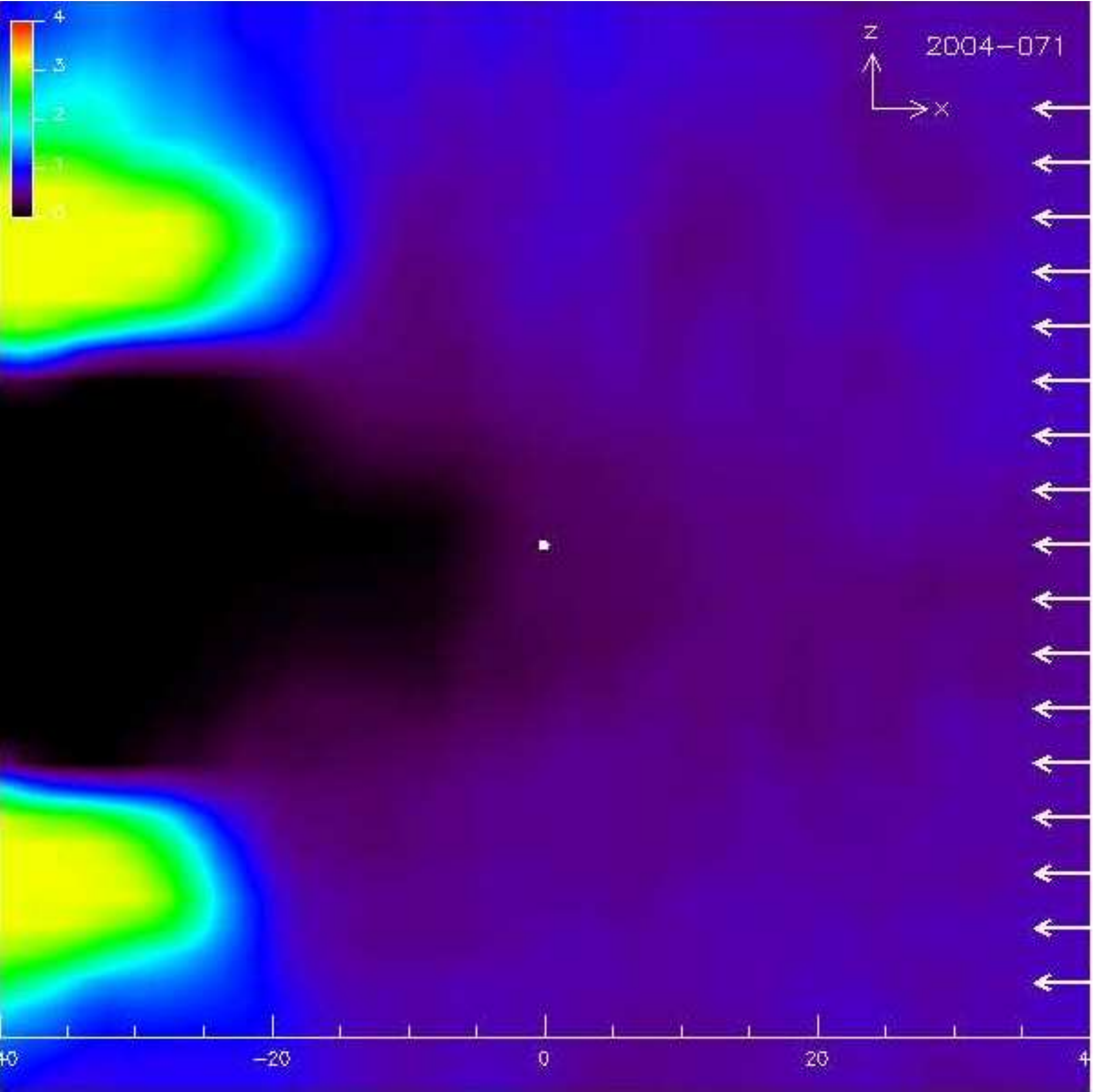}
\includegraphics[scale=0.17]{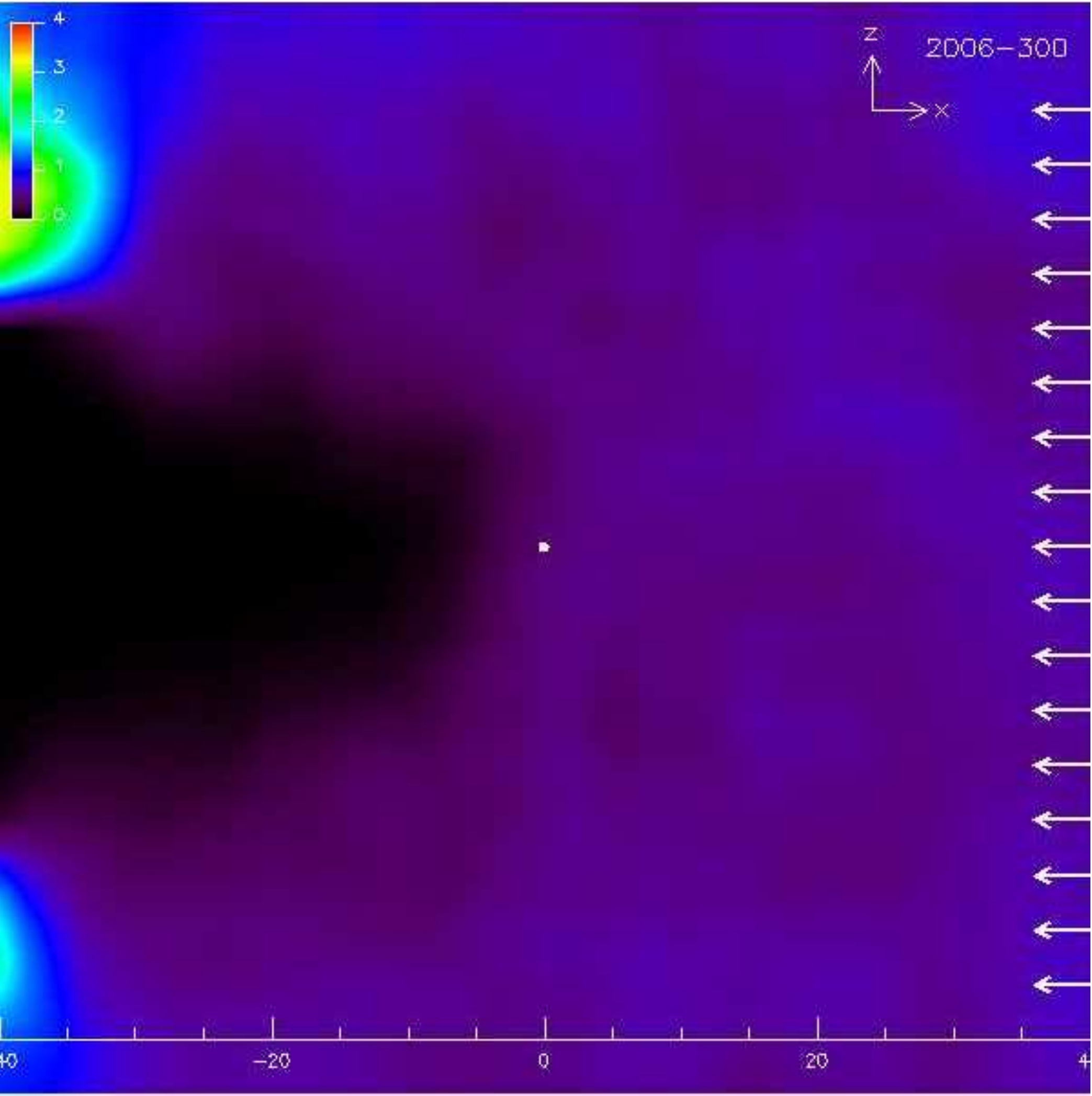}
\includegraphics[scale=0.17]{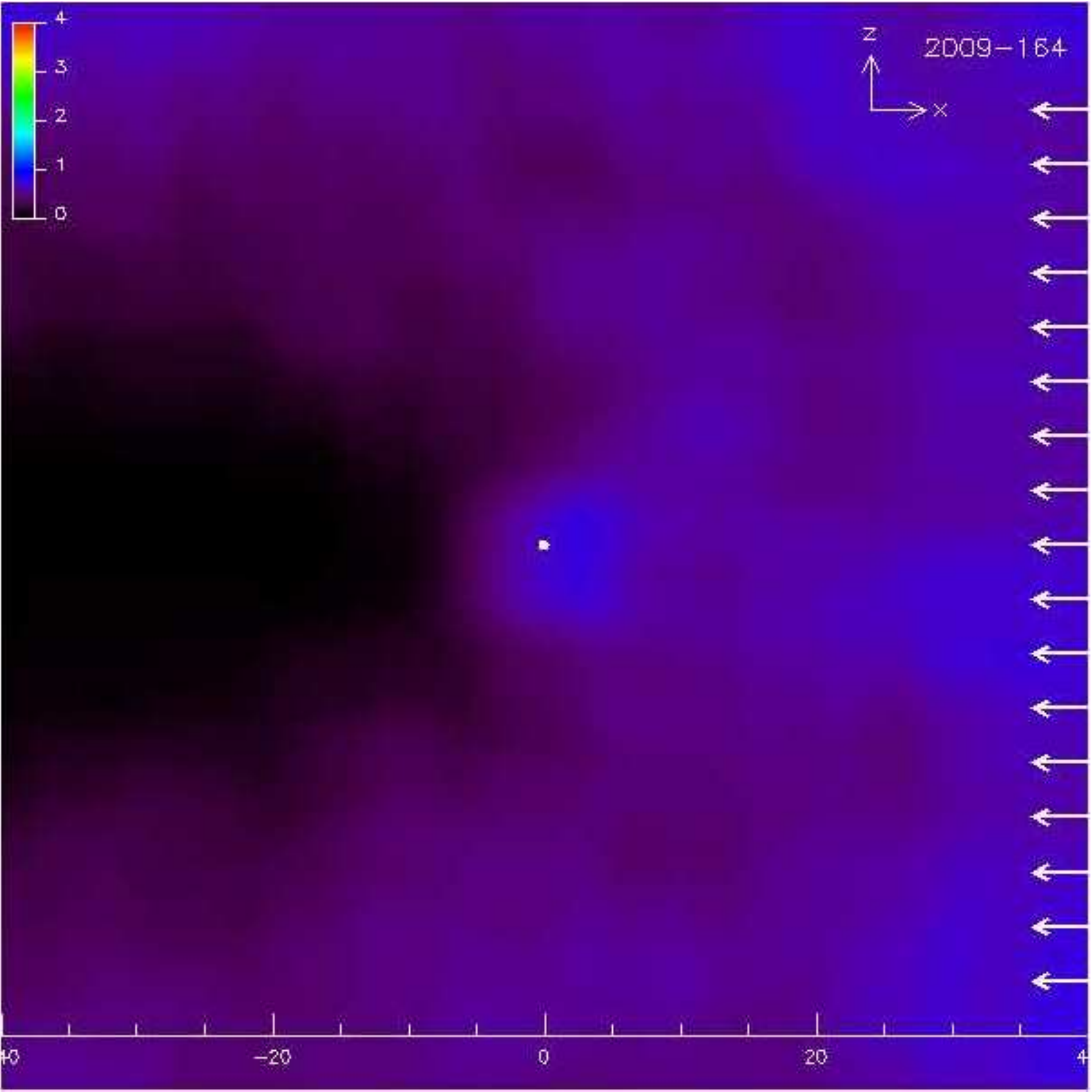}
\end{center}
\caption{Simulations by \protect\citet{Landgraf_2000} of the distribution of interstellar dust grains
with radii $a \sim 0.2$ \micron\ in the X-Z plane (see below) are shown for the
years 1996, 1998.5, 2001, 2003.5, 2006, 2008.5 (upper left through
lower right).  The 1996 figure (upper left)
shows the grain distribution during solar minimum defocusing
conditions, while the 2006 figure (lower middle) shows the distribution
during solar minimum focusing conditions.  ISDGs
move through the heliosphere at velocities of about $5 $ AU/yr, so the dust distribution in each panel reflects the
influence of the solar cycle over several years prior to that date.
The 1996 solar cycle phase is similar to that expected for a 2018
CMBPol launch date.  Individual plots are $80$~AU $\times$ $80$~AU, with the
Sun at the center.  The X-axis is directed
towards the upwind direction of the CHISM flowing through the
heliosphere, while the Z-axis is perpendicular to the X-axis and nominally
shows the solar rotation axis since the upwind direction is $5^\circ$
from the ecliptic plane.  The color scale is given in units of the
dust density in the diffuse interstellar medium, and ranges from 0 (deep purple) to 4 (red).
\label{f:hde}
}
\end{figure}
\vspace{-0.2cm}

The dependence of grain propagation on $Q/m$ and $\beta$ leads to
several characteristics of the spatial distribution of ISDGs in the
heliosphere
\citep{Kimura+Mann_1998,Frisch+Dorschner+Geiss_etal_1999,Landgraf+Kruger+Altobelli_etal_2003,Frisch_2007,Kruger+Grun_2008}:
({\bf a}) There is a severe reduction in the number of grains with radii $ a
\lesssim 0.2$ \micron\ penetrating to within 5 AU of the Sun when compared
to the initial interstellar power law distribution (see, e.g., the simulations shown in Fig.~\ref{f:hde}).
({\bf b})
Large grains, $a \gtsim 0.5 $ \micron, on hyperbolic orbits are
gravitationally focused into a tail downwind of the Sun
\citep{Grogan+Dermott+Gustafson_1996,Landgraf_2000}.
({\bf c}) A gap in the distribution
of ISDGs with mass between $10^{-14}$ and $10^{-13}$~g is found $\sim 2$ -- $4$~AU
from the Sun due to the effect of radiation pressure \citep{Landgraf+Augustsson+Grun_etal_1999}.
({\bf d})
Small grains with masses $ m \lesssim 10^{-13}$~g are alternately focused
towards, or away from, the ecliptic plane depending on solar magnetic
polarity \citep{Grogan+Dermott+Gustafson_1996,Landgraf_2000}.
({\bf e}) There is a time
lag of $\sim 4$ -- $6$ years between solar cycle phase and its influence on
inner heliosphere ISDGs because of the nominal 5 AU/yr grain
propagation velocity through the heliosphere.
({\bf f}) The CHISM contains
large ISDGs, $m \gtrsim 10^{-12}$~g, that are difficult to explain in the
context of the extinction properties of more heavily obscured
sightlines \citep{Frisch+Dorschner+Geiss_etal_1999,Landgraf+Baggaley+Grun_etal_2000,Draine_2008}.  Only
properties ({\bf b}), ({\bf e}), and~({\bf f}) above are relatively invariant with the solar
cycle phase.  

Several current or proposed spacecraft missions have the potential to
return data on \nobreak{ISDGs} in the heliosphere by the time CMBPol flies.  The only spacecraft presently capable of making in situ
measurements of ISDGs beyond 1 AU from the Sun is \Cassini.  In situ data collected
during the same time period by \Ulysses~and \Cassini~find consistent
results, detecting grains with radii $\sim 0.4$ \micron\ on hyperbolic
orbits with fluxes $\lesssim 2 \times 10^{-9}$ \cmtwo\ s$^{-1}$
\citep[see the analysis by][]{Altobelli+Dikarev+Kempf_etal_2007}. The \Cassini~mission has now been
extended to 2010; the further extension of this mission to $\sim 2016$
(depending on spacecraft consumables) would provide in situ data on
\nobreak{ISDGs} for a similar solar magnetic polarity as expected for the CMBPol
launch time.  Proposed interstellar dust missions include DUst Near
Earth (DUNE) \citep{Grun+Srama_etal_2006} and SAmple Return of Interstellar Matter
(SARIM) \citep{Srama+Stephan+Grun_etal_2008}. DUNE will have the capability of
acquiring the mass distribution, charge state, elemental and isotopic
composition, and velocity vectors of ISDGs near 1 AU, while SARIM will
return interstellar and interplanetary dust grains for laboratory
analysis, providing ``ground-truth" on ISDGs.

\subsection{Emission Characteristics and CMBPol Observations}
\label{s:hde_cmbpol}

The heating of interstellar dust grains $1$ -- $90$ AU from the Sun is
dominated by solar radiation, yielding a grain temperature in K at distance
$R$ from the Sun of $ \sim 278 ~ R^{-0.5}$, where $R$ is in units of AU.  The emission of these
grains is expected to be $\sim 10^{-3}$ of zodiacal light emission
\citep{Mukai_1981,Frisch+Dorschner+Geiss_etal_1999}.  For some inner heliosphere viewing
configurations and solar cycle phases, strong asymmetries in ISDG
emissions are expected for leading versus trailing sightlines with
respect to the Earth's orbital motion, and high versus low ecliptic
latitude lines-of-sight.
\citet{Grogan+Dermott+Gustafson_1996} predict emissions of $\sim
0.1$ MJy/sr at 12 \micron\ for sightlines that view the hotter
ISDGs near the Sun during a focusing phase.
Smaller grains at low solar elongation angles dominate this emission.  
\btdred{\citet{Fixsen+Dwek_2002} find t}he ratio of the 12 $\mu$m to 340 $\mu$m  flux
to be $I_{12}/I_{340} \sim 500$ for zodiacal light.
Full models of the emissions from heliospheric interstellar dust are under development (Brahman et al., in preparation), but, based on 
\btdred{this value of $I_{12}/I_{340}$},
we expect that the emission from interstellar dust grains in the heliosphere 
\btdred{could be as bright as}
\btdred{0.2~kJy/sr} during a focusing cycle.

The distribution of ISDGs inside of the heliosphere is highly
asymmetric around the Sun.
Figure~\ref{f:hde} 
shows simulations of the distribution of $a = 0.2$~\micron~ISDGs in
the X-Z plane (see caption for definition) at $\sim 2.5$~year intervals starting in 1996 \citep{Landgraf_2000}.
During the year 2018, which could see the launch of the CMBPol satellite, conditions similar to those shown for the year 1996 are expected.  
For comparison, the COBE data were collected at
the end of the 1980's focusing cycle, while \WMAP, launched in 2001, has witnessed the phases shown in the last four panels of Fig~\ref{f:hde} (starting by the top right panel). The flux from 0.2 \micron\ ISDGs in the inner
heliosphere varies by over an order of magnitude between solar minimum
and solar maximum conditions \citep{Landgraf+Kruger+Altobelli_etal_2003}.  As a result, a CMBPol mission is likely to detect time variability in the emission from heliospheric dust if it is operated for a long enough time.  This result would be interesting by itself, since it would probe both the magnetic conditions in the heliosphere and the properties of ISDGs, but it would also be useful to the CMB analysis by signaling a possible contamination by heliospheric dust emission.

\section{Anomalous Emissions}
\label{s:had}

\subsection{Anomalous Dust Emission}
\label{s:ade}

Prior to the launch of \WMAP, detailed intensity maps of the 
Galaxy were available at 408~MHz \citep{Haslam+Salter+Stoffel_etal_1982} and
$100~\um$ \citep{Schlegel+Finkbeiner+Davis_1998}.  The morphologies 
of these maps were notably different, reflecting the 
different emission mechanisms that are dominant at those 
frequencies:  at 408~MHz, the diffuse Galactic emission is 
dominated by synchrotron, with large characteristic 
morphological features such as Loop I (the North Polar 
Spur); at $100~\um$, the diffuse emission is generated 
primarily by dust grains emitting blackbody radiation as 
they are heated by the ambient interstellar radiation field.
When these two maps were extrapolated to \WMAP~frequencies, assuming reasonable power law models, it 
was thought that the 94~GHz map would look morphologically 
very similar to the $100~\um$ map since thermal emission from interstellar dust dominates the diffuse Galactic signal at these frequencies, while the 23~GHz map 
(away from bright HII regions and filaments where free-free 
emission dominates) would closely resemble the synchrotron dominated 408~MHz map.  However, the \WMAP~23~GHz map looks much more like the $100~\um$ map than like the 408~MHz map \citep{Hinshaw+Weiland+Hill_etal_2008}.

The origin of this dust-correlated emission in the low 
frequency \WMAP~channels is still not completely certain, 
though there is significant evidence that it arises from the 
smallest interstellar dust grains excited into rotational modes 
and emitting \btdred{electric} dipole radiation \citep{Draine+Lazarian_1998a,Draine+Lazarian_1998b}.  This 
``spinning dust'' model is currently the leading hypothesis,
although ``magnetic dust'' emission  \citep{Draine+Lazarian_1999} and ``dust-correlated 
synchrotron'' emission \citep{Bennett+Hill+Hinshaw_etal_2003} are sometimes offered as alternative explanations.
Both total intensity and polarization observations can provide useful information as to which model can best explain the observed anomalous dust-correlated emission in the \WMAP~data.

\subsubsection{Sorting through Candidates}
\label{s:ad_candidates}

\paragraph{\it Total Intensity}

Prior to \WMAP's launch, there was already evidence against 
the hypothesis of dust-correlated synchrotron emission.  Using the 
Tenerife 10 and 15~GHz data along with the COBE/DMR maps, 
\citet{de-Oliveira-Costa+Tegmark+Gutierrez_etal_1999} showed that the dust-correlated emission does 
not rise with decreasing frequency as expected for 
synchrotron, but rather falls off below 30~GHz, a feature 
which is characteristic of spinning and magnetic dust emissions.  
Furthermore, \citet{Finkbeiner+Schlegel+Frank_etal_2002} identified at least one confirmed 
example, L1622, of a cloud in which the spectrum of emission 
from 5 to 10~GHz resembles a spinning dust spectrum.

As part of the first-year \WMAP~data release, \citet{Bennett+Hill+Hinshaw_etal_2003} 
presented maps of the spectral index of synchrotron emission derived 
under the assumption that all of the anomalous emission is 
synchrotron.  These maps have large diffuse regions in which 
the spectral index of synchrotron is $\beta \gtrsim -2.6$,
whereas, if the synchrotron emitting electrons are accelerated through~Fermi acceleration by 
supernova shocks, the injected spectral index should be 
$\beta \sim -2.5$ \citep[see, e.g.,][]{Gieseler+Jones_2000}, which softens to $\beta \sim -3$ 
when taking into account diffusion energy losses (the total 
energy losses from synchrotron and inverse Compton 
scattering are proportional to $E^2$).  Such a discrepancy seems to invalidate their initial assumption, a conclusion reinforced by the fact that a spectral 
index of $\beta \sim -3$ has been found in more rigid fits 
to the \WMAP~temperature data \citep{Dobler+Finkbeiner_2008a}, by combining observations at  408~MHz and 1.42~GHz to predict the emission at frequencies above 20~GHz  \citep{La-Porta+Burigana+Reich_etal_2008}, and by using only low 
frequency \WMAP~polarization data \citep[see][and the {\it Polarization} discussion below for more polarization related~arguments]{Kogut+Dunkley+Bennett_etal_2007}.

Targeted 
observations of individual diffuse clouds have also proven 
fruitful \citep[see, e.g.,][]{Dickinson+Casassus+Pineda_etal_2006}, and there 
is evidence that when combining \WMAP~data with data sets at 
other frequencies, diffuse, inverted spectrum emission 
(as opposed to the 
monotonically decreasing spectrum expected for synchrotron) is seen 
\citep{Boughn+Pober_2007}.  Additionally, the Green Bank Galactic 
plane survey \citep{Finkbeiner+Langston+Minter_2004} presented very strong evidence for 
diffuse inverted spectrum emission between 8.35 and 14.35 
GHz outside HII regions.

Finally, \citet{Dobler+Finkbeiner_2008b} \gdred{and \citet{Dobler+Draine+Finkbeiner_2008}} showed that a spinning dust spectrum 
is completely recoverable
\gdred{within}
the \WMAP~\gdred{frequency range}.  However, 
this \gdred{WIM~(``warm ionized medium")} spinning dust emission correlates \gdred{more strongly} with maps of 
H$\alpha$ emission \citep{Finkbeiner_2003}
than
\gdred{with maps of} 100 $\mu$m emission.  Since H$\alpha$ maps represent an emission 
measure,
and since spinning dust emission is also a density 
squared process \gdred{\citep{Dobler+Draine+Finkbeiner_2008}}, H$\alpha$ should trace 
spinning dust emission as well.  Magnetic~dust emission, on 
the other hand, should be proportional to density and 
would not be traced by~H$\alpha$.

\paragraph{\it Polarization}

Perhaps the strongest evidence against the dust-correlated 
synchrotron hypothesis is derived from the \WMAP~polarization 
maps presented by \citet{Page+Hinshaw+Komatsu_etal_2007} and \citet{Kogut+Dunkley+Bennett_etal_2007}.  
Synchrotron emission is thought to be highly polarized, typically
$\sim$ 40 -- 70\% in the presence of reasonably ordered 
magnetic fields, although the polarization signal can effectively be eliminated for lines-of-sight going through regions with 
turbulent magnetic fields \citep[see Sec.~\ref{ss:gmf_cmb} and, e.g.,][]{Miville-Deschenes+Ysard+Lavabre_etal_2008}. The polarized fraction of spinning dust emission, on the other hand, is expected to be smaller than 3\% at the lowest \WMAP~frequencies \citep{Lazarian+Draine_2000}.  However, the \WMAP~maps of the polarized 
emission at 23 GHz look morphologically very different from 
the corresponding total intensity map.  In particular, the 
polarization map more closely resembles (but is not 
perfectly traced by) the 408 MHz map and does not 
correlate very strongly with the $100~\um$ dust map.  The 
implication is that there is a highly polarized component 
that morphologically looks very much like synchrotron and a 
low polarization component that looks very much like dust.  
The polarization properties of 
magnetic dust are not well known \citep[see][]{Draine+Lazarian_1999}.

\subsubsection{Towards a Spectral Identification}
\label{s:ad_identification}

Presently, the study of anomalous dust emission is mostly 
aimed at targeted searches of promising regions (e.g., dust 
clouds and HII regions) that may exhibit evidence for 
anomalous emission spectrum.  These studies are being 
undertaken in both intensity \citep{Dickinson+Casassus+Pineda_etal_2006} and 
polarization \citep{Mason+Robishaw+Finkbeiner_2008}.  The polarization studies in 
particular will prove very useful in determining just how 
much of an issue certain emission mechanisms will present 
when attempting to clean polarization maps to get a 
cosmological signal.  However, we stress that determining 
the polarization properties of spinning dust does 
not exclude the possibility that other emission mechanisms 
(such as magnetic dust) might exist with very significant 
polarization fractions exhibiting a strong wavelength dependence \citep{Draine+Lazarian_1999}.

With the H$\alpha$-correlated spinning dust signal detected in the 
\WMAP~data (see Sec.~\ref{s:ad_candidates}), progress \gdred{has been made} towards measuring specific parameters of 
the dust grains and environment
\gdred{\citep{Dobler+Draine+Finkbeiner_2008}}, but \gdred{is} limited by the relatively narrow 
frequency coverage of \WMAP.  Near-term and future missions 
will address this issue.  For example, the C-Band 
All-Sky 
Survey \citep{Pearson_2007} will 
map the sky in temperature and polarization at 5 GHz.  This 
survey should determine if the dust-correlated 
emission continues to rise from 23~GHz down to 5~GHz 
(indicating synchrotron) or if it has fallen (indicating 
spinning or magnetic~dust).

The \Planck~satellite will 
now launch shortly and with it will come high resolution 
maps at 545 and 857~GHz.  At these frequencies, 
the foreground emission is totally dominated by thermal dust 
emission and not only will this give insight into the 
polarization properties of dust, but it will also significantly facilitate 
foreground cleaning, especially if the polarized fraction of dust emission is independent of wavelength (see Sec.~\ref{s:pde}).  Obtaining cleaner maps is essential to deriving an 
unbiased spectrum of foreground emission at anomalous 
emission frequencies, i.e., in the range 20 -- 100~GHz 
\citep{Hinshaw+Nolta+Bennett_etal_2007,Dobler+Finkbeiner_2008a}.

Additionally, if the Absolute Spectrum Polarimeter
(ASP, PI: A.~Kogut) is funded, it will provide excellent 
frequency coverage from 30~GHz to 5~THz, with over 300 spectral bins dedicated to characterizing foregrounds.  
Many spectral channels are highly desirable since the 
primary goal in foreground science is to characterize the 
spectrum of each emission component.  Furthermore, the more 
well known the foreground spectra, the easier it will be to 
clean maps of the foreground emission to measure the 
cosmological signal.  The shortcoming of ASP is that the 
angular resolution is a 2$^\circ$ tophat, a resolution 
considerably lower than even \WMAP's.

\subsection{The \WMAP~``Haze''}
\label{s:haze}

In an analysis of the \WMAP~\gdred{temperature} maps, \citet{Finkbeiner_2004} detected an excess of diffuse 
microwaves \gdred{coming from an extended region around}
the Galactic center.  \gdtwo{This emission has also been detected in the three-year temperature data by \citet{Dobler+Finkbeiner_2008a} and \citet{Bottino+Banday+Maino_2008}.}  \gdred{When CMB polarization data is not taken into account, t}his excess emission, coined ``the haze," exhibit\gdred{s} a spectrum too soft to be free-free 
emission, but too hard to be synchrotron from supernova 
shock accelerated electrons \gdred{when diffusion losses are accounted for} \citep{Dobler+Finkbeiner_2008a}.  \gdred{It was therefore proposed that the haze could be synchrotron emission from a population of hard electrons surrounding the Galactic center.}
\gdred{However, \citet{Gold+Bennett+Hill_etal_2008} do not find any evidence of this hardening in 
\WMAP's observations of the polarized Galactic synchrotron emission (see, e.g., their Fig.~16).  Although the \WMAP~polarization data is not yet sufficient to convincingly rule out the synchrotron nature of the haze,~this~discrepancy~is~worth~investigating.}

As with all inferences of foreground spectra from the 
\WMAP~data, there is substantial systematic uncertainty in 
the haze spectrum due to the CMB cross-correlation bias.
Any CMB estimator is indeed contaminated by some 
residual foregrounds that then get subtracted off of each 
channel prior to the foreground analysis.  As a result, the inferred foreground spectra are in turn contaminated by 
a CMB spectrum.  The amplitude of the bias can be hard to estimate given that 
it is related to the chance morphological 
correlation of the foregrounds with the CMB, and, consequently, the haze 
spectrum is relatively uncertain.
\gdred{M}easurements at more 
frequencies will
therefore 
prove crucial to determining the 
fundamental nature of the~haze.

Thus, 
the upcoming experiments discussed in the context of spinning dust emission
can also play a role in characterizing the haze.
In particular, with its improved foreground removal abilities, \Planck~will lead to a significant decrease in the amplitude of the CMB cross-correlation bias.  Moreover, the amplitude of 
the haze in the C-BASS 5 GHz maps will distinguish between 
spinning dust and synchrotron type spectrum for this excess emission.  If these two surveys confirm that the haze is compatible with hard synchrotron emission, the Fermi satellite
should see 
$\gamma$-rays from inverse Compton scattered starlight (Dobler, in preparation), which would confirm the synchrotron nature of the haze.  Lastly, the many frequency channels of ASP may 
allow us to trace out the synchrotron spectrum (and thus the
energy spectrum of the electrons) as a function of position 
on the sky with a resolution of a couple of degrees.

\vspace{0.5cm}
\subsection{CMBPol as a Morphological Tracer}
\label{s:morphol_trace}

The CMBPol satellite has the potential to combine the most 
promising aspects of upcoming experiments and shed new light 
on both anomalous dust-correlated emission and the 
haze.  Although the final design may only have a limited number of frequency bands, the large-scale spectral information provided by a survey like the ASP would indeed enable the separation of foreground emission and cosmological signal with high accuracy, and in particular, with insignificant CMB cross-correlation bias.
Thanks to the expected very high spatial 
resolution of CMBPol, one could then trace out the spectrum of individual 
components morphologically, which not 
only could provide clues regarding the origin of the haze, 
but is essential to constraining the parameters of an environment-dependent emission such as spinning dust.

\vspace{0.5cm}
\section{Foregrounds and CMB Lensing}
\label{s:egs}

Foregrounds are a potential source of systematic error for lensing of the CMB.  
Quadratic lensing reconstruction with CMB polarization indeed generally involves taking products of the $E$ and $B$-mode polarization signals to compute the integral \citep{Hu+Okamoto_2002}
\begin{equation}
\hat\phi({\bf l}) \equiv \int w({\bf l}',{\bf l}-{\bf l}')\, E({\bf l}')\, B({\bf l}-{\bf l}')\, \d^2{\bf l}'\,,
\label{eq:hatphi1}
\end{equation}
where $\hat\phi({\bf l})$ is the estimator for the lensing potential $\phi({\bf l})$, $w$ a weight function, and ${\bf l}$ the wavevector.  There are other reconstruction techniques that give smaller error bars \citep[see, e.g.,][]{Hirata+Seljak_2003}, but, for the purpose of this discussion, we focus on the quadratic methods as they are easier to analyze.
In general, three types of measurements concern us:
\vspace{-0.1cm}
\begin{itemize}
\item the galaxy-lensing potential cross-power spectrum,
\vspace{-0.25cm}
\item the lensing potential auto-power spectrum, and
\vspace{-0.25cm}
\item the low-$l$ $B$-mode power spectrum after correcting for lensing.
\end{itemize}
\vspace{-0.05cm}
The last of these is the most directly important for the community's quest to detect a primordial gravitational wave signal, while the first two are relevant for low-redshift (e.g., dark energy and neutrino mass) studies .  The companion paper by \citet{CMBPol_Smith_2008}
focuses on the lensing auto-power spectrum.  We assess the foreground situation for this particular~case.

As Eq.~(\ref{eq:hatphi1}) involves a product of $E$ and $B$-mode polarizations, the estimator's power spectrum is a trispectrum containing two factors of the $E$-mode polarization and two of the~$B$, with both ``disconnected'' terms (involving the product of two power spectra) and ``connected'' (i.e., non-Gaussian) terms.  The disconnected terms depend on the product of the CMB $C_l^{EE}$ and the foreground power spectrum $C_l^{BB}$.  The latter are much harder to estimate as they depend on the polarized foreground trispectrum, which is mostly unknown at the present time.

Since lensing reconstruction uses mainly the small-scale polarization, our expectation is that extragalactic foregrounds (point sources) will dominate.  In this sense we are fortunate, in that for point sources with random polarization angles it is possible to write the relevant trispectrum in terms of three quantities: (i) the polarized flux-squared-weighted integral of the unmasked source counts, $I_2$; (ii) the fourth-power-weighted integral, $I_4$; and (iii) the angular clustering power spectrum of the point sources, $C_l^{\rm s}$.  While the polarized source counts at relevant frequencies are presently unknown, we can estimate reasonable ranges for these parameters.  More importantly, the dependence on a single function raises the hope that determination of the foreground parameters may be possible by examining the full configuration dependence of the trispectrum, a possibility yet to be investigated.
This rather optimistic take on the foreground trispectrum should however be tempered by some ``known unknowns.''
\vspace{-0.1cm}
\begin{list}{$\bullet$}{}
\item Point sources have a different spectrum than the CMB, which means that in a broadband CMB map they are not truly ``pointlike.''
\vspace{-0.25cm}
\item Extragalactic microwave sources may not have random polarization angles.  For example, tidal torques may align disk galaxies with each other, resulting in a net alignment of magnetic fields.  The current data are insufficient to constrain such a contamination.
\vspace{-0.25cm}
\item Even in ``clean'' parts of the sky, there is Galactic emission which may contain non-Gaussian small-scale polarized structure with a complicated configuration dependence.
\end{list}
\vspace{-0.05cm}

These problems will have to be continuously re-assessed as we continue with small-scale CMB polarization experiments.  For example, \Planck~will measure the small angular scale Galactic dust emission and empirically constrain its $E/B$ structure and configuration dependence, which cannot be accurately predicted from theory or measured with \WMAP.  We also expect that prior to CMBPol, a number of ground-based experiments will make the first characterization of the lensing $B$-mode peak, and provide information on the polarization fractions and polarized SEDs of point sources at relevant frequencies.  Because of the frequency dependence of the beams, this information will be crucial even if the use of angular information (configuration dependence) turns out to be a viable approach to lensing foreground removal.

\section{Summary and Conclusions}
\label{s:conclusion}

The main conclusions of this paper are as follows:

\begin{enumerate}

\item Our current knowledge of the large-scale structure of the Galactic magnetic field comes mostly from radio observations of the Galactic synchrotron emission (Sec.~\ref{s:synch}) and Faraday rotation (Sec.~\ref{ss:faraday}), with the former providing information about the structure of the field in the plane of the sky, while the latter encodes information about the line-of-sight component of the field.  Other techniques (Sec.~\ref{s:other}), such as observations of polarized dust emission, starlight polarization, and Zeeman splitting, complete this large-scale picture by probing the field on smaller scales, e.g., in molecular clouds, massive star forming regions, and low-dust extinction environments.  Despite this combination of techniques, the large-scale structure of the field is still poorly constrained (Sec.~\ref{ss:lsgmf}), with, in particular, no real agreement as to the number and location of large-scale reversals (Fig.~\ref{arrows}).
\jacbred{In addition, very} little is \jacbred{currently} known about
both 
the small-scale field (Sec.~\ref{ss:randomgmf})
and 
the halo field (Sec.~\ref{ss:hgmf}).

\item Combining multiple probes of the Galactic magnetic field, such as observations of total and polarized synchrotron emissions, free-free emission, and Faraday rotation, to simultaneously fit  for various parameters describing the magnetized interstellar medium can lift degeneracies between them (Sec.~\ref{s:mim}), thereby improving constraints on their values.

\item Knowing the structure and strength of the Galactic magnetic field is a pre-requisite to understanding both the total and polarized synchrotron emissions (Sec.~\ref{s:synch}) and the polarized dust emission (Sec.~\ref{s:starlight}).  The progress we will make towards characterizing the field before CMBPol flies will therefore be of tremendous help in separating Galactic foregrounds from cosmological signals. Combining RM synthesis (Sec.~\ref{s:rmsyn}) and radio polarimetric surveys with wide frequency and spatial coverage will greatly improve our knowledge of the field by mapping it in 3 dimensions and enabling the use of RM grids (Sec.~\ref{s:radio_next}).  In particular, detailed studies of the number and locations of large-scale field reversals, the possible inclination of the field with respect to the Galactic plane, the vertical structure of the field, and the properties of the turbulent field on a wide range of scales will be carried out by the end of the next decade. Data from the Auger experiment \rjred{may} additionally allow probing the large-scale structure of the Galactic magnetic field without having to rely on a model for the electron density in our galaxy (Sec.~\ref{s:other}).

\item CMBPol will shed new light on the structure of the Galactic magnetic field through higher resolution and sensitivity
observations of the
total and polarized synchrotron emissions from the Galaxy than \WMAP\dnsred{~and \Planck}~(Sec.~\ref{ss:gmf_cmb}).  These observations are unaffected by Faraday rotation 
and therefore complement RM data by probing the sky-projected component of field.  However, 
\dnsred{the sensitivities of \Planck~and \WMAP~}will not be high enough for
\dnsred{them} to accurately determine the characteristics of the polarized synchrotron emission at intermediate and high Galactic latitudes.  With 10 times higher sensitivity, CMBPol would be able to constrain the field at these latitudes, and in particular its vertical structure and its halo component.  Higher resolution will also allow CMBPol to study turbulent structures in the magneto-ionized medium, and to potentially trace magnetic fields from the diffuse ISM into small-scale structures.

\item Models of interstellar dust (Sec.~\ref{ss:dust_models}) based on observations of the Galactic extinction and polarization of starlight (Sec.~\ref{s:obs_ext_pol_star}) can be used to predict the diffuse polarized far-infrared and submillimeter dust mission from our galaxy (Sec.~\ref{ss:firpol}).  However, predictions are different for models involving different species of polarization-contributing dust grains (Fig.~\ref{f:firpol_predictions}).  Although this might make separating polarized dust emission from cosmological signal more complicated (Sec.~\ref{s:dust_forecasts}), it also offers the opportunity of improving our knowledge of the properties of interstellar dust grains with CMBPol observations~(Sec.~\ref{s:constraints_pde}).

\item Starlight polarization data can provide useful Galactic magnetic field templates to trace the polarized dust emission (Sec.~\ref{s:starlight}), but few data points are available at latitudes of interest to CMB studies, typically $|b| \gtrsim10^\circ$.  No currently underway or plan\pcfred{n}ed experiment addresses this issue, since all of them focus either on regions of the Galactic plane or on dense small-scale environments.  We believe obtaining this data by the time CMBPol flies should be a high priority item.

\item Alignment of dust grains with the Galactic magnetic field is the physical process behind both polarized dust emission and starlight polarization. The radiative torque theory of grain alignment (Sec.~\ref{s:gam}) seems to be able to explain why and how dust grains in the interstellar medium align the way observations indicate (Sec.~\ref{s:why}), but the most exciting feature of this theory is its ability to make testable predictions (Sec.~\ref{predictions}) thanks to simple models accurately reproducing complex numerical calculations (Sec.~\ref{s:model}).  CMBPol's high sensitivity and resolution, combined with starlight polarization information, will therefore allow testing the theory of radiative torques by observing  
the diffuse polarized dust emission from the Galaxy in numerous independent patches of the sky (Sec.~\ref{s:constraints_pde}).

\item The distribution of dust grains in the heliosphere, and therefore the corresponding far-infrared and submillimeter dust emission, vary cyclically due to the varying magnetic conditions originating from the solar activity cycle (Fig.~\ref{f:hde}).  Although more precise estimates are needed, CMBPol could detect these fluctuations (Sec.~\ref{s:hde_cmbpol}), thereby constraining the magnetic conditions in the heliosphere and the properties of heliospheric~dust.

\item Two anomalous emissions were detected in the \WMAP~data.  Dust-correlated emission in its low frequency channels is likely to be spinning or magnetic dust emission (Sec.~\ref{s:ade}), whereas the origin of an excess emission of microwaves towards the Galactic center, dubbed ``the haze" (Sec.~\ref{s:haze}), is still unknown but could be synchrotron radiation from a population of hard electrons.  A definitive identification of these two emissions likely requires observing their spectra, an effort ground-based experiments, e.g., C-BASS, and dedicated satellite missions, such as the ASP, will contribute to (Sec.~\ref{s:ad_identification}). CMBPol's high resolution will enable us to trace these spectra morphologically (Sec.~\ref{s:morphol_trace}).  This is essential to determining the parameters of spinning dust emission, which strongly depend on environment, and it might give us clues as to the possible origin of the haze.

\item Finally, extra-Galactic sources are a potential contaminant for studies of CMB lensing (Sec.~\ref{s:egs}), but not much is currently known about the polarized source counts at CMB frequencies, or the spectra and intrinsic polarizations of these sources.  The input of ground-based experiments on these questions will be essential to improve our measurements of CMB lensing \citep[see more details in the companion paper by][]{CMBPol_Smith_2008}.

\end{enumerate}

Detecting a primordial B-mode polarization signal would have dramatic consequences for our understanding of the phenomena that took place in the early Universe and led to the necessary conditions for it to evolve into the Universe we observe today \citep[see the companion paper by][]{CMBPol_Baumann_2008}.  Being able to do so will not only require advanced CMB data analysis techniques \citep[see the companion paper by][]{CMBPol_Dunkley_2008}, but also an~in-depth understanding of how Galactic [and, for some purposes \citep[see, e.g., the companion paper by][]{CMBPol_Smith_2008}, extra-Galactic] foregrounds behave.  Taking this synergy to the next level necessitates coordinated and collaborative efforts by both the CMB and the ISM communities, each of which, as we hope to have shown throughout this document, will highly benefit~from.

\vspace{-0.1cm}
\section*{Acknowledgments}
\addcontentsline{toc}{section}{Acknowledgments} 

\sdred{This research was partly funded by NASA Mission Concept Study award NNX08AT71G S01.  We also acknowledge the organizational work of the Primordial Polarization Program Definition Team (PPPDT).}
AAF was supported by NSF grant AST-0707932 and Princeton University.  \pcfred{PCF acknowledges support from NASA grant NNX08AJ33G}, AL \pcfred{from} NSF grant  AST-0507164 and the NSF Center for Magnetic Self-Organization in Laboratory and Astrophysical
Plasmas\ammred{, and AMM from FAPESP and CNPq}\pcfred{.}
AAF
would like to thank
D.~P.~Finkbeiner and J.~E.~Vaillancourt for their contributions to the CMBPol Theory Workshop held at Fermilab in June 2008,
L.~A.~Page for stimulating discussions, D.~N.~Spergel for comments on a draft version of this paper, and 
R.~H.~Lupton for making the SM plotting program available to him.
\ammred{AMM is grateful to all members of the IAG Polarimetry Group for their continued support.}
Typesetting of this document was made easier by the use of NASA's Astrophysics Data System Bibliographic Services and of the AAS\TeX~package.  All the \Planck-related numbers and information referred to in this document are from the \Planck~scientific program (also known as ``blue book") available online in PDF format at \url{http://www.rssd.esa.int/planck/}.

\vspace{-0.1cm}
\addcontentsline{toc}{section}{References}
\small
\setlength{\bibsep}{0.2cm}
\bibliography{biblio_int,biblio_gmf,biblio_mim,biblio_pde,biblio_epl,biblio_gam,biblio_hde,biblio_had,biblio_egs,biblio_ccs}

\begin{thebibliography}{188}
\expandafter\ifx\csname natexlab\endcsname\relax\def\natexlab#1{#1}\fi

\bibitem[{{Abbas} {et~al.}(2004){Abbas}, {Craven}, {Spann}, {Tankosic},
  {LeClair}, {Gallagher}, {West}, {Weingartner}, {Witherow}, \&
  {Tielens}}]{Abbas_Craven_Spann_etal_2004}
{Abbas}, M.~M. {et~al.} 2004, \apj, 614, 781

\bibitem[{{Abraham} {et~al.}(2007){Abraham}, {Abreu}, {Aglietta}, {Aguirre},
  {Allard}, {Allekotte}, {Allen}, {Allison}, {Alvarez}, {Alvarez-Mu{\~n}iz},
  {Anchordoqui}, {Andringa}, {Anzalone}, {Argir{\`o}}, {Arisaka}, {Armengaud},
  {Arneodo}, {Arqueros}, {Asch}, {Asorey}, {Assis}, {Atulugama}, {Aublin},
  {Ave}, {Avila}, {B{\"a}cker}, {Badagnani}, {Barbosa}, {Barnhill}, {Barroso},
  {Bauleo}, {Beatty}, {Beau}, {Becker}, {Becker}, {Bellido}, {BenZvi}, {Berat},
  {Bergmann}, {Bernardini}, {Bertou}, {Biermann}, {Billoir}, {Blanch-Bigas},
  {Blanco}, {Blasi}, {Bleve}, {Bl{\"u}mer}, {Boh{\'a}cov{\'a}}, {Bonifazi},
  {Bonino}, {Boratav}, {Brack}, {Brogueira}, {Brown}, {Buchholz}, {Bueno},
  {Busca}, {Caballero-Mora}, {Cai}, {Camin}, {Caruso}, {Carvalho},
  {Castellina}, {Catalano}, {Cataldi}, {Caz{\'o}n-Boado}, {Cester}, {Chauvin},
  {Chiavassa}, {Chinellato}, {Chou}, {Chye}, {Clark}, {Clay}, {Colombo},
  {Concei{\c c}{\~a}o}, {Connolly}, {Contreras}, {Coppens}, {Cordier}, {Cotti},
  {Coutu}, {Covault}, {Creusot}, {Cronin}, {Dagoret-Campagne}, {Daumiller},
  {Dawson}, {de Almeida}, {De Donato}, {de Jong}, {De La Vega}, {de Mello
  Junior}, {de Mello Neto}, {De Mitri}, {de Souza}, {del Peral}, {Deligny},
  {Della Selva}, {Delle Fratte}, {Dembinski}, {Di Giulio}, {Diaz},
  {Dobrigkeit}, {D'Olivo}, {Dornic}, {Dorofeev}, {dos Anjos}, {Dova}, {D'Urso},
  {DuVernois}, {Engel}, {Epele}, {Erdmann}, {Escobar}, {Etchegoyen}, {Facal San
  Luis}, {Falcke}, {Farrar}, {Fauth}, {Fazzini}, {Fern{\'a}ndez}, {Ferrer},
  {Ferry}, {Fick}, {Filevich}, {Filipcic}, {Fleck}, {Fonte}, {Fracchiolla},
  {Fulgione}, {Garc{\'{\i}}a}, {Garc{\'{\i}}a G{\'a}mez}, {Garcia-Pinto},
  {Garrido}, {Geenen}, {Gelmini}, {Gemmeke}, {Ghia}, {Giller}, {Glass}, {Gold},
  {Golup}, {Gomez Albarracin}, {G{\'o}mez Berisso}, {G{\'o}mez Herrero},
  {Gon{\c c}alves}, {Gon{\c c}alves do Amaral}, {Gonzalez}, {Gonzalez},
  {Gonz{\'a}lez}, {G{\'o}ra}, {Gorgi}, {Gouffon}, {Grassi}, {Grillo},
  {Grunfeld}, {Guardincerri}, {Guarino}, {Guedes}, {Guti{\'e}rrez}, {Hague},
  {Hamilton}, {Hansen}, {Harari}, {Harmsma}, {Harton}, {Haungs}, {Hauschildt},
  {Healy}, {Hebbeker}, {Heck}, {Hojvat}, {Holmes}, {Homola}, {H{\"o}randel},
  {Horneffer}, {Horvat}, {Hrabovsky}, {Huege}, {Iarlori}, {Insolia}, {Ionita},
  {Italiano}, {Kaducak}, {Kampert}, {Keilhauer}, {Kemp}, {Kieckhafer},
  {Klages}, {Kleifges}, {Kleinfeller}, {Knapik}, {Knapp}, {Koang}, {Kopmann},
  {Krieger}, {Kr{\"o}mer}, {K{\"u}mpel}, {Kunka}, {Kusenko}, {La Rosa},
  {Lachaud}, {Lago}, {Lebrun}, {LeBrun}, {Lee}, {Leigui de Oliveira},
  {Letessier-Selvon}, {Leuthold}, {Lhenry-Yvon}, {L{\'o}pez}, {Lopez
  Ag{\"u}era}, {Lozano Bahilo}, {Maccarone}, {Macolino}, {Maldera}, {Malek},
  {Mancarella}, {Mance{\~n}ido}, {Mandat}, {Mantsch}, {Mariazzi}, {Maris},
  {Martello}, {Mart{\'{\i}}nez}, {Mart{\'{\i}}nez Bravo}, {Mathes}, {Matthews},
  {Matthews}, {Matthiae}, {Maurizio}, {Mazur}, {McCauley}, {McEwen}, {McNeil},
  {Medina}, {Medina-Tanco}, {Meli}, {Melo}, {Menichetti}, {Menschikov},
  {Meurer}, {Meyhandan}, {Micheletti}, {Miele}, {Miller}, {Mollerach},
  {Monasor}, {Monnier Ragaigne}, {Montanet}, {Morales}, {Morello}, {Moreno},
  {Moreno}, {Morris}, {Mostaf{\'a}}, {Muller}, {Mussa}, {Navarra}, {Navarro},
  {Navas}, {Nellen}, {Newman-Holmes}, {Newton}, {Nguyen Thi},
  {Nierstenh{\"o}fer}, {Nitz}, {Nosek}, {Nozka}, {Oehlschl{\"a}ger}, {Ohnuki},
  {Olinto}, {Olmos-Gilbaja}, {Ortiz}, {Ostapchenko}, {Otero}, {Pakk Selmi-Dei},
  {Palatka}, {Pallotta}, {Parente}, {Parizot}, {Parlati}, {Pastor}, {Patel},
  {Paul}, {Pavlidou}, {Payet}, {Pech}, {P{\c c}kala}, {Pelayo}, {Pepe},
  {Perrone}, {Petrera}, {Petrinca}, {Petrov}, {Ngoc}, {Ngoc}, {Pham Thi},
  {Pichel}, {Piegaia}, {Pierog}, {Pimenta}, {Pinto}, {Pirronello}, {Pisanti},
  {Platino}, {Pochon}, {Porter}, {Privitera}, {Prouza}, {Quel}, {Rautenberg},
  {Reucroft}, {Revenu}, {Rezende}, {R{\'{\i}}dky}, {Riggi}, {Risse},
  {Rivi{\`e}re}, {Rizi}, {Roberts}, {Robledo}, {Rodriguez}, {Rodr{\'{\i}}guez
  Fr{\'{\i}}as}, {Rodriguez Martino}, {Rodriguez Rojo}, {Rodriguez-Cabo},
  {Ros}, {Rosado}, {Roth}, {Rouill{\'e}-d'Orfeuil}, {Roulet}, {Rovero},
  {Salamida}, {Salazar}, {Salina}, {S{\'a}nchez}, {Santander}, {Santo},
  {Santos}, {Sarazin}, {Sarkar}, {Sato}, {Scherini}, {Schieler}, {Schmidt},
  {Schmidt}, {Scholten}, {Schov{\'a}nek}, {Sch{\"u}ssler}, {Sciutto},
  {Scuderi}, {Segreto}, {Semikoz}, {Settimo}, {Shellard}, {Sidelnik},
  {Siffert}, {Sigl}, {Smetniansky De Grande}, {Smialkowski}, {Sm{\'{\i}}da},
  {Smith}, {Smith}, {Snow}, {Sokolsky}, {Sommers}, {Sorokin}, {Spinka},
  {Squartini}, {Strazzeri}, {Stutz}, {Suarez}, {Suomij{\"a}rvi}, {Supanitsky},
  {Sutherland}, {Swain}, {Szadkowski}, {Takahashi}, {Tamashiro}, {Tamburro},
  {Tascau}, {Tcaciuc}, {Thomas}, {Ticona}, {Tiffenberg}, {Timmermans},
  {Tkaczyk}, {Todero Peixoto}, {Tom{\'e}}, {Tonachini}, {Torresi}, {Travnicek},
  {Tripathi}, {Tristram}, {Tscherniakhovski}, {Tueros}, {Tunnicliffe},
  {Ulrich}, {Unger}, {Urban}, {Vald{\'e}s Galicia}, {Vali{\~n}o}, {Valore},
  {van den Berg}, {van Elewyck}, {V{\'a}zquez}, {Veberic}, {Veiga}, {Velarde},
  {Venters}, {Verzi}, {Videla}, {Villase{\~n}or}, {Vorobiov}, {Voyvodic},
  {Wahlberg}, {Wainberg}, {Waldenmaier}, {Walker}, {Warner}, {Watson},
  {Westerhoff}, {Wieczorek}, {Wiencke}, {Wilczynska}, {Wilczynski}, {Wileman},
  {Winnick}, {Wu}, {Wundheiler}, {Xu}, {Yamamoto}, {Younk}, {Zas}, {Zavrtanik},
  {Zavrtanik}, {Zech}, {Zepeda}, \& {Ziolkowski}}]{Auger_2007}
{Abraham}, J. {et~al.} 2007, Science, 318, 938

\bibitem[{{Altobelli} {et~al.}(2007){Altobelli}, {Dikarev}, {Kempf}, {Srama},
  {Helfert}, {Moragas-Klostermeyer}, {Roy}, \&
  {Gr{\"u}n}}]{Altobelli+Dikarev+Kempf_etal_2007}
{Altobelli}, N., {Dikarev}, V., {Kempf}, S., {Srama}, R., {Helfert}, S.,
  {Moragas-Klostermeyer}, G., {Roy}, M., \& {Gr{\"u}n}, E. 2007, Journal of
  Geophysical Research (Space Physics), 112, 7105

\bibitem[{{Altobelli} {et~al.}(2006){Altobelli}, {Gr{\"u}n}, \&
  {Landgraf}}]{Altobelli+Grun+Landgraf_2006}
{Altobelli}, N., {Gr{\"u}n}, E., \& {Landgraf}, M. 2006, \aap, 448, 243

\bibitem[{{Anderson} {et~al.}(1996){Anderson}, {Weitenbeck}, {Code},
  {Nordsieck}, {Meade}, {Babler}, {Zellner}, {Bjorkman}, {Fox}, {Johnson},
  {Sanders}, {Lupie}, \& {Edgar}}]{Anderson+Weitenbeck+Code_etal_1996}
{Anderson}, C.~M. {et~al.} 1996, \aj, 112, 2726

\bibitem[{{Andersson} \& {Potter}(2007)}]{Andersson+Potter_2007}
{Andersson}, B.-G., \& {Potter}, S.~B. 2007, \apj, 665, 369

\bibitem[{{Andreasyan} \& {Makarov}(1988)}]{Andreasyan+Makarov_1988}
{Andreasyan}, R.~R., \& {Makarov}, A.~N. 1988, Astrophysics, 28, 247

\bibitem[{{Arendt} {et~al.}(1998){Arendt}, {Odegard}, {Weiland}, {Sodroski},
  {Hauser}, {Dwek}, {Kelsall}, {Moseley}, {Silverberg}, {Leisawitz},
  {Mitchell}, {Reach}, \& {Wright}}]{Arendt+Odegard+Weiland_etal_1998}
{Arendt}, R.~G. {et~al.} 1998, \apj, 508, 74

\bibitem[{{Baumann} {et~al.}(2008{\natexlab{a}}){Baumann}, {Komatsu}, {Nolta},
  {Spergel}, {Larson}, {Hinshaw}, {Page}, {Bennett}, {Gold}, {Jarosik},
  {Weiland}, {Halpern}, {Hill}, {Kogut}, {Limon}, {Meyer}, {Tucker}, {Wollack},
  \& {Wright}}]{CMBPol_Baumann_2008}
{Baumann}, D. {et~al.} 2008{\natexlab{a}}, ArXiv e-prints, 0811.3919

\bibitem[{{Baumann} {et~al.}(2008{\natexlab{b}}){Baumann}, {Komatsu}, {Nolta},
  {Spergel}, {Larson}, {Hinshaw}, {Page}, {Bennett}, {Gold}, {Jarosik},
  {Weiland}, {Halpern}, {Hill}, {Kogut}, {Limon}, {Meyer}, {Tucker}, {Wollack},
  \& {Wright}}]{CMBPol_Summary_2008}
------. 2008{\natexlab{b}}, ArXiv e-prints, 0811.3911

\bibitem[{{Beck}(2001)}]{Beck_2001}
{Beck}, R. 2001, Space Science Reviews, 99, 243

\bibitem[{{Beck}(2007)}]{Beck_2007}
{Beck}, R. 2007, in EAS Publications Series, Vol.~23, Sky Polarisation at
  Far-Infrared to Radio Wavelengths, ed. M.-A. {Miville-Desch{\^e}nes} \&
  F.~{Boulanger}, 19--36

\bibitem[{{Beck} {et~al.}(1996){Beck}, {Brandenburg}, {Moss}, {Shukurov}, \&
  {Sokoloff}}]{Beck+Brandenburg+Moss_etal_1996}
{Beck}, R., {Brandenburg}, A., {Moss}, D., {Shukurov}, A., \& {Sokoloff}, D.
  1996, \araa, 34, 155

\bibitem[{{Beck} \& {Gaensler}(2004)}]{Beck+Gaensler_2004}
{Beck}, R., \& {Gaensler}, B.~M. 2004, New Astronomy Review, 48, 1289

\bibitem[{{Beck} \& {Krause}(2005)}]{Beck+Krause_2005}
{Beck}, R., \& {Krause}, M. 2005, Astronomische Nachrichten, 326, 414

\bibitem[{{Beck} {et~al.}(2003){Beck}, {Shukurov}, {Sokoloff}, \&
  {Wielebinski}}]{Beck+Shukurov+Sokoloff_2003}
{Beck}, R., {Shukurov}, A., {Sokoloff}, D., \& {Wielebinski}, R. 2003, \aap,
  411, 99

\bibitem[{{Bennett} {et~al.}(2003){Bennett}, {Hill}, {Hinshaw}, {Nolta},
  {Odegard}, {Page}, {Spergel}, {Weiland}, {Wright}, {Halpern}, {Jarosik},
  {Kogut}, {Limon}, {Meyer}, {Tucker}, \&
  {Wollack}}]{Bennett+Hill+Hinshaw_etal_2003}
{Bennett}, C.~L. {et~al.} 2003, \apjs, 148, 97

\bibitem[{{Beno{\^i}t} {et~al.}(2004){Beno{\^i}t}, {Ade}, {Amblard}, {Ansari},
  {Aubourg}, {Bargot}, {Bartlett}, {Bernard}, {Bhatia}, {Blanchard}, {Bock},
  {Boscaleri}, {Bouchet}, {Bourrachot}, {Camus}, {Couchot}, {de Bernardis},
  {Delabrouille}, {D{\'e}sert}, {Dor{\'e}}, {Douspis}, {Dumoulin}, {Dupac},
  {Filliatre}, {Fosalba}, {Ganga}, {Gannaway}, {Gautier}, {Giard},
  {Giraud-H{\'e}raud}, {Gispert}, {Guglielmi}, {Hamilton}, {Hanany},
  {Henrot-Versill{\'e}}, {Kaplan}, {Lagache}, {Lamarre}, {Lange},
  {Mac{\'{\i}}as-P{\'e}rez}, {Madet}, {Maffei}, {Magneville}, {Marrone},
  {Masi}, {Mayet}, {Murphy}, {Naraghi}, {Nati}, {Patanchon}, {Perrin}, {Piat},
  {Ponthieu}, {Prunet}, {Puget}, {Renault}, {Rosset}, {Santos}, {Starobinsky},
  {Strukov}, {Sudiwala}, {Teyssier}, {Tristram}, {Tucker}, {Vanel}, {Vibert},
  {Wakui}, \& {Yvon}}]{Benoit+Ade+Amblard_etal_2004}
{Beno{\^i}t}, A. {et~al.} 2004, \aap, 424, 571

\bibitem[{{Berkhuijsen} {et~al.}(1964){Berkhuijsen}, {Brouw}, {Muller}, \&
  {Tinbergen}}]{Berkhuijsen+Brouw+Muller_etal_1964}
{Berkhuijsen}, E.~M., {Brouw}, W.~N., {Muller}, C.~A., \& {Tinbergen}, J. 1964,
  \bain, 17, 465

\bibitem[{{Berkhuijsen} {et~al.}(2006){Berkhuijsen}, {Mitra}, \&
  {Mueller}}]{Berkhuijsen+Mitra+Mueller_2006}
{Berkhuijsen}, E.~M., {Mitra}, D., \& {Mueller}, P. 2006, Astronomische
  Nachrichten, 327, 82

\bibitem[{{Bock} {et~al.}(2008){Bock}, {Cooray}, {Hanany}, {Keating}, {Lee},
  {Matsumura}, {Milligan}, {Ponthieu}, {Renbarger}, \&
  {Tran}}]{Bock+Cooray+Hanany_etal_2008}
{Bock}, J. {et~al.} 2008, ArXiv e-prints, 0805.4207

\bibitem[{{Bottino} {et~al.}(2008){Bottino}, {Banday}, \&
  {Maino}}]{Bottino+Banday+Maino_2008}
{Bottino}, M., {Banday}, A.~J., \& {Maino}, D. 2008, \mnras, 389, 1190

\bibitem[{{Boughn} \& {Pober}(2007)}]{Boughn+Pober_2007}
{Boughn}, S.~P., \& {Pober}, J.~C. 2007, \apj, 661, 938

\bibitem[{{Boulares} \& {Cox}(1990)}]{Boulares+Cox_1990}
{Boulares}, A., \& {Cox}, D.~P. 1990, \apj, 365, 544

\bibitem[{{Bradley}(1994)}]{Bradley_1994}
{Bradley}, J.~P. 1994, Science, 265, 925

\bibitem[{{Brentjens} \& {de Bruyn}(2005)}]{Brentjens+de-Bruyn_2005}
{Brentjens}, M.~A., \& {de Bruyn}, A.~G. 2005, \aap, 441, 1217

\bibitem[{{Brouw} \& {Spoelstra}(1976)}]{Brouw+Spoelstra_1976}
{Brouw}, W.~N., \& {Spoelstra}, T.~A.~T. 1976, \aaps, 26, 129

\bibitem[{{Brown} {et~al.}(2007){Brown}, {Haverkorn}, {Gaensler}, {Taylor},
  {Bizunok}, {McClure-Griffiths}, {Dickey}, \&
  {Green}}]{Brown+Haverkorn+Gaensler_etal_2007}
{Brown}, J.~C., {Haverkorn}, M., {Gaensler}, B.~M., {Taylor}, A.~R., {Bizunok},
  N.~S., {McClure-Griffiths}, N.~M., {Dickey}, J.~M., \& {Green}, A.~J. 2007,
  \apj, 663, 258

\bibitem[{{Brown} {et~al.}(2003{\natexlab{a}}){Brown}, {Taylor}, \&
  {Jackel}}]{Brown+Taylor+Jackel_2003}
{Brown}, J.~C., {Taylor}, A.~R., \& {Jackel}, B.~J. 2003{\natexlab{a}}, \apjs,
  145, 213

\bibitem[{{Brown} {et~al.}(2003{\natexlab{b}}){Brown}, {Taylor}, {Wielebinski},
  \& {Mueller}}]{Brown+Taylor+Wielebinski_etal_2003}
{Brown}, J.~C., {Taylor}, A.~R., {Wielebinski}, R., \& {Mueller}, P.
  2003{\natexlab{b}}, \apjl, 592, L29

\bibitem[{{Burn}(1966)}]{Burn_1966}
{Burn}, B.~J. 1966, \mnras, 133, 67

\bibitem[{{Burstein} \& {Heiles}(1982)}]{Burstein+Heiles_1982}
{Burstein}, D., \& {Heiles}, C. 1982, \aj, 87, 1165

\bibitem[{{Carciofi} {et~al.}(2007){Carciofi}, {Magalh{\~a}es}, {Leister},
  {Bjorkman}, \& {Levenhagen}}]{Carciofi+Magalhaes+Leister_etal_2007}
{Carciofi}, A.~C., {Magalh{\~a}es}, A.~M., {Leister}, N.~V., {Bjorkman}, J.~E.,
  \& {Levenhagen}, R.~S. 2007, \apjl, 671, L49

\bibitem[{{Chandrasekhar} \& {Fermi}(1953)}]{Chandrasekhar+Fermi_1953}
{Chandrasekhar}, S., \& {Fermi}, E. 1953, \apj, 118, 113

\bibitem[{{Chiar} {et~al.}(2006){Chiar}, {Adamson}, {Whittet}, {Chrysostomou},
  {Hough}, {Kerr}, {Mason}, {Roche}, \&
  {Wright}}]{Chiar+Adamson+Whittet_etal_2006}
{Chiar}, J.~E. {et~al.} 2006, \apj, 651, 268

\bibitem[{{Clayton} {et~al.}(1992){Clayton}, {Anderson}, {Magalhaes}, {Code},
  {Nordsieck}, {Meade}, {Wolff}, {Babler}, {Bjorkman}, {Schulte-Ladbeck},
  {Taylor}, \& {Whitney}}]{Clayton+Anderson+Magalhaes_etal_1992}
{Clayton}, G.~C. {et~al.} 1992, \apjl, 385, L53

\bibitem[{{Clemens} {et~al.}(2007){Clemens}, {Pinnick}, {Pavel}, {Jameson},
  {Carveth}, \& {Taylor}}]{Clemens_Pinnick_Pavel_etal_2007}
{Clemens}, D.~P., {Pinnick}, A., {Pavel}, M., {Jameson}, K., {Carveth}, C., \&
  {Taylor}, B. 2007, in Bulletin of the American Astronomical Society, Vol.~38,
  761

\bibitem[{{Cordes} \& {Lazio}(2002)}]{Cordes+Lazio_2002}
{Cordes}, J.~M., \& {Lazio}, T.~J.~W. 2002, ArXiv Astrophysics e-prints,
  astro-ph/0207156

\bibitem[{{Cordes} \& {Lazio}(2003)}]{Cordes+Lazio_2003}
------. 2003, ArXiv Astrophysics e-prints, astro-ph/0301598

\bibitem[{{Crutcher} {et~al.}(2003){Crutcher}, {Heiles}, \&
  {Troland}}]{Crutcher+Heiles+Troland_2003}
{Crutcher}, R., {Heiles}, C., \& {Troland}, T. 2003, in Lecture Notes in
  Physics, Vol. 614, Turbulence and Magnetic Fields in Astrophysics, ed.
  E.~{Falgarone} \& T.~{Passot}, 155

\bibitem[{{Davies}(2007)}]{Davies_2007}
{Davies}, R.~D. 2007, Astronomische Nachrichten, 328, 436

\bibitem[{{Davies} {et~al.}(1960){Davies}, {Slater}, {Shuter}, \&
  {Wild}}]{Davies+Slater+Shuter_etal_1960}
{Davies}, R.~D., {Slater}, C.~H., {Shuter}, W.~L.~H., \& {Wild}, P.~A.~T. 1960,
  \nat, 187, 1088

\bibitem[{{Davis} \& {Greenstein}(1951)}]{Davis+Greenstein_1951}
{Davis}, L.~J., \& {Greenstein}, J.~L. 1951, \apj, 114, 206

\bibitem[{{de Oliveira-Costa} {et~al.}(2008){de Oliveira-Costa}, {Tegmark},
  {Gaensler}, {Jonas}, {Landecker}, \&
  {Reich}}]{de_Oliveira-Costa+Tegmark+Gaensler_etal_2008}
{de Oliveira-Costa}, A., {Tegmark}, M., {Gaensler}, B.~M., {Jonas}, J.,
  {Landecker}, T.~L., \& {Reich}, P. 2008, \mnras, 388, 247

\bibitem[{{de Oliveira-Costa} {et~al.}(1999){de Oliveira-Costa}, {Tegmark},
  {Gutierrez}, {Jones}, {Davies}, {Lasenby}, {Rebolo}, \&
  {Watson}}]{de-Oliveira-Costa+Tegmark+Gutierrez_etal_1999}
{de Oliveira-Costa}, A., {Tegmark}, M., {Gutierrez}, C.~M., {Jones}, A.~W.,
  {Davies}, R.~D., {Lasenby}, A.~N., {Rebolo}, R., \& {Watson}, R.~A. 1999,
  \apjl, 527, L9

\bibitem[{{Dickinson} {et~al.}(2006){Dickinson}, {Casassus}, {Pineda},
  {Pearson}, {Readhead}, \& {Davies}}]{Dickinson+Casassus+Pineda_etal_2006}
{Dickinson}, C., {Casassus}, S., {Pineda}, J.~L., {Pearson}, T.~J., {Readhead},
  A.~C.~S., \& {Davies}, R.~D. 2006, \apjl, 643, L111

\bibitem[{{Dobler} {et~al.}(2008){Dobler}, {Draine}, \&
  {Finkbeiner}}]{Dobler+Draine+Finkbeiner_2008}
{Dobler}, G., {Draine}, B.~T., \& {Finkbeiner}, D.~P. 2008, ArXiv e-prints,
  0811.1040

\bibitem[{{Dobler} \&
  {Finkbeiner}(2008{\natexlab{a}})}]{Dobler+Finkbeiner_2008a}
{Dobler}, G., \& {Finkbeiner}, D.~P. 2008{\natexlab{a}}, \apj, 680, 1222

\bibitem[{{Dobler} \&
  {Finkbeiner}(2008{\natexlab{b}})}]{Dobler+Finkbeiner_2008b}
------. 2008{\natexlab{b}}, \apj, 680, 1235

\bibitem[{{Dolginov} \& {Mitrofanov}(1976)}]{Dolginov+Mitrofanov_1976}
{Dolginov}, A.~Z., \& {Mitrofanov}, I.~G. 1976, \apss, 43, 291

\bibitem[{{Draine}(2003)}]{Draine_2003}
{Draine}, B.~T. 2003, \araa, 41, 241

\bibitem[{{Draine}(2008)}]{Draine_2008}
------. 2008, Space Science Reviews

\bibitem[{{Draine} \& {Flatau}(1994)}]{Draine+Flatau_1994}
{Draine}, B.~T., \& {Flatau}, P.~J. 1994, Journal of the Optical Society of
  America A, 11, 1491

\bibitem[{{Draine} \& {Fraisse}(2008)}]{Draine+Fraisse_2008}
{Draine}, B.~T., \& {Fraisse}, A.~A. 2008, ArXiv e-prints, 0809.2094

\bibitem[{{Draine} \& {Lazarian}(1998{\natexlab{a}})}]{Draine+Lazarian_1998a}
{Draine}, B.~T., \& {Lazarian}, A. 1998{\natexlab{a}}, \apjl, 494, L19

\bibitem[{{Draine} \& {Lazarian}(1998{\natexlab{b}})}]{Draine+Lazarian_1998b}
------. 1998{\natexlab{b}}, \apj, 508, 157

\bibitem[{{Draine} \& {Lazarian}(1999)}]{Draine+Lazarian_1999}
------. 1999, \apj, 512, 740

\bibitem[{{Draine} \& {Weingartner}(1996)}]{Draine+Weingartner_1996}
{Draine}, B.~T., \& {Weingartner}, J.~C. 1996, \apj, 470, 551

\bibitem[{{Draine} \& {Weingartner}(1997)}]{Draine+Weingartner_1997}
------. 1997, \apj, 480, 633

\bibitem[{{Dumke} {et~al.}(1995){Dumke}, {Krause}, {Wielebinski}, \&
  {Klein}}]{Dumke+Krause+Wielebinski_etal_1995}
{Dumke}, M., {Krause}, M., {Wielebinski}, R., \& {Klein}, U. 1995, \aap, 302,
  691

\bibitem[{{Dunkley} {et~al.}(2008{\natexlab{a}}){Dunkley}, {Komatsu}, {Nolta},
  {Spergel}, {Larson}, {Hinshaw}, {Page}, {Bennett}, {Gold}, {Jarosik},
  {Weiland}, {Halpern}, {Hill}, {Kogut}, {Limon}, {Meyer}, {Tucker}, {Wollack},
  \& {Wright}}]{CMBPol_Dunkley_2008}
{Dunkley}, J. {et~al.} 2008{\natexlab{a}}, ArXiv e-prints, 0811.3915

\bibitem[{{Dunkley} {et~al.}(2008{\natexlab{b}}){Dunkley}, {Komatsu}, {Nolta},
  {Spergel}, {Larson}, {Hinshaw}, {Page}, {Bennett}, {Gold}, {Jarosik},
  {Weiland}, {Halpern}, {Hill}, {Kogut}, {Limon}, {Meyer}, {Tucker}, {Wollack},
  \& {Wright}}]{Dunkley+Komatsu+Nolta_etal_2008}
------. 2008{\natexlab{b}}, ArXiv e-prints, 0803.0586

\bibitem[{{Dwek} {et~al.}(1997){Dwek}, {Arendt}, {Fixsen}, {Sodroski},
  {Odegard}, {Weiland}, {Reach}, {Hauser}, {Kelsall}, {Moseley}, {Silverberg},
  {Shafer}, {Ballester}, {Bazell}, \&
  {Isaacman}}]{Dwek+Arendt+Fixsen_etal_1997}
{Dwek}, E. {et~al.} 1997, \apj, 475, 565

\bibitem[{{Fermi}(1949)}]{Fermi_1949}
{Fermi}, E. 1949, Physical Review, 75, 1169

\bibitem[{{Fink\-bei\-ner}(2004)}]{Finkbeiner_2004}
{Fink\-bei\-ner}, D.~P. 2004, \apj, 614, 186

\bibitem[{{Finkbeiner}(2003)}]{Finkbeiner_2003}
{Finkbeiner}, D.~P. 2003, \apjs, 146, 407

\bibitem[{{Finkbeiner} {et~al.}(2004){Finkbeiner}, {Langston}, \&
  {Minter}}]{Finkbeiner+Langston+Minter_2004}
{Finkbeiner}, D.~P., {Langston}, G.~I., \& {Minter}, A.~H. 2004, \apj, 617, 350

\bibitem[{{Finkbeiner} {et~al.}(2002){Finkbeiner}, {Schlegel}, {Frank}, \&
  {Heiles}}]{Finkbeiner+Schlegel+Frank_etal_2002}
{Finkbeiner}, D.~P., {Schlegel}, D.~J., {Frank}, C., \& {Heiles}, C. 2002,
  \apj, 566, 898

\bibitem[{{Fish} {et~al.}(2003){Fish}, {Reid}, {Argon}, \&
  {Menten}}]{Fish+Reid+Argon_etal_2003}
{Fish}, V.~L., {Reid}, M.~J., {Argon}, A.~L., \& {Menten}, K.~M. 2003, \apj,
  596, 328

\bibitem[{{Fitzpatrick}(1999)}]{Fitzpatrick_1999}
{Fitzpatrick}, E.~L. 1999, \pasp, 111, 63

\bibitem[{{Fitzpatrick}(2004)}]{Fitzpatrick_2004}
{Fitzpatrick}, E.~L. 2004, in Astrophysics of Dust, ed. A.~N. {Witt}, G.~C.
  {Clayton}, \& B.~T. {Draine}, ASP Conference Series 309, 33

\bibitem[{{Fixsen} \& {Dwek}(2002)}]{Fixsen+Dwek_2002}
{Fixsen}, D.~J., \& {Dwek}, E. 2002, \apj, 578, 1009

\bibitem[{{Fosalba} {et~al.}(2002){Fosalba}, {Lazarian}, {Prunet}, \&
  {Tauber}}]{Fosalba+Lazarian+Prunet_etal_2002}
{Fosalba}, P., {Lazarian}, A., {Prunet}, S., \& {Tauber}, J.~A. 2002, \apj,
  564, 762

\bibitem[{{Frick} {et~al.}(2001){Frick}, {Stepanov}, {Shukurov}, \&
  {Sokoloff}}]{Frick+Stepanov+Shukurov_etal_2001}
{Frick}, P., {Stepanov}, R., {Shukurov}, A., \& {Sokoloff}, D. 2001, \mnras,
  325, 649

\bibitem[{{Frisch}(2007)}]{Frisch_2007}
{Frisch}, P.~C. 2007, ArXiv e-prints, 0707.2970

\bibitem[{{Frisch} {et~al.}(1999){Frisch}, {Dorschner}, {Geiss}, {Greenberg},
  {Gr{\"u}n}, {Landgraf}, {Hoppe}, {Jones}, {Kr{\"a}tschmer}, {Linde},
  {Morfill}, {Reach}, {Slavin}, {Svestka}, {Witt}, \&
  {Zank}}]{Frisch+Dorschner+Geiss_etal_1999}
{Frisch}, P.~C. {et~al.} 1999, \apj, 525, 492

\bibitem[{{Gaensler} {et~al.}(2005){Gaensler}, {Haverkorn}, {Staveley-Smith},
  {Dickey}, {McClure-Griffiths}, {Dickel}, \&
  {Wolleben}}]{Gaensler+Haverkorn+Staveley-Smith_etal_2005}
{Gaensler}, B.~M., {Haverkorn}, M., {Staveley-Smith}, L., {Dickey}, J.~M.,
  {McClure-Griffiths}, N.~M., {Dickel}, J.~R., \& {Wolleben}, M. 2005, Science,
  307, 1610

\bibitem[{{Gaensler} {et~al.}(2008){Gaensler}, {Madsen}, {Chatterjee}, \&
  {Mao}}]{Gaensler+Madsen+Chatterjee_etal_2008}
{Gaensler}, B.~M., {Madsen}, G.~J., {Chatterjee}, S., \& {Mao}, S.~A. 2008,
  ArXiv e-prints, 0808.2550

\bibitem[{{Gieseler} \& {Jones}(2000)}]{Gieseler+Jones_2000}
{Gieseler}, U.~D.~J., \& {Jones}, T.~W. 2000, \aap, 357, 1133

\bibitem[{{Gold} {et~al.}(2008){Gold}, {Bennett}, {Hill}, {Hinshaw}, {Odegard},
  {Page}, {Spergel}, {Weiland}, {Dunkley}, {Halpern}, {Jarosik}, {Kogut},
  {Komatsu}, {Larson}, {Meyer}, {Nolta}, {Wollack}, \&
  {Wright}}]{Gold+Bennett+Hill_etal_2008}
{Gold}, B. {et~al.} 2008, ArXiv e-prints, 0803.0715

\bibitem[{{Goldreich} \& {Kylafis}(1981)}]{Goldreich+Kylafis_1981}
{Goldreich}, P., \& {Kylafis}, N.~D. 1981, \apjl, 243, L75

\bibitem[{{Goodman} \& {Whittet}(1995)}]{Goodman+Whittet_1995}
{Goodman}, A.~A., \& {Whittet}, D.~C.~B. 1995, \apjl, 455, L181

\bibitem[{{Grogan} {et~al.}(1996){Grogan}, {Dermott}, \&
  {Gustafson}}]{Grogan+Dermott+Gustafson_1996}
{Grogan}, K., {Dermott}, S.~F., \& {Gustafson}, B.~A.~S. 1996, \apj, 472, 812

\bibitem[{{Gr{\"u}n} {et~al.}(1994){Gr{\"u}n}, {Gustafson}, {Mann}, {Baguhl},
  {Morfill}, {Staubach}, {Taylor}, \& {Zook}}]{Grun+Gustafson+Mann_etal_1994}
{Gr{\"u}n}, E., {Gustafson}, B., {Mann}, I., {Baguhl}, M., {Morfill}, G.~E.,
  {Staubach}, P., {Taylor}, A., \& {Zook}, H.~A. 1994, \aap, 286, 915

\bibitem[{{Grun} \& {Srama}(2006)}]{Grun+Srama_etal_2006}
{Grun}, E., \& {Srama}, R. 2006, in European Planetary Science Congress 2006,
  292

\bibitem[{{Guth}(1981)}]{Guth_1981}
{Guth}, A.~H. 1981, \prd, 23, 347

\bibitem[{{Hall}(1949)}]{Hall_1949}
{Hall}, J.~S. 1949, Science, 109, 166

\bibitem[{{Han} {et~al.}(1997){Han}, {Manchester}, {Berkhuijsen}, \&
  {Beck}}]{Han+Manchester+Berkhuijsen_etal_1997}
{Han}, J.~L., {Manchester}, R.~N., {Berkhuijsen}, E.~M., \& {Beck}, R. 1997,
  \aap, 322, 98

\bibitem[{{Han} {et~al.}(2006){Han}, {Manchester}, {Lyne}, {Qiao}, \& {van
  Straten}}]{Han+Manchester+Lyne_etal_2006}
{Han}, J.~L., {Manchester}, R.~N., {Lyne}, A.~G., {Qiao}, G.~J., \& {van
  Straten}, W. 2006, \apj, 642, 868

\bibitem[{{Han} \& {Qiao}(1994)}]{Han+Qiao_1994}
{Han}, J.~L., \& {Qiao}, G.~J. 1994, \aap, 288, 759

\bibitem[{{Han} \& {Zhang}(2007)}]{Han+Zhang_2007}
{Han}, J.~L., \& {Zhang}, J.~S. 2007, \aap, 464, 609

\bibitem[{{Hanbury Brown} {et~al.}(1960){Hanbury Brown}, {Davies}, \&
  {Hazard}}]{Hanbury-Brown+Davies+Hazard_1960}
{Hanbury Brown}, R., {Davies}, R.~D., \& {Hazard}, C. 1960, The Observatory,
  80, 191

\bibitem[{{Haslam} {et~al.}(1981){Haslam}, {Klein}, {Salter}, {Stoffel},
  {Wilson}, {Cleary}, {Cooke}, \& {Thomasson}}]{Haslam+Klein+Salter_etal_1981}
{Haslam}, C.~G.~T., {Klein}, U., {Salter}, C.~J., {Stoffel}, H., {Wilson},
  W.~E., {Cleary}, M.~N., {Cooke}, D.~J., \& {Thomasson}, P. 1981, \aap, 100,
  209

\bibitem[{{Haslam} {et~al.}(1982){Haslam}, {Salter}, {Stoffel}, \&
  {Wilson}}]{Haslam+Salter+Stoffel_etal_1982}
{Haslam}, C.~G.~T., {Salter}, C.~J., {Stoffel}, H., \& {Wilson}, W.~E. 1982,
  \aaps, 47, 1

\bibitem[{{Haverkorn}(2005)}]{Haverkorn_2005}
{Haverkorn}, M. 2005, in American Institute of Physics Conference Series, Vol.
  784, Magnetic Fields in the Universe: From Laboratory and Stars to Primordial
  Structures., ed. E.~M. {de Gouveia dal Pino}, G.~{Lugones}, \& A.~{Lazarian},
  308--317

\bibitem[{{Haverkorn} {et~al.}(2008){Haverkorn}, {Brown}, {Gaensler}, \&
  {McClure-Griffiths}}]{Haverkorn+Brown+Gaensler_etal_2008}
{Haverkorn}, M., {Brown}, J.~C., {Gaensler}, B.~M., \& {McClure-Griffiths},
  N.~M. 2008, \apj, 680, 362

\bibitem[{{Haverkorn} {et~al.}(2006){Haverkorn}, {Gaensler},
  {McClure-Griffiths}, {Dickey}, \&
  {Green}}]{Haverkorn+Gaensler+McClure-Griffiths_etal_2006}
{Haverkorn}, M., {Gaensler}, B.~M., {McClure-Griffiths}, N.~M., {Dickey},
  J.~M., \& {Green}, A.~J. 2006, \apjs, 167, 230

\bibitem[{{Heiles}(1995)}]{Heiles_1995}
{Heiles}, C. 1995, in Astronomical Society of the Pacific Conference Series,
  Vol.~80, The Physics of the Interstellar Medium and Intergalactic Medium, ed.
  A.~{Ferrara}, C.~F. {McKee}, C.~{Heiles}, \& P.~R. {Shapiro}, 507

\bibitem[{{Heiles}(1996)}]{Heiles_1996_a}
{Heiles}, C. 1996, in Astronomical Society of the Pacific Conference Series,
  Vol.~97, Polarimetry of the Interstellar Medium, ed. W.~G. {Roberge} \&
  D.~C.~B. {Whittet}, 457

\bibitem[{{Heiles}(2000)}]{Heiles_2000}
{Heiles}, C. 2000, \aj, 119, 923

\bibitem[{{Heitsch} {et~al.}(2001){Heitsch}, {Zweibel}, {Mac Low}, {Li}, \&
  {Norman}}]{Heitsch+Zweibel+Mac-Low_etal_2001}
{Heitsch}, F., {Zweibel}, E.~G., {Mac Low}, M.-M., {Li}, P., \& {Norman}, M.~L.
  2001, \apj, 561, 800

\bibitem[{{Hiltner}(1949)}]{Hiltner_1949}
{Hiltner}, W.~A. 1949, \apj, 109, 471

\bibitem[{{Hinshaw} {et~al.}(2007){Hinshaw}, {Nolta}, {Bennett}, {Bean},
  {Dor{\'e}}, {Greason}, {Halpern}, {Hill}, {Jarosik}, {Kogut}, {Komatsu},
  {Limon}, {Odegard}, {Meyer}, {Page}, {Peiris}, {Spergel}, {Tucker}, {Verde},
  {Weiland}, {Wollack}, \& {Wright}}]{Hinshaw+Nolta+Bennett_etal_2007}
{Hinshaw}, G. {et~al.} 2007, \apjs, 170, 288

\bibitem[{{Hinshaw} {et~al.}(2008){Hinshaw}, {Weiland}, {Hill}, {Odegard},
  {Larson}, {Bennett}, {Dunkley}, {Gold}, {Greason}, {Jarosik}, {Komatsu},
  {Nolta}, {Page}, {Spergel}, {Wollack}, {Halpern}, {Kogut}, {Limon}, {Meyer},
  {Tucker}, \& {Wright}}]{Hinshaw+Weiland+Hill_etal_2008}
------. 2008, ArXiv e-prints, 0803.0732

\bibitem[{{Hirata} \& {Seljak}(2003)}]{Hirata+Seljak_2003}
{Hirata}, C.~M., \& {Seljak}, U. 2003, \prd, 68, 083002

\bibitem[{{Hoang} \& {Lazarian}(2008)}]{Hoang+Lazarian_2008b}
{Hoang}, T., \& {Lazarian}, A. 2008, ArXiv e-prints, 0801.0266

\bibitem[{{Hu} \& {Okamoto}(2002)}]{Hu+Okamoto_2002}
{Hu}, W., \& {Okamoto}, T. 2002, \apj, 574, 566

\bibitem[{{Jansson} {et~al.}(2007){Jansson}, {Farrar}, {Waelkens}, \&
  {Ensslin}}]{Jansson+Farrar+Waelkens_etal_2007}
{Jansson}, R., {Farrar}, G.~R., {Waelkens}, A., \& {Ensslin}, T.~A. 2007, ArXiv
  e-prints, 0708.2714

\bibitem[{{Jenkins}(2004)}]{Jenkins_2004}
{Jenkins}, E.~B. 2004, in Origin and Evolution of the Elements, ed. {McWilliam}
  \& {Rauch}, 336

\bibitem[{{Johnston} {et~al.}(2007){Johnston}, {Bailes}, {Bartel}, {Baugh},
  {Bietenholz}, {Blake}, {Braun}, {Brown}, {Chatterjee}, {Darling}, {Deller},
  {Dodson}, {Edwards}, {Ekers}, {Ellingsen}, {Feain}, {Gaensler}, {Haverkorn},
  {Hobbs}, {Hopkins}, {Jackson}, {James}, {Joncas}, {Kaspi}, {Kilborn},
  {Koribalski}, {Kothes}, {Landecker}, {Lenc}, {Lovell}, {Macquart},
  {Manchester}, {Matthews}, {McClure-Griffiths}, {Norris}, {Pen}, {Phillips},
  {Power}, {Protheroe}, {Sadler}, {Schmidt}, {Stairs}, {Staveley-Smith},
  {Stil}, {Taylor}, {Tingay}, {Tzioumis}, {Walker}, {Wall}, \&
  {Wolleben}}]{Johnston+Bailes+Bartel_etal_2007}
{Johnston}, S. {et~al.} 2007, Publications of the Astronomical Society of
  Australia, 24, 174

\bibitem[{{Jones} \& {Spitzer}(1967)}]{Jones+Spitzer_1967}
{Jones}, R.~V., \& {Spitzer}, L.~J. 1967, \apj, 147, 943

\bibitem[{{Kim} {et~al.}(1996){Kim}, {Olinto}, \&
  {Rosner}}]{Kim+Olinto+Rosner_1996}
{Kim}, E.-J., {Olinto}, A.~V., \& {Rosner}, R. 1996, \apj, 468, 28

\bibitem[{{Kim} \& {Martin}(1995)}]{Kim+Martin_1995}
{Kim}, S.-H., \& {Martin}, P.~G. 1995, \apj, 444, 293

\bibitem[{{Kimura} \& {Mann}(1998)}]{Kimura+Mann_1998}
{Kimura}, H., \& {Mann}, I. 1998, \apj, 499, 454

\bibitem[{{Kogut} {et~al.}(2007){Kogut}, {Dunkley}, {Bennett}, {Dor{\'e}},
  {Gold}, {Halpern}, {Hinshaw}, {Jarosik}, {Komatsu}, {Nolta}, {Odegard},
  {Page}, {Spergel}, {Tucker}, {Weiland}, {Wollack}, \&
  {Wright}}]{Kogut+Dunkley+Bennett_etal_2007}
{Kogut}, A. {et~al.} 2007, \apj, 665, 355

\bibitem[{{Komatsu} {et~al.}(2008){Komatsu}, {Dunkley}, {Nolta}, {Bennett},
  {Gold}, {Hinshaw}, {Jarosik}, {Larson}, {Limon}, {Page}, {Spergel},
  {Halpern}, {Hill}, {Kogut}, {Meyer}, {Tucker}, {Weiland}, {Wollack}, \&
  {Wright}}]{Komatsu+Dunkley+Nolta_etal_2008}
{Komatsu}, E. {et~al.} 2008, ArXiv e-prints, 0803.0547

\bibitem[{{Kr{\"u}ger} \& {Gr{\"u}n}(2008)}]{Kruger+Grun_2008}
{Kr{\"u}ger}, H., \& {Gr{\"u}n}, E. 2008, ArXiv e-prints, 0802.3787

\bibitem[{{La Porta} {et~al.}(2008){La Porta}, {Burigana}, {Reich}, \&
  {Reich}}]{La-Porta+Burigana+Reich_etal_2008}
{La Porta}, L., {Burigana}, C., {Reich}, W., \& {Reich}, P. 2008, \aap, 479,
  641

\bibitem[{{Landgraf}(2000)}]{Landgraf_2000}
{Landgraf}, M. 2000, \jgr, 105, 10303

\bibitem[{{Landgraf} {et~al.}(1999){Landgraf}, {Augustsson}, {Gr{\"u}n}, \&
  {Gustafson}}]{Landgraf+Augustsson+Grun_etal_1999}
{Landgraf}, M., {Augustsson}, K., {Gr{\"u}n}, E., \& {Gustafson}, B.~{\AA}.~S.
  1999, Science, 286, 2319

\bibitem[{{Landgraf} {et~al.}(2000){Landgraf}, {Baggaley}, {Gr{\"u}n},
  {Kr{\"u}ger}, \& {Linkert}}]{Landgraf+Baggaley+Grun_etal_2000}
{Landgraf}, M., {Baggaley}, W.~J., {Gr{\"u}n}, E., {Kr{\"u}ger}, H., \&
  {Linkert}, G. 2000, \jgr, 105, 10343

\bibitem[{{Landgraf} {et~al.}(2003){Landgraf}, {Kr{\"u}ger}, {Altobelli}, \&
  {Gr{\"u}n}}]{Landgraf+Kruger+Altobelli_etal_2003}
{Landgraf}, M., {Kr{\"u}ger}, H., {Altobelli}, N., \& {Gr{\"u}n}, E. 2003,
  Journal of Geophysical Research (Space Physics), 108, 8030

\bibitem[{{Lazarian}(1994)}]{Lazarian_1994}
{Lazarian}, A. 1994, \mnras, 268, 713

\bibitem[{{Lazarian}(1995)}]{Lazarian_1995}
------. 1995, \mnras, 277, 1235

\bibitem[{{Lazarian}(1997)}]{Lazarian_1997}
------. 1997, \mnras, 288, 609

\bibitem[{{Lazarian}(2007)}]{Lazarian_2007}
------. 2007, Journal of Quantitative Spectroscopy and Radiative Transfer, 106,
  225

\bibitem[{{Lazarian} \& {Draine}(1997)}]{Lazarian+Draine_1997}
{Lazarian}, A., \& {Draine}, B.~T. 1997, \apj, 487, 248

\bibitem[{{Lazarian} \& {Draine}(2000)}]{Lazarian+Draine_2000}
------. 2000, \apjl, 536, L15

\bibitem[{{Lazarian} \& {Hoang}(2007)}]{Lazarian+Hoang_2007a}
{Lazarian}, A., \& {Hoang}, T. 2007, \mnras, 378, 910

\bibitem[{{Lazarian} \& {Roberge}(1997)}]{Lazarian+Roberge_1997}
{Lazarian}, A., \& {Roberge}, W.~G. 1997, \apj, 484, 230

\bibitem[{{Li} {et~al.}(1991){Li}, {Wheeler}, {Bash}, \&
  {Jefferys}}]{Li+Wheeler+Bash_etal_1991}
{Li}, Z., {Wheeler}, J.~C., {Bash}, F.~N., \& {Jefferys}, W.~H. 1991, \apj,
  378, 93

\bibitem[{{Magalh{\~a}es} {et~al.}(2005){Magalh{\~a}es}, {Pereyra},
  {Melgarejo}, {de Matos}, {Carciofi}, {Benedito}, {Valentim}, {Vidotto}, {da
  Silva}, {de Souza}, {Faria}, \&
  {Gabriel}}]{Magalhaes+Pereyra+Melgarejo_etal_2005}
{Magalh{\~a}es}, A.~M. {et~al.} 2005, in Astronomical Society of the Pacific
  Conference Series, Vol. 343, Astronomical Polarimetry: Current Status and
  Future Directions, ed. A.~{Adamson}, C.~{Aspin}, C.~{Davis}, \&
  T.~{Fujiyoshi}, 305

\bibitem[{{Martin}(1995)}]{Martin_1995}
{Martin}, P.~G. 1995, \apjl, 445, L63

\bibitem[{{Martin} {et~al.}(1992){Martin}, {Adamson}, {Whittet}, {Hough},
  {Bailey}, {Kim}, {Sato}, {Tamura}, \&
  {Yamashita}}]{Martin+Adamson+Whittet_etal_1992}
{Martin}, P.~G. {et~al.} 1992, \apj, 392, 691

\bibitem[{{Martin} {et~al.}(1999){Martin}, {Clayton}, \&
  {Wolff}}]{Martin+Clayton+Wolff_1999}
{Martin}, P.~G., {Clayton}, G.~C., \& {Wolff}, M.~J. 1999, \apj, 510, 905

\bibitem[{{Mason} {et~al.}(2008){Mason}, {Robishaw}, \&
  {Finkbeiner}}]{Mason+Robishaw+Finkbeiner_2008}
{Mason}, B., {Robishaw}, T., \& {Finkbeiner}, D. 2008, in Astronomical Society
  of the Pacific Conference Series, Vol. 395, Frontiers of Astrophysics: A
  Celebration of NRAO's 50th Anniversary, ed. A.~H. {Bridle}, J.~J. {Condon},
  \& G.~C. {Hunt}, 373

\bibitem[{{Mathis}(1986)}]{Mathis_1986}
{Mathis}, J.~S. 1986, \apj, 308, 281

\bibitem[{{Mathis} {et~al.}(1983){Mathis}, {Mezger}, \&
  {Panagia}}]{Mathis+Mezger+Panagia_1983}
{Mathis}, J.~S., {Mezger}, P.~G., \& {Panagia}, N. 1983, \aap, 128, 212

\bibitem[{{McClure-Griffiths} {et~al.}(2005){McClure-Griffiths}, {Dickey},
  {Gaensler}, {Green}, {Haverkorn}, \&
  {Strasser}}]{McClure-Griffiths+Dickey+Gaensler_etal_2005}
{McClure-Griffiths}, N.~M., {Dickey}, J.~M., {Gaensler}, B.~M., {Green}, A.~J.,
  {Haverkorn}, M., \& {Strasser}, S. 2005, \apjs, 158, 178

\bibitem[{{Men} {et~al.}(2008){Men}, {Ferri{\`e}re}, \&
  {Han}}]{Men+Ferriere+Han_2008}
{Men}, H., {Ferri{\`e}re}, K., \& {Han}, J.~L. 2008, \aap, 486, 819

\bibitem[{{Minter} \& {Spangler}(1996)}]{Minter+Spangler_1996}
{Minter}, A.~H., \& {Spangler}, S.~R. 1996, \apj, 458, 194

\bibitem[{{Miville-Deschenes} {et~al.}(2008){Miville-Deschenes}, {Ysard},
  {Lavabre}, {Ponthieu}, {Macias-Perez}, {Aumont}, \&
  {Bernard}}]{Miville-Deschenes+Ysard+Lavabre_etal_2008}
{Miville-Deschenes}, M.-A., {Ysard}, N., {Lavabre}, A., {Ponthieu}, N.,
  {Macias-Perez}, J.~F., {Aumont}, J., \& {Bernard}, J.~P. 2008, ArXiv
  e-prints, 0802.3345

\bibitem[{{Mukai}(1981)}]{Mukai_1981}
{Mukai}, T. 1981, \aap, 99, 1

\bibitem[{{Nishiyama} {et~al.}(2008){Nishiyama}, {Tamura}, {Hatano}, {Kanai},
  {Kurita}, {Sato}, {Matsunaga}, {Nagata}, {Nagayama}, {Kandori}, {Nakajima},
  {Kusakabe}, {Sato}, {Hough}, {Sugitani}, \&
  {Okuda}}]{Nishiyama+Tamura+Hatano_etal_2008}
{Nishiyama}, S. {et~al.} 2008, ArXiv e-prints, 0809.3089

\bibitem[{{Noutsos} {et~al.}(2008){Noutsos}, {Johnston}, {Kramer}, \&
  {Karastergiou}}]{Noutsos+Johnston+Kramer_etal_2008}
{Noutsos}, A., {Johnston}, S., {Kramer}, M., \& {Karastergiou}, A. 2008,
  \mnras, 386, 1881

\bibitem[{{Ohno} \& {Shibata}(1993)}]{Ohno+Shibata_1993}
{Ohno}, H., \& {Shibata}, S. 1993, \mnras, 262, 953

\bibitem[{{Page} {et~al.}(2007){Page}, {Hinshaw}, {Komatsu}, {Nolta},
  {Spergel}, {Bennett}, {Barnes}, {Bean}, {Dor{\'e}}, {Dunkley}, {Halpern},
  {Hill}, {Jarosik}, {Kogut}, {Limon}, {Meyer}, {Odegard}, {Peiris}, {Tucker},
  {Verde}, {Weiland}, {Wollack}, \& {Wright}}]{Page+Hinshaw+Komatsu_etal_2007}
{Page}, L. {et~al.} 2007, \apjs, 170, 335

\bibitem[{{Pearson}(2007)}]{Pearson_2007}
{Pearson}, T.~J. 2007, in Bulletin of the American Astronomical Society,
  Vol.~38, 883

\bibitem[{{Peterson} \& {Webber}(2002)}]{Peterson+Webber_2002}
{Peterson}, J.~D., \& {Webber}, W.~R. 2002, \apj, 575, 217

\bibitem[{{Phillipps} {et~al.}(1981){Phillipps}, {Kearsey}, {Osborne},
  {Haslam}, \& {Stoffel}}]{Phillipps+Kearsey+Osborne_etal_1981}
{Phillipps}, S., {Kearsey}, S., {Osborne}, J.~L., {Haslam}, C.~G.~T., \&
  {Stoffel}, H. 1981, \aap, 98, 286

\bibitem[{{Ponthieu} {et~al.}(2005){Ponthieu}, {Mac{\'{\i}}as-P{\'e}rez},
  {Tristram}, {Ade}, {Amblard}, {Ansari}, {Aumont}, {Aubourg}, {Beno{\^i}t},
  {Bernard}, {Blanchard}, {Bock}, {Bouchet}, {Bourrachot}, {Camus}, {Cardoso},
  {Couchot}, {de Bernardis}, {Delabrouille}, {D{\'e}sert}, {Douspis},
  {Dumoulin}, {Filliatre}, {Fosalba}, {Giard}, {Giraud-H{\'e}raud}, {Gispert},
  {Grain}, {Guglielmi}, {Hamilton}, {Hanany}, {Henrot-Versill{\'e}}, {Kaplan},
  {Lagache}, {Lange}, {Madet}, {Maffei}, {Masi}, {Mayet}, {Nati}, {Patanchon},
  {Perdereau}, {Plaszczynski}, {Piat}, {Prunet}, {Puget}, {Renault}, {Rosset},
  {Santos}, {Vibert}, \& {Yvon}}]{Ponthieu+Macias-Perez+Tristram_etal_2005}
{Ponthieu}, N. {et~al.} 2005, \aap, 444, 327

\bibitem[{{Purcell}(1979)}]{Purcell_1979}
{Purcell}, E.~M. 1979, \apj, 231, 404

\bibitem[{{Quireza} {et~al.}(2006){Quireza}, {Rood}, {Bania}, {Balser}, \&
  {Maciel}}]{Quireza+Rood+Bania_etal_2006}
{Quireza}, C., {Rood}, R.~T., {Bania}, T.~M., {Balser}, D.~S., \& {Maciel},
  W.~J. 2006, \apj, 653, 1226

\bibitem[{{Rand} \& {Kulkarni}(1989)}]{Rand+Kulkarni_1989}
{Rand}, R.~J., \& {Kulkarni}, S.~R. 1989, \apj, 343, 760

\bibitem[{{Rand} \& {Lyne}(1994)}]{Rand+Lyne_1994}
{Rand}, R.~J., \& {Lyne}, A.~G. 1994, \mnras, 268, 497

\bibitem[{{Ratcliffe}(1959)}]{Ratcliffe_1959}
{Ratcliffe}, J.~A. 1959, {The Magneto-Ionic Theory and Its Applications to the
  Ionosphere} (Cambridge University Press)

\bibitem[{{Reynolds} {et~al.}(1999){Reynolds}, {Haffner}, \&
  {Tufte}}]{Reynolds+Haffner+Tufte_1999}
{Reynolds}, R.~J., {Haffner}, L.~M., \& {Tufte}, S.~L. 1999, \apjl, 525, L21

\bibitem[{{Roberge} {et~al.}(1993){Roberge}, {Degraff}, \&
  {Flaherty}}]{Roberge+Degraff+Flaherty_1993}
{Roberge}, W.~G., {Degraff}, T.~A., \& {Flaherty}, J.~E. 1993, \apj, 418, 287

\bibitem[{{Roberge} \& {Lazarian}(1999)}]{Roberge+Lazarian_1999}
{Roberge}, W.~G., \& {Lazarian}, A. 1999, \mnras, 305, 615

\bibitem[{{Rybicki} \& {Lightman}(1979)}]{Rybicki+Lightman_1979}
{Rybicki}, G.~B., \& {Lightman}, A.~P. 1979, {Radiative Processes in
  Astrophysics} (New York, Wiley-Interscience)

\bibitem[{{Schlegel} {et~al.}(1998){Schlegel}, {Finkbeiner}, \&
  {Davis}}]{Schlegel+Finkbeiner+Davis_1998}
{Schlegel}, D.~J., {Finkbeiner}, D.~P., \& {Davis}, M. 1998, \apj, 500, 525

\bibitem[{{Serkowski}(1973)}]{Serkowski_1973}
{Serkowski}, K. 1973, in Interstellar Dust and Related Topics, ed. J.~M.
  {Greenberg} \& H.~C. {van de Hulst}, IAU Symposium 52, 145

\bibitem[{{Simard-Normandin} \&
  {Kronberg}(1979)}]{Simard-Normandin+Kronberg_1979}
{Simard-Normandin}, M., \& {Kronberg}, P.~P. 1979, \nat, 279, 115

\bibitem[{{Simard-Normandin} \&
  {Kronberg}(1980)}]{Simard-Normandin+Kronberg_1980}
------. 1980, \apj, 242, 74

\bibitem[{{Slavin} \& {Frisch}(2008)}]{Slavin+Frisch_2008}
{Slavin}, J.~D., \& {Frisch}, P.~C. 2008, \aap, 491, 53

\bibitem[{{Smith} {et~al.}(2008){Smith}, {Komatsu}, {Nolta}, {Spergel},
  {Larson}, {Hinshaw}, {Page}, {Bennett}, {Gold}, {Jarosik}, {Weiland},
  {Halpern}, {Hill}, {Kogut}, {Limon}, {Meyer}, {Tucker}, {Wollack}, \&
  {Wright}}]{CMBPol_Smith_2008}
{Smith}, K.~M. {et~al.} 2008, ArXiv e-prints, 0811.3916

\bibitem[{{Smoot} {et~al.}(1992){Smoot}, {Bennett}, {Kogut}, {Wright}, {Aymon},
  {Boggess}, {Cheng}, {de Amici}, {Gulkis}, {Hauser}, {Hinshaw}, {Jackson},
  {Janssen}, {Kaita}, {Kelsall}, {Keegstra}, {Lineweaver}, {Loewenstein},
  {Lubin}, {Mather}, {Meyer}, {Moseley}, {Murdock}, {Rokke}, {Silverberg},
  {Tenorio}, {Weiss}, \& {Wilkinson}}]{Smoot+Bennett+Kogut_etal_1992}
{Smoot}, G.~F. {et~al.} 1992, \apjl, 396, L1

\bibitem[{{Spitzer} \& {McGlynn}(1979)}]{Spitzer+McGlynn_1979}
{Spitzer}, Jr., L., \& {McGlynn}, T.~A. 1979, \apj, 231, 417

\bibitem[{{Srama} {et~al.}(2008){Srama}, {Stephan}, {Gr{\"u}n}, {Pailer},
  {Kearsley}, {Graps}, {Laufer}, {Ehrenfreund}, {Altobelli}, {Altwegg}, {Auer},
  {Baggaley}, {Burchell}, {Carpenter}, {Colangeli}, {Esposito}, {Green},
  {Henkel}, {Horanyi}, {J{\"a}ckel}, {Kempf}, {McBride},
  {Moragas-Klostermeyer}, {Kr{\"u}ger}, {Palumbo}, {Srowig}, {Trieloff},
  {Tsou}, {Sternovsky}, {Zeile}, \& {R{\"o}ser}}]{Srama+Stephan+Grun_etal_2008}
{Srama}, R. {et~al.} 2008, Experimental Astronomy, 35

\bibitem[{{Strong} {et~al.}(2007){Strong}, {Moskalenko}, \&
  {Ptuskin}}]{Strong+Moskalenko+Ptuskin_2007}
{Strong}, A.~W., {Moskalenko}, I.~V., \& {Ptuskin}, V.~S. 2007, Annual Review
  of Nuclear and Particle Science, 57, 285

\bibitem[{{Strong} {et~al.}(2000){Strong}, {Moskalenko}, \&
  {Reimer}}]{Strong+Moskalenko+Reimer_2000}
{Strong}, A.~W., {Moskalenko}, I.~V., \& {Reimer}, O. 2000, \apj, 537, 763

\bibitem[{{Sun} {et~al.}(2008){Sun}, {Reich}, {Waelkens}, \&
  {En{\ss}lin}}]{Sun+Reich+Waelkens_etal_2008}
{Sun}, X.~H., {Reich}, W., {Waelkens}, A., \& {En{\ss}lin}, T.~A. 2008, \aap,
  477, 573

\bibitem[{{Taylor} {et~al.}(2003){Taylor}, {Gibson}, {Peracaula}, {Martin},
  {Landecker}, {Brunt}, {Dewdney}, {Dougherty}, {Gray}, {Higgs}, {Kerton},
  {Knee}, {Kothes}, {Purton}, {Uyaniker}, {Wallace}, {Willis}, \&
  {Durand}}]{Taylor+Gibson+Peracaula_etal_2003}
{Taylor}, A.~R. {et~al.} 2003, \aj, 125, 3145

\bibitem[{{Tegmark} \& {Efstathiou}(1996)}]{Tegmark+Efstathiou_1996}
{Tegmark}, M., \& {Efstathiou}, G. 1996, \mnras, 281, 1297

\bibitem[{{Thomson} \& {Nelson}(1980)}]{Thomson+Nelson_1980}
{Thomson}, R.~C., \& {Nelson}, A.~H. 1980, \mnras, 191, 863

\bibitem[{{Vaillancourt}(2007)}]{Vaillancourt_2007}
{Vaillancourt}, J.~E. 2007, in EAS Publications Series, Vol.~23, Sky
  Polarisation at Far-Infrared to Radio Wavelengths, ed. M.-A.
  {Miville-Desch{\^e}nes} \& F.~{Boulanger}, 147--164

\bibitem[{{Vall{\'e}e}(2002)}]{Vallee_2002}
{Vall{\'e}e}, J.~P. 2002, \apj, 566, 261

\bibitem[{{Waelkens} {et~al.}(2008){Waelkens}, {Jaffe}, {Reinecke}, {Kitaura},
  \& {Ensslin}}]{Waelkens+Jaffe+Reinecke_etal_2008}
{Waelkens}, A., {Jaffe}, T., {Reinecke}, M., {Kitaura}, F.~S., \& {Ensslin},
  T.~A. 2008, ArXiv e-prints, 0807.2262

\bibitem[{{Weingartner} \& {Draine}(2003)}]{Weingartner+Draine_2003}
{Weingartner}, J.~C., \& {Draine}, B.~T. 2003, \apj, 589, 289

\bibitem[{{Weisberg} {et~al.}(2004){Weisberg}, {Cordes}, {Kuan}, {Devine},
  {Green}, \& {Backer}}]{Weisberg+Cordes+Kuan_etal_2004}
{Weisberg}, J.~M., {Cordes}, J.~M., {Kuan}, B., {Devine}, K.~E., {Green},
  J.~T., \& {Backer}, D.~C. 2004, \apjs, 150, 317

\bibitem[{{Whittet}(2003)}]{Whittet_2003}
{Whittet}, D.~C.~B. 2003, {Dust in the Galactic Environment} (Bristol: IOP
  Publishing)

\bibitem[{{Whittet} {et~al.}(2008){Whittet}, {Hough}, {Lazarian}, \&
  {Hoang}}]{Whittet+Hough+Lazarian_etal_2008}
{Whittet}, D.~C.~B., {Hough}, J.~H., {Lazarian}, A., \& {Hoang}, T. 2008, \apj,
  674, 304

\bibitem[{{Witte}(2004)}]{Witte_2004}
{Witte}, M. 2004, \aap, 426, 835

\bibitem[{{Wolleben}(2007)}]{Wolleben_2007}
{Wolleben}, M. 2007, \apj, 664, 349

\bibitem[{{Wolleben} {et~al.}(2006){Wolleben}, {Landecker}, {Reich}, \&
  {Wielebinski}}]{Wolleben+Landecker+Reich_etal_2006}
{Wolleben}, M., {Landecker}, T.~L., {Reich}, W., \& {Wielebinski}, R. 2006,
  \aap, 448, 411

\bibitem[{{Yusef-Zadeh} {et~al.}(1996){Yusef-Zadeh}, {Roberts}, {Goss},
  {Frail}, \& {Green}}]{Yusef-Zadeh+Roberts+Goss_etal_1996}
{Yusef-Zadeh}, F., {Roberts}, D.~A., {Goss}, W.~M., {Frail}, D.~A., \& {Green},
  A.~J. 1996, \apjl, 466, L25

\bibitem[{{Zaldarriaga} {et~al.}(2008){Zaldarriaga}, {Komatsu}, {Nolta},
  {Spergel}, {Larson}, {Hinshaw}, {Page}, {Bennett}, {Gold}, {Jarosik},
  {Weiland}, {Halpern}, {Hill}, {Kogut}, {Limon}, {Meyer}, {Tucker}, {Wollack},
  \& {Wright}}]{CMBPol_Zaldarriaga_2008}
{Zaldarriaga}, M. {et~al.} 2008, ArXiv e-prints, 0811.3918

\bibitem[{{Zweibel} \& {Heiles}(1997)}]{Zweibel+Heiles_1997}
{Zweibel}, E.~G., \& {Heiles}, C. 1997, \nat, 385, 131

\end{thebibliography}
\bibliographystyle{hapj}

\end{document}